\documentclass[%
 reprint,
superscriptaddress,
 amsmath,amssymb,
 aps,pre,hyphens,floatfix
]{revtex4-2}

\usepackage[]{amsmath}
\usepackage{amssymb}
\usepackage{enumerate}
\usepackage[usenames,dvipsnames,svgnames,table]{xcolor}
\usepackage{IEEEtrantools}
\usepackage{graphicx}
\usepackage{array}
\usepackage{wrapfig}
\usepackage{fullpage}
\usepackage{overpic}
\usepackage{verbatim}
\usepackage{float}
\usepackage{neuralnetwork}
\usepackage[caption=false]{subfig}
\usepackage{multirow}
\usepackage{diagbox}

\usepackage[normalem]{ulem}  

\newcommand{\tcb}[1]{\textcolor{blue}{\em #1}}

\begin{document}

\preprint{APS/123-QED}

\title{Machine learning of (1+1)-dimensional directed percolation based on raw and shuffled configurations}

\author{Jianmin Shen}
\affiliation{College of engineering and technology, Baoshan University, Baoshan 678000, China}
\affiliation{Key Laboratory of Quark and Lepton Physics (MOE) and Institute of Particle Physics, Central China Normal University, Wuhan 430079, China}
\author{Wei Li}
\email[]{liw@mail.ccnu.edu.cn}
\affiliation{Key Laboratory of Quark and Lepton Physics (MOE) and Institute of Particle Physics, Central China Normal University, Wuhan 430079, China}
\author{Dian Xu}
\email[]{xudian.work@mails.ccnu.edu.cn}
\author{Yuxiang Yang}
\author{Yanyang Wang}
\author{Feng Gao}
\author{Shanshan Wang}
\affiliation{Key Laboratory of Quark and Lepton Physics (MOE) and Institute of Particle Physics, Central China Normal University, Wuhan 430079, China}
\author{Yueying Zhu}
\affiliation{Research Center of Applied Mathematics and Interdisciplinary Science, Wuhan Textile University, Wuhan 430073, China}
\author{Kui Tuo}
\affiliation{Key Laboratory of Quark and Lepton Physics (MOE) and Institute of Particle Physics, Central China Normal University, Wuhan 430079, China}

\date{\today}

\begin{abstract}
Machine learning (ML) can process large sets of data generated from complex systems, which is ideal for classification tasks as often appeared in critical phenomena. Meanwhile ML techniques have been found effective in detecting critical points, or in a broader sense phase separation, and extracting critical exponents. But there are still many unsolved issues with the ML, one of which is the meaning of hidden variables of unsupervised learning. Some say that the hidden variables and the principal component may contain basic information regarding the order parameter of the system of interest, which sounds plausible but lacks evidence. This study aims at searching for evidence supporting the conjecture that the autoencoder's (AE) single latent variable and PCA's first principal component can only serve as signals related to particle density, which happens to be the order parameter of the non-equilibrium DP model. Indeed, in some phase transition (PT) models the order parameter is the particle density, whereas in some PT models it is not. Having conducted a certain degree of random shuffling on the DP configurations, which are then fed to the neural networks as input, we find that AE's single latent variable and PCA's first principal component can indeed represent particle density. It is found that shuffling does affect the size of maximum cluster in the system, which suggests that the second principal component of the PCA is related to the maximal cluster. This has been supported by changes in the correlation length of the transition system with variations in the shuffle ratio.

\tcb{}
\end{abstract}


\maketitle

\section{intorduction}

Machine learning (ML) \cite{jordan2015machine,goodfellow2016machine} has exhibited significant potential in diverse research areas and has been extensively utilized in various branches of physics. A comprehensive overview of ML applications in physics can be obtained from literature sources \cite{carleo2019machine,mehta2019high}. Many physical systems that are of interest to scientific researchers have been investigated through ML methods, including classical \cite{carrasquilla2017machine,wang2016discovering,zhang2019machine,li2018applications}, nonequilibrium \cite{tang2024learning,jo2021absorbing,shen2021supervised,shen2022transfer,shen2022machine}, quantum \cite{carrasquilla2020machine,sarma2019machine}, and topological \cite{rodriguez2019identifying,deng2017machine} phase transitions. The study of phase transition systems has thrived with ML research, and physical methods and theories have complemented the study of ML principles \cite{huang2021statistical,karniadakis2021physics}. 

Typically, the search for physical quantities that can characterize phase transitions, also known as order parameters, is required when examining phase transition systems. In certain ML cases, neural networks or other algorithms are capable of detecting or obtaining information about the order parameters of phase transition systems \cite{carrasquilla2017machine,wang2016discovering,wetzel2017unsupervised,ponte2017kernel,wetzel2017machine,jadrich2018unsupervised}. However, certain phase transition systems do not possess an apparent (explicit) order parameter (e.g. disordered systems, topological phase transitions), and the order parameter of some systems is not merely a matter of occupied sites or particle density. Hence, the information captured by ML algorithms must be sufficiently described, and systems described by varying order parameters should be handled differently during ML research to ensure the effectiveness and accuracy of the ML methods for the exploration of phase transition problems.

The hidden variables or neurons of autoencoder (AE) neural networks \cite{bourlard1988auto,hinton1994autoencoders,hinton2006reducing} are very useful, which is why neural networks have black box properties. Equipped with so many layers and neurons, autoencoder certainly generates nonlinear activation. However, by extracting neurons encoded from hidden layers, such as one, two, or three dimensions, we may obtain a clustering representation or critical information from the original data. In many studies, some suggest that hidden variables may contain information about the order parameters of physical systems, which still lacks evidence. Wang's paper \cite{wang2016discovering} have shown that the first principal component of Principal Component Analysis (PCA) \cite{pearson1901liii,abdi2010principal} could represent the order parameters of Ising model. After feeding the effective input data, the trained neural network can effectively capture the order parameters of Ising model, $M = \sum_i s_{i}/N$. 

By learning the functioning of autoencoder' single latent variable and PCA' first principal component, we conjecture that for a phase transition model, if its order parameter is particle density in lattice model, the information contained in the hidden neurons of the autoencoder neural network and first principal component of PCA is very likely related to the order parameter. The main objective of this paper is to study the hidden information of neurons of AE or PCA through the well-known model of directed percolation (DP) \cite{hinrichsen2000non,lubeck2004universal} which is a prototype in non-equilibrium phase transition. The order parameter of DP can be expressed by the steady-state particle density, $\rho = \sum_i s_{i}/N$. 

The DP model is a highly regarded non-equilibrium phase transition \cite{hinrichsen2000non,lubeck2004universal,henkel2008non} that is associated with various critical phenomena including epidemic spreading \cite{dickman2008nature,mata2021overview} and catalytic reactions \cite{ziff1986kinetic,grinstein1989critical}, all of which can be classified as the DP universality class. Two models are considered to fall under the same universality class if they possess the same set of critical exponents and scaling functions. In non-equilibrium phase transitions DP plays a role similar to the one Ising model does in equilibrium phase transitions. The most notable characteristic of non-equilibrium phase transitions is the additional time dimension. DP universality class has been identified for non-equilibrium phase transitions, although the analytical solution for (1+1)-dimensional DP is yet to be attainable \cite{hinrichsen2000non,lubeck2004universal}. The study of the DP model has been approached using numerous methods including machine learning. In this article, we will primarily focus on the machine learning aspects of DP and how the hidden information will be explored.

Initially, the autoencoder and PCA techniques will be utilized for the purpose of feature extraction of the original configurations of the (1+1)-dimensional DP, with the objective of obtaining the representation of the single latent variable and the first principal component. It is conjectured, based on the outcome of the computation, that the single latent variable and the first principal component signify particle densities. To validate this conjecture, random shuffling, with a certain ratio, will be made on DP's raw configurations. Thereafter, the detailed results of the autoencoder and principal component analysis of the DP configurations following the shuffle will be compared. The second principal component is observed to change with the shuffle ratio, which can be attributed to changes in the correlation length of the system. In a bid to explore this phenomenon further, we propose retaining the maximum clusters for the DP configurations of shuffle, followed by conducting autoencoder and PCA analysis on the maximum clusters. This approach not only serves to verify that shuffling alters the correlation length of the system, but also effectively validates that the autoencoder's single latent variable and PCA's first principal component can successfully extract the particle density of the DP model. The results obtained lend credence to the notion that the autoencoder's single latent variable and the PCA's first principal component depict the particle density, and can prove to be highly beneficial in studying various systems with Ising- and DP-like phase transitions.

The main structure of this paper is as follows. In Sec.~\ref{Models}, the DP model of interest will be introduced. Sec.~\ref{Methods} gives the two methods used in this paper. Sec.~\ref{Ml_results} includes all the data sets and machine learning results of the DP model. Sec.~\ref{Conclusion} is a summary of this work.

\section{Machine learning of DP}

\subsection{The DP Model}
\label{Models}

DP is a percolation process that takes place in systems where there is a preferred direction of growth or flow. It has been extensively studied in various contexts, including epidemics, traffic flow, and social networks. The fundamental concept of DP involves the creation of clusters comprising particles or individuals that move in a directed manner through a medium. These clusters are expanded by incorporating new particles in the direction of the preferred flow, and any particles that do not propagate may be removed. With time, the clusters may fuse together to form larger ones, and eventually, a spanning cluster may emerge, which is a sign of a phase transition.

One of the primary challenges in studying DP during absorbing phase transitions \cite{hinrichsen2000non,lubeck2004universal,henkel2008non} is to comprehend the correlation between the particle density and the critical behavior of the system. The rate of cluster growth and coalescence is determined by the particle density, which consequently impacts the emergence of a spanning cluster. Moreover, the critical behavior of DP can also be influenced by the dynamics of particle evolution. Hence, it is essential to comprehend the processes of particle density and evolution to investigate DP in absorbing phase transitions.

\begin{figure*}[t]
\begin{tabular}{ccc}
    \includegraphics[width=0.45\textwidth]{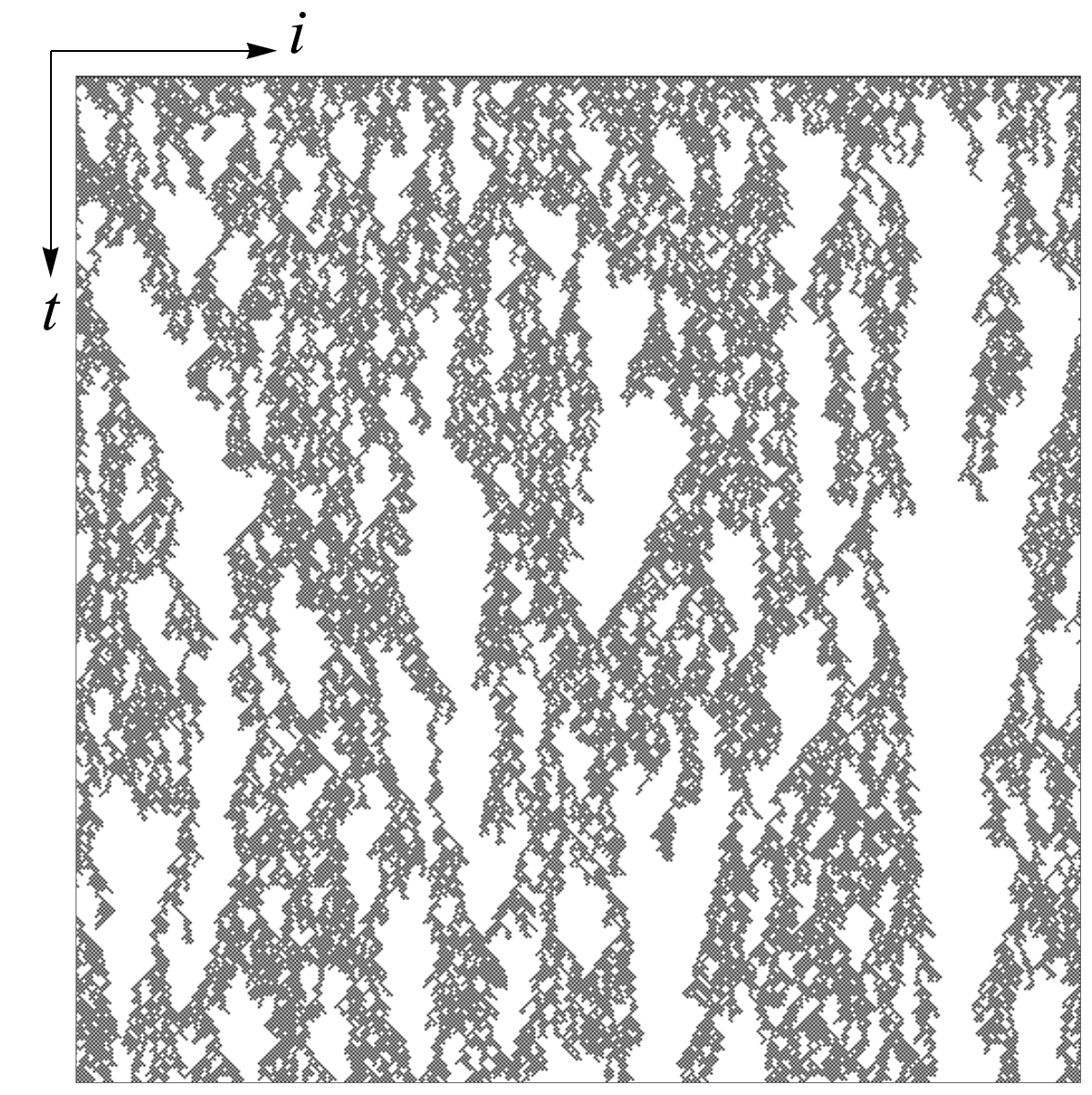} &
    \includegraphics[width=0.45\textwidth]{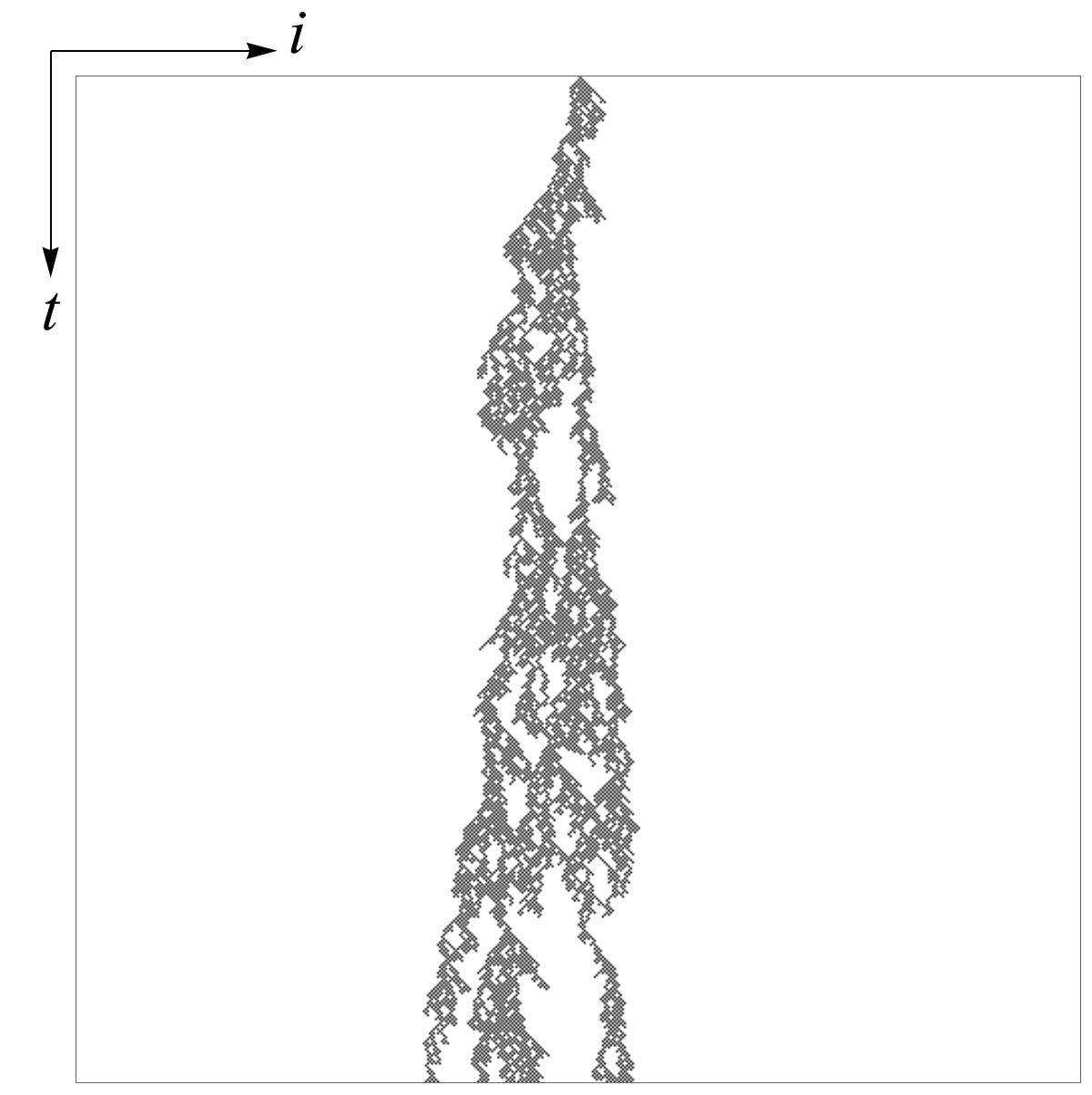} \\
     Fully occupied lattice &  Single active seed
\end{tabular}
\caption{The critical configurations of bond DP in (1+1) dimensions, starting from fully occupied lattice (left panel) and a single active seed (right panel), respectively, where $L = 500$, the time step is $500$ and the bond probabilities $p$ in both panels are $0.6447$.}
\label{DP_configuration}
\end{figure*}

Fig. \ref{DP_configuration} shows the evolution patterns of the (1+1)-dimensional DP model under two different initial conditions, with fully occupied lattice and a single active seed respectively. The absorbing state of DP is that all lattice sites are empty at one single time step. DP is a special example of reaction-diffusion processes of interacting particles where there is no diffusion. Take (1+1)-dimensional DP as an example, its evolution mechanism is a competition between proliferation and annihilation among lattice sites. The specific process of DP model evolution over time can be represented by the following equation,
\begin{equation}
  s_{i}(t+1)=\left\{
\begin{array}{rcl}
1     &      & {if \quad s_{i-1}(t) = 1 \quad and \quad z_{i}^{- } < p,}\\
1     &      & {if \quad s_{i+1}(t) = 1 \quad and \quad z_{i}^{+ } < p,}\\
0     &      & {otherwise,}
\end{array} \right.
\end{equation}
where $s_{i}(t+1)$ represents the state of node $i$ at time $t+1$, $z_{i}^{\pm} \in (0, 1) $ are two random numbers.

Order parameters are important physical quantities in describing the behavior of phase transitions. The order parameter of DP model can be expressed by percolation probability or the steady-state density,
\begin{equation}
  P_{perc}(p)\widetilde{\propto}(p-p_{c})^{\beta^{^{\prime}}},\quad
 \rho_{a}(p)\widetilde{\propto}(p-p_{c})^{\beta}
\end{equation}
where $P_{perc}(p)$ denotes the probability that a site belongs to a percolating cluster, $\rho_{a}$ denotes particle density, and $\beta=\beta^{\prime}$ is resulted by rapidity-reversal symmetry. The simulation proves that the characteristic time of DP evolution in (1+1) dimensions obeys the following relation, 
\begin{equation}
t_c \sim L^z,
\end{equation}
where $L$ denotes lattice size, $z \simeq 1.58$. Note that the particle density at a single time step can be expressed as 
\begin{equation}
\rho_t = \sum_i s_{i}/N.
\label{stational}
\end{equation}
While in the Monte Carlo (MC) \cite{hammersley2013monte} simulation of this paper, it refers to the particle density in the (1+1)-dimensional lattice 
\begin{equation}
\rho_{all} = \sum_i s_{i}/(N*t),
\label{all}
\end{equation}
which contains all time steps.

\subsection{Methods}
\label{Methods}

\subsubsection{Autoencoder}
\begin{figure}
\centering
 \includegraphics[width=0.4\textwidth]{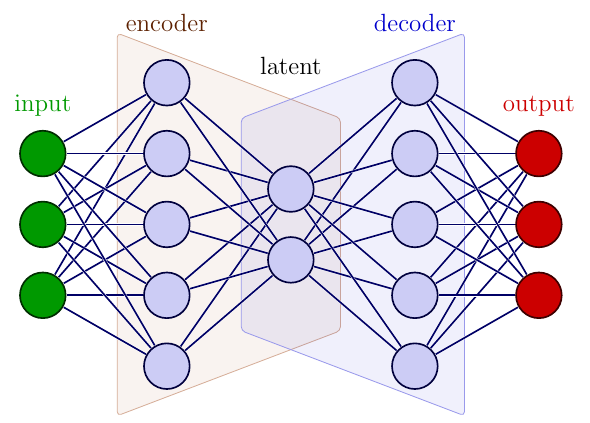}
\caption{ Neural network schematic structure of autoencoder.}
\label{autostruc}
\end{figure}

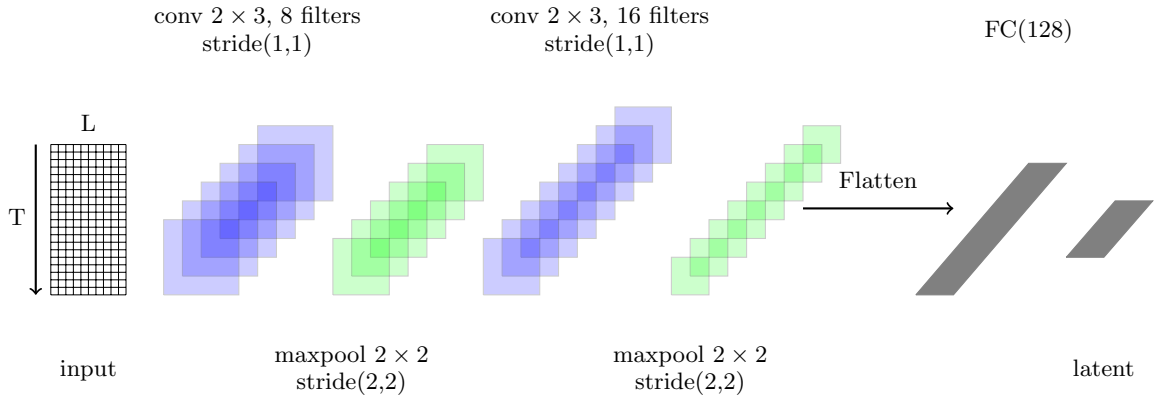
\begin{figure*}[t!]
	\centering
	\begin{tikzpicture}
		\node at (0.5,-1){\begin{tabular}{c} input \end{tabular}};
		\draw[step=0.1cm,black,very thin] (0,0) grid (1,2);
        \node at (0.5,2.25){\begin{tabular}{c} L \end{tabular}};
        \draw[thick,->] (-0.2,2) -- (-0.2,0);
        \node at (-0.45,1){\begin{tabular}{c} T \end{tabular}};
		
		\node at (2.75,3.5){\begin{tabular}{c} conv $2 \times 3$, 8 filters \\ stride(1,1) \end{tabular}};
		
		\draw[fill=blue,opacity=0.2,draw=black] (2.75,1.25) -- (3.75,1.25) -- (3.75,2.25) -- (2.75,2.25) -- (2.75,1.25);
		\draw[fill=blue,opacity=0.2,draw=black] (2.5,1) -- (3.5,1) -- (3.5,2) -- (2.5,2) -- (2.5,1);
		\draw[fill=blue,opacity=0.2,draw=black] (2.25,0.75) -- (3.25,0.75) -- (3.25,1.75) -- (2.25,1.75) -- (2.25,0.75);
		\draw[fill=blue,opacity=0.2,draw=black] (2,0.5) -- (3,0.5) -- (3,1.5) -- (2,1.5) -- (2,0.5);
		\draw[fill=blue,opacity=0.2,draw=black] (1.75,0.25) -- (2.75,0.25) -- (2.75,1.25) -- (1.75,1.25) -- (1.75,0.25);
		\draw[fill=blue,opacity=0.2,draw=black] (1.5,0) -- (2.5,0) -- (2.5,1) -- (1.5,1) -- (1.5,0);
		
		\node at (4.0,-1){\begin{tabular}{c} maxpool $2 \times 2$ \\ stride(2,2)\end{tabular}};
		
		\draw[fill=green,opacity=0.2,draw=black] (5,1.25) -- (5.75,1.25) -- (5.75,2) -- (5,2) -- (5,1.25);
		\draw[fill=green,opacity=0.2,draw=black] (4.75,1) -- (5.5,1) -- (5.5,1.75) -- (4.75,1.75) -- (4.75,1);
		\draw[fill=green,opacity=0.2,draw=black] (4.5,0.75) -- (5.25,0.75) -- (5.25,1.5) -- (4.5,1.5) -- (4.5,0.75);
		\draw[fill=green,opacity=0.2,draw=black] (4.25,0.5) -- (5,0.5) -- (5,1.25) -- (4.25,1.25) -- (4.25,0.5);
		\draw[fill=green,opacity=0.2,draw=black] (4,0.25) -- (4.75,0.25) -- (4.75,1) -- (4,1) -- (4,0.25);
		\draw[fill=green,opacity=0.2,draw=black] (3.75,0) -- (4.5,0) -- (4.5,0.75) -- (3.75,0.75) -- (3.75,0);
		
		\node at (7.3,3.5){\begin{tabular}{c} conv $2 \times 3$, 16 filters \\ stride(1,1) \end{tabular}};
		
		\draw[fill=blue,opacity=0.2,draw=black] (7.5,1.75) -- (8.25,1.75) -- (8.25,2.5) -- (7.5,2.5) -- (7.5,1.75);
		\draw[fill=blue,opacity=0.2,draw=black] (7.25,1.5) -- (8,1.5) -- (8,2.25) -- (7.25,2.25) -- (7.25,1.5);
		\draw[fill=blue,opacity=0.2,draw=black] (7,1.25) -- (7.75,1.25) -- (7.75,2) -- (7,2) -- (7,1.25);
		\draw[fill=blue,opacity=0.2,draw=black] (6.75,1) -- (7.5,1) -- (7.5,1.75) -- (6.75,1.75) -- (6.75,1);
		\draw[fill=blue,opacity=0.2,draw=black] (6.5,0.75) -- (7.25,0.75) -- (7.25,1.5) -- (6.5,1.5) -- (6.5,0.75);
		\draw[fill=blue,opacity=0.2,draw=black] (6.25,0.5) -- (7,0.5) -- (7,1.25) -- (6.25,1.25) -- (6.25,0.5);
		\draw[fill=blue,opacity=0.2,draw=black] (6,0.25) -- (6.75,0.25) -- (6.75,1) -- (6,1) -- (6,0.25);
		\draw[fill=blue,opacity=0.2,draw=black] (5.75,0) -- (6.5,0) -- (6.5,0.75) -- (5.75,0.75) -- (5.75,0);
		
		\node at (8.5,-1){\begin{tabular}{c} maxpool $2 \times 2$ \\ stride(2,2) \end{tabular}};
		
		\draw[fill=green,opacity=0.2,draw=black] (10,1.75) -- (10.5,1.75) -- (10.5,2.25) -- (10,2.25) -- (10,1.75);
		\draw[fill=green,opacity=0.2,draw=black] (9.75,1.5) -- (10.25,1.5) -- (10.25,2) -- (9.75,2) -- (9.75,1.5);
		\draw[fill=green,opacity=0.2,draw=black] (9.5,1.25) -- (10,1.25) -- (10,1.75) -- (9.5,1.75) -- (9.5,1.25);
		\draw[fill=green,opacity=0.2,draw=black] (9.25,1) -- (9.75,1) -- (9.75,1.5) -- (9.25,1.5) -- (9.25,1);
		\draw[fill=green,opacity=0.2,draw=black] (9,0.75) -- (9.5,0.75) -- (9.5,1.25) -- (9,1.25) -- (9,0.75);
		\draw[fill=green,opacity=0.2,draw=black] (8.75,0.5) -- (9.25,0.5) -- (9.25,1) -- (8.75,1) -- (8.75,0.5);
		\draw[fill=green,opacity=0.2,draw=black] (8.5,0.25) -- (9,0.25) -- (9,0.75) -- (8.5,0.75) -- (8.5,0.25);
		\draw[fill=green,opacity=0.2,draw=black] (8.25,0) -- (8.75,0) -- (8.75,0.5) -- (8.25,0.5) -- (8.25,0);
		
		\node at (11,1.5){\begin{tabular}{c} Flatten \end{tabular}};
		\draw[thick,->] (10.,1.15) -- (12.,1.15);
		
		\node at (13,3.5){\begin{tabular}{c} FC(128) \end{tabular}};
		
		\draw[fill=black,draw=black,opacity=0.5] (11.5,0) -- (12.,0) -- (13.5,1.75) -- (13,1.75) -- (11.5,0);
		
		\node at (14,-1){\begin{tabular}{c} latent \end{tabular}};
		
		\draw[fill=black,draw=black,opacity=0.5] (13.5,0.5) -- (14,0.5) -- (14.65,1.25) -- (14.15,1.25) -- (13.5,0.5);
	\end{tikzpicture}
	\caption{Architecture of the encoding process in the autoencoder neural network. The feature maps of the final max-pooling layer are then flattened into a fully connected layer. The last layer is the single latent variable encoded.}
	\label{the_encoding_network}
\end{figure*}

\par The autoencoder (AE) is a neural network that is capable of learning how to compress and reconstruct data. Its fundamental principle involves reducing the input data to a lower-dimensional representation, which can then be utilized for data reconstruction. This representation is known as the "latent space" and it can be viewed as a compressed version of the input data that retains the most significant features.

Autoencoders (AEs) are versatile tools with applications in various fields such as computer vision \cite{voulodimos2018deep}, natural language processing \cite{freitag2018unsupervised}, and anomaly detection \cite{chen2018autoencoder}. In the study of phase transitions, AEs have been utilized to analyze and forecast the behavior of physical systems \cite{shen2021supervised,wetzel2017unsupervised,alexandrou2019unsupervised}. In particular, AEs have been proven useful in identifying the critical properties of phase transitions, which describe how a system's behavior changes in the vicinity of the critical point.

A commonly used method involves acquiring knowledge of the latent space of a physical system via AE. In this approach, the latent variables are believed to be representative of the system's order parameter and other pertinent variables. By scrutinizing the structure of the latent space, one can attain valuable understanding of the system's behavior and even forecast its phase transitions.

\par Fig. \ref{autostruc} provides an overview of the autoencoder structure, which comprises several neural network layers: the input, encoding, hidden, decoding, and output layers. In order to create an effective autoencoder, appropriate neural networks, such as Fully Connected Neural Network (FCN), Convolutionnal Neural Network (CNN), and Long Short Term Memory (LSTM) \cite{goodfellow2016deep,hochreiter1997long}, can be utilized for the encoding and decoding layers. The decoding process of an autoencoder can be perceived as the reverse of the encoding process, and Fig. \ref{the_encoding_network} demonstrates a detailed flow chart of the encoding process used in this study. By minimizing a specific quantity linked to the input and output layers, referred to as the cross-entropy, the original data is compressed and stored in potential neurons. Ultimately, via the latent neurons, a compressed representation of the original configurations is obtained.

Within our autoencoder neural network, each configuration featuring a single bond probability corresponds to respective sample volumes for our training set, verification set, and test set; namely, 2500, 200, and 200. Our verification set serves to finely tune the hyper-parameters, permitting the model to reach its optimal performance.

\par To avoid over-fitting as suggested by Glassner \cite{glassner2018deep}, we incorporate the L2-norm ($\lambda/(2N) \sum\limits_{i} w^{2}_{i}$) into the loss function. To enhance the training of our neural networks, we employ the AdamOptimizer. TensorFlow 1.15 forms the basis for our machine learning implementation.

\subsubsection{PCA}

PCA has been widely applied to phase transitions in various fields, such as physics, chemistry, and materials science. In the context of phase transitions, PCA can be used to extract the most relevant features that capture the critical behavior of the system \cite{wang2016discovering,wetzel2017machine,hu2017discovering}. For example, in the study of Ising model phase transitions, PCA can be used to extract the critical information that governs the phase transition.

Principal Component Analysis (PCA) is a linear unsupervised learning method that aims to find a lower-dimensional representation of high-dimensional data while preserving the most significant variability in the data. PCA is based on the eigen-decomposition of the covariance matrix of the data, where the eigenvectors represent the directions of maximal variance and the eigenvalues represent the amount of variance along each eigenvector. By projecting the data onto a subspace spanned by the first few principal components with the largest eigenvalues, one can obtain a lower-dimensional representation of the data that retains the most relevant information.


\subsection{Machine learning results}
\label{Ml_results}
\begin{figure}
\centering
\includegraphics[width=0.4\textwidth]{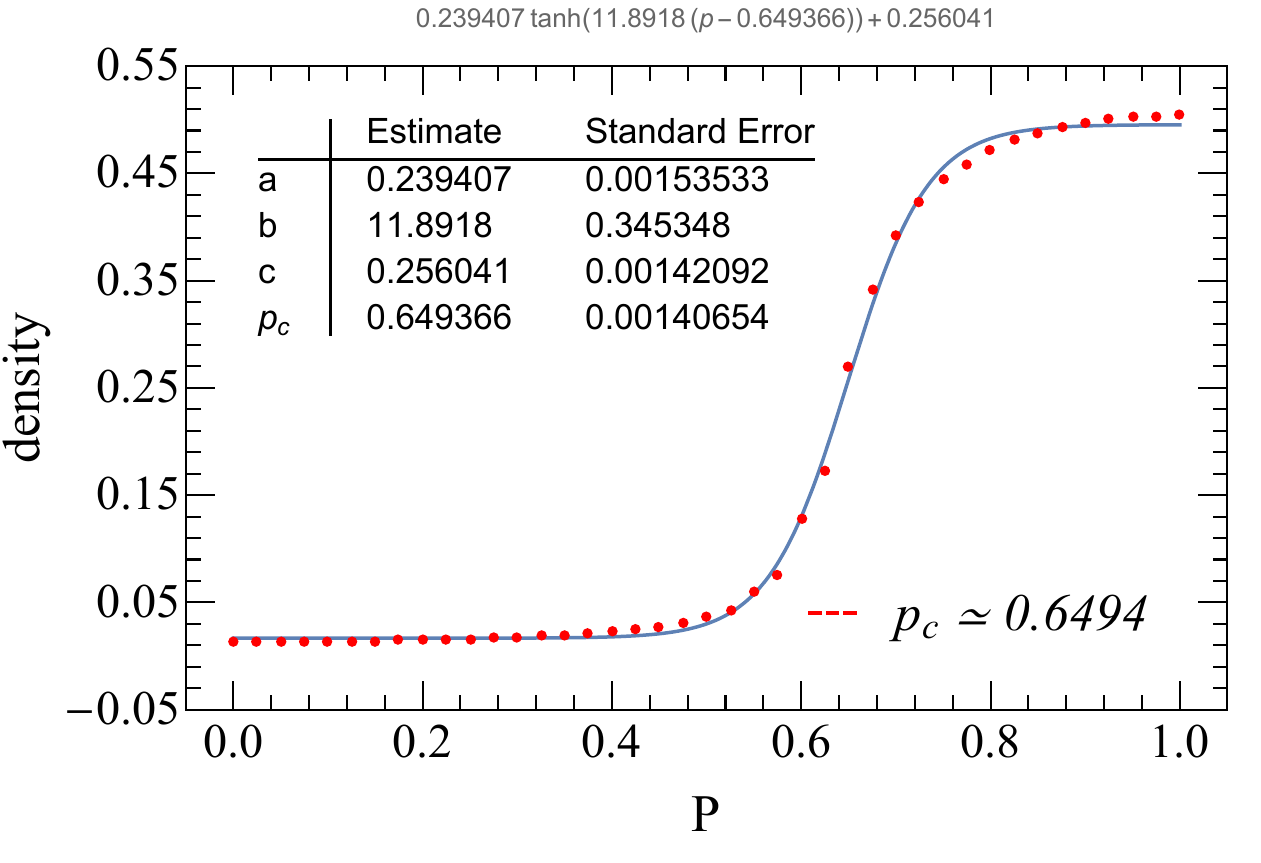}
\caption{Monte Carlo simulations are employed to determine the relationship between the sum of particle densities, $\rho_{all}$, and bond probability $p$ in (1+1)-dimensional bond DP. The simulations are conducted with lattice size $N = 16$, utilizing a total of $120$ time steps for averaging and $100$ ensemble averages.}
\label{mc_result}
\end{figure}

To conduct unsupervised learning of the DP model, it is necessary to first acquire the MC simulation results. Specifically, the stationary particle density $\rho$ can be obtained from a single time step. By varying the parameter $p$, a sum of particle densities containing multiple time steps can be generated, as demonstrated in Fig. \ref{mc_result}. This plot reveals a clear transition behavior occurring at the critical point, which is precisely specified by $p_{c} \simeq 0.649(1)$ for (1+1)-dimensional DP.

\subsubsection{Learning with raw configurations}

\begin{figure*}[h]
\begin{tabular}{cccc}
    \includegraphics[width=0.15\textwidth]{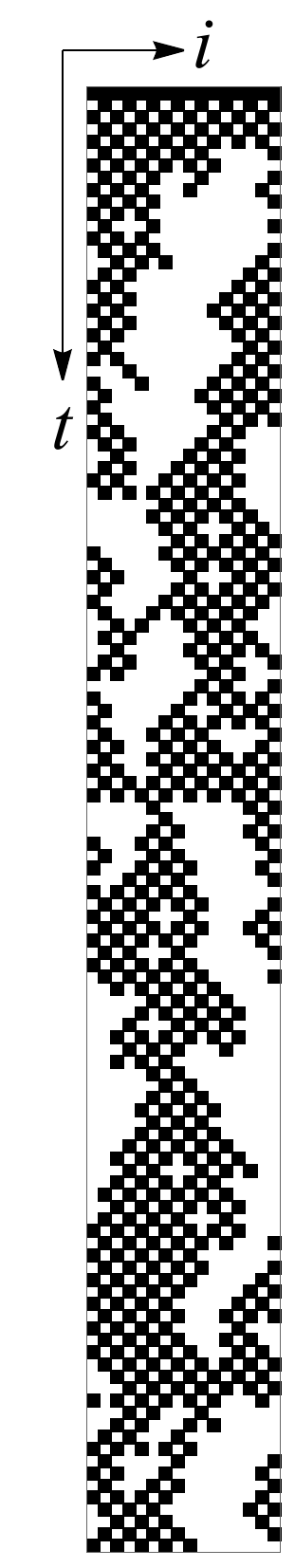} &
    \includegraphics[width=0.15\textwidth]{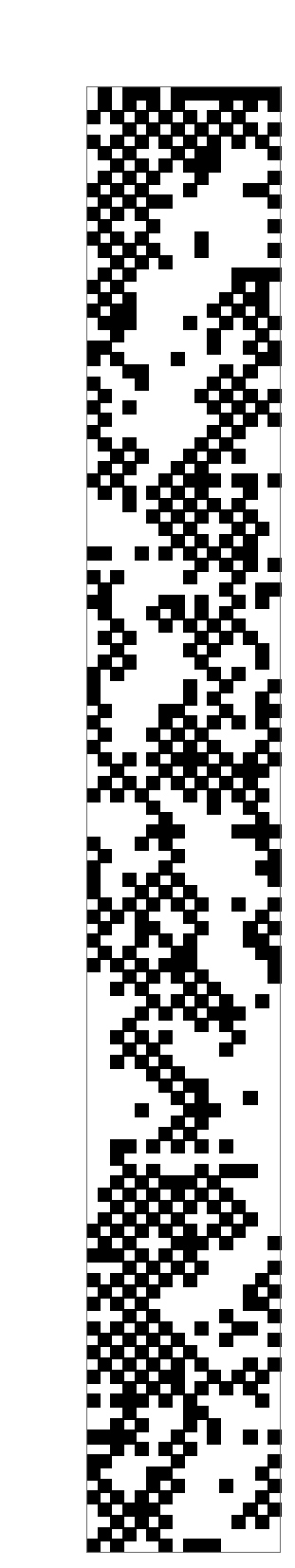} &
    \includegraphics[width=0.15\textwidth]{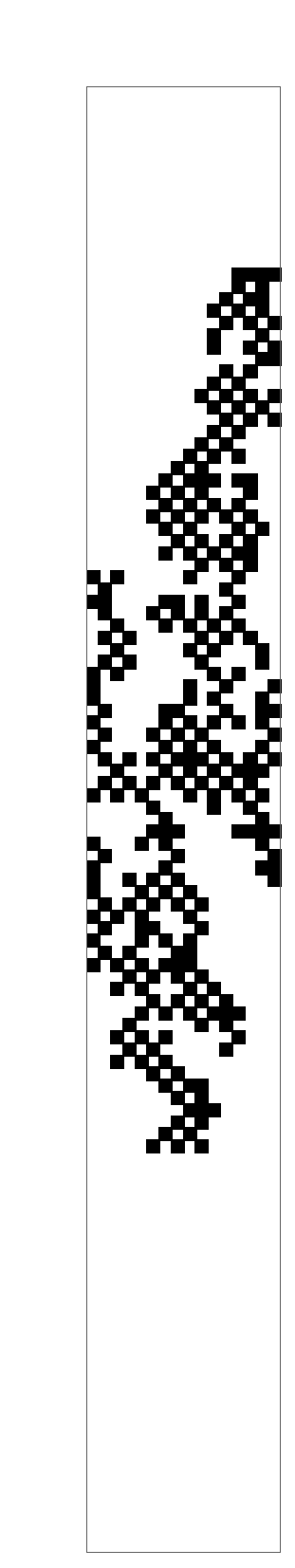} \\
     (a) &  (b) & (c)
\end{tabular}
\caption{Panel \textbf{a} is an example of stochastic critical configurations of bond DP in (1+1) dimensions starting from a fully occupied lattice, where $L = 16$, the time length is $120$ and the corresponding bond probability $p$ is $0.6447$. Panel \textbf{b} is a shuffled version of \textbf{a} with a shuffle ratio $0.2$, where the particle density remains unchanged. Panel \textbf{c} corresponds to the maximum cluster of \textbf{b}.}
\label{raw_shuffle_max_config}
\end{figure*}

In article \cite{shen2021supervised}, it has been illustrated that the (1+1)-dimensional bond DP generates a configuration with a fully occupied lattice as the initial condition. It is important to note that in non-equilibrium lattice models, the system's characteristic time $t_{c}$ is estimated to be $t_{c} \sim L^{z}$, where $z \sim 1.58$ for the (1+1)-DP model. This time scale, for the system to reach the stationary state, is typically much longer than the lattice size. For this reason, we select a lattice size of $L=16$ in our simulations. According to the dynamical exponent, the characteristic time $T_{c} \simeq L^{z} \simeq 79$ can be roughly estimated. However, since the DP is a random sequence updating process, there is no guarantee that the system will reach an absorbing state even if $t \geq T_{c}$, as the particle configuration may display fluctuating behavior even close to the critical point. To allow for better particle information learning, we opt to utilize a time length of $t=120$ for machine learning in (1+1)-dimensional bond DP. Fig. \ref{raw_shuffle_max_config} (a) displays the raw configurations that have been used.

\begin{figure*}[htbp]
\begin{tabular}{cc}
    \includegraphics[width=0.45\textwidth]{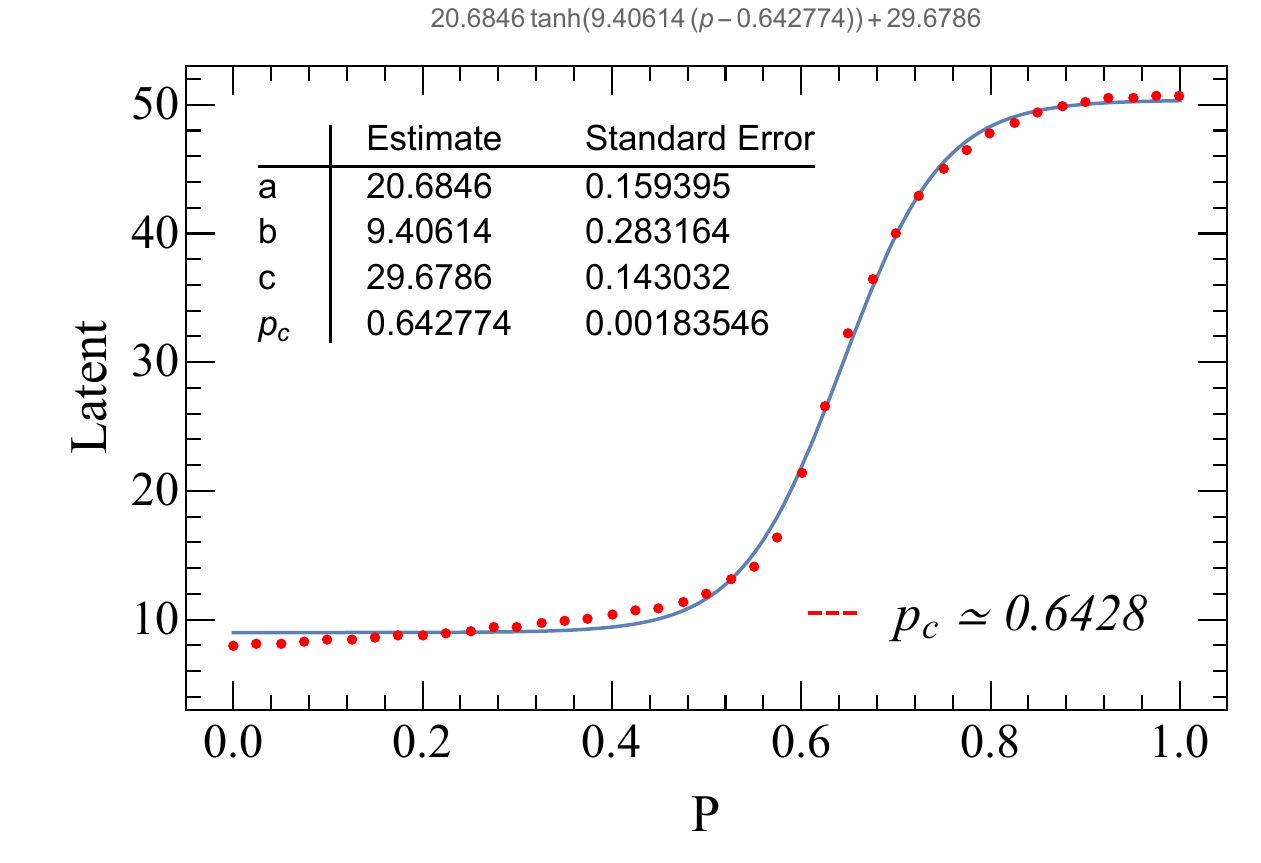} &
    $\qquad$\includegraphics[width=0.45\textwidth]{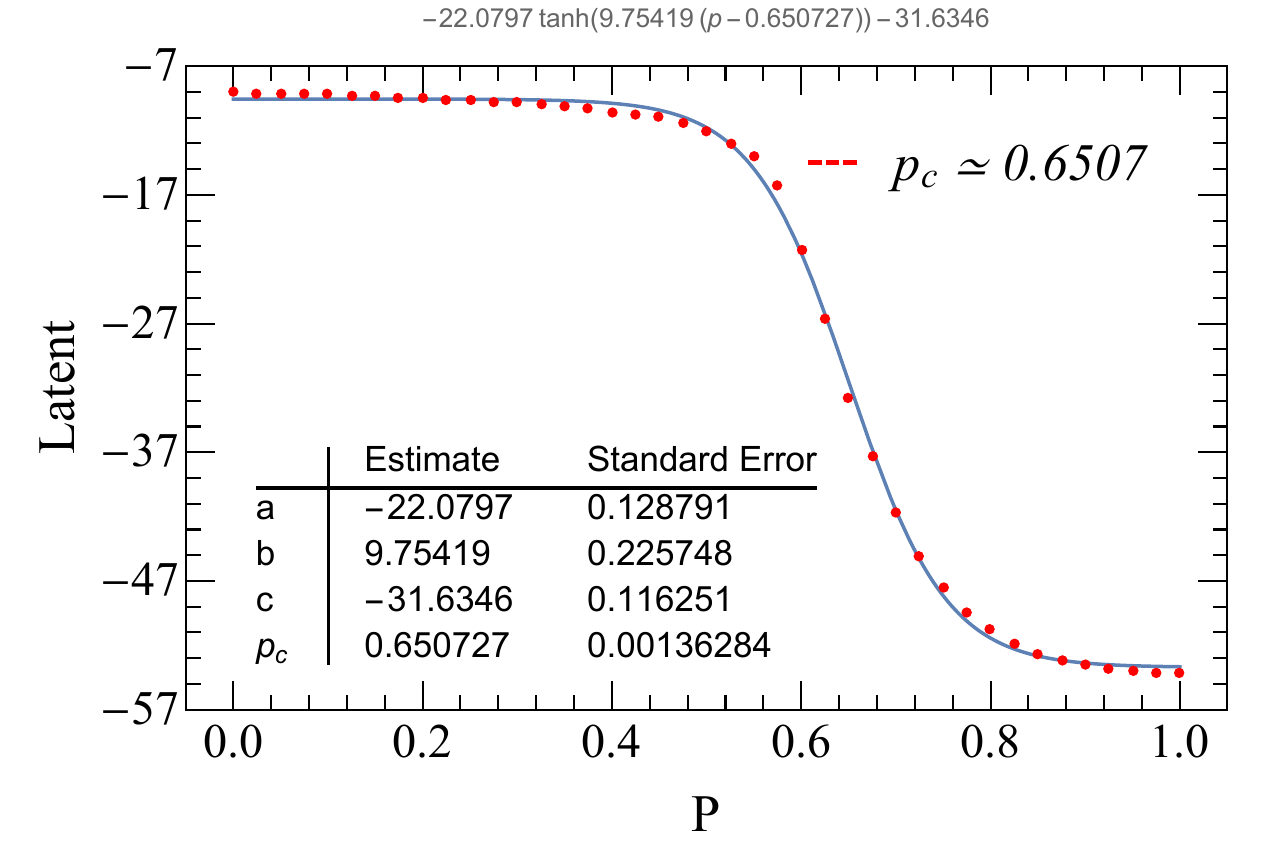} \\
    (a) & $\qquad$ (b)\\
    \includegraphics[width=0.45\textwidth]{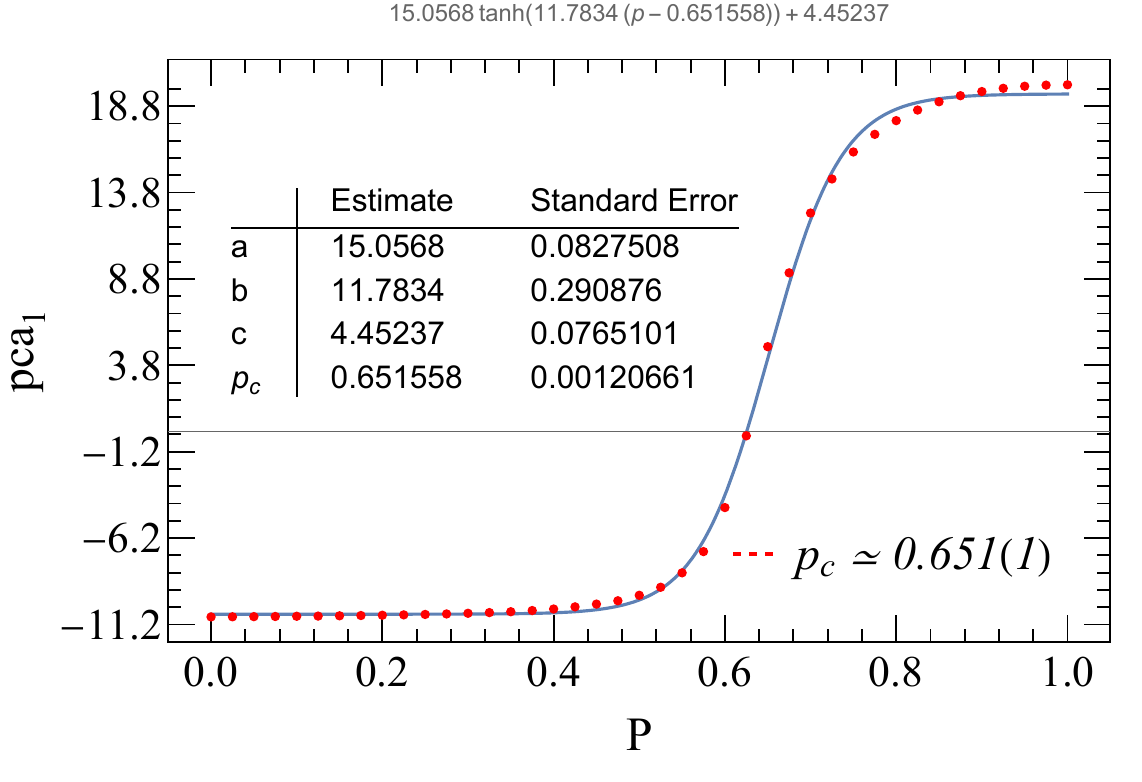} &
    $\qquad$ \includegraphics[width=0.45\textwidth]{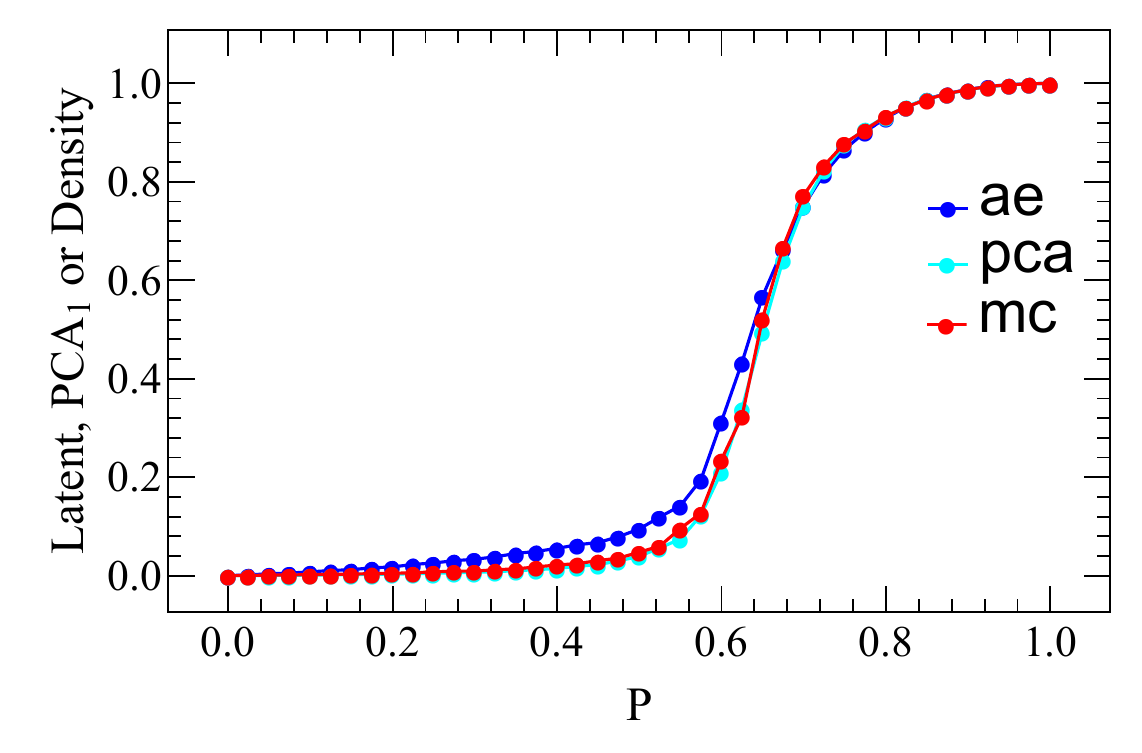} \\
    (c) & $\qquad$(d)
\end{tabular}
\caption{ML raw configurations of (1+1)-dimensional DP by 
autoencoder and PCA. \textbf{a} and \textbf{b} are two similar autoencoder results of (1+1)-dimensional bond DP, using the same data set and convolutional autoencoder neural network. Encoding of the raw DP configurations with a single hidden neuron activation $Latent$ as a function of the bond probability. Each data point of the bond probability is averaged over $200$ testing samples. \textbf{c} and is the PCA result of (1+1)-dimensional DP from the first principal component of the raw DP configurations as a function of the bond probability. \textbf{d}, The rescaled hidden variable $h$ of autoencoder, first principal component of PCA and $rho$ of MC simulations versus $p$, represented by blue, cyan and red colors, respectively. }
\label{ae_pca_mc_dp_raw}
\end{figure*}

As witnessed in previous research, DP configurations may be clustered by an autoencoder through the representation of two hidden neurons, though this is not the primary focus of this work. Rather, we aim to explore the potential of a single hidden neuron in gathering information pertaining to the phase transition. The upper row of Fig. \ref{ae_pca_mc_dp_raw} depicts the autoencoder outcomes of (1+1)-dimensional bond DP's raw configurations. Specifically, Fig. \ref{ae_pca_mc_dp_raw} (a) and Fig. \ref{ae_pca_mc_dp_raw} (b) represent two distinct outputs generated from the same data set and hyper-parameters of the convolutional autoencoder neural network. Both models yield the critical value of DP model through hyperbolic tangent fittings, i.e. $p_{c} \simeq 0.643(2)$ and $p_{c} \simeq 0.651(1)$ for Fig. \ref{ae_pca_mc_dp_raw} (a) and (b), respectively. These predictions conforms to the commonly accepted value of $p_{c} \simeq 0.6447$ \cite{henkel2008non}, and the error falls within 2\textperthousand. These results demonstrate the reliability of the autoencoder approach for identifying the critical point. While finite-size scaling could further enhance measurement accuracy, this technique's nuances will not be discussed further in this work.

The autoencoder neural network is a highly prevalent technique in the field of unsupervised learning. However, with the utilization of neural networks, the outcome of our analysis relies on the intricate tuning of an array of parameters or hyper-parameters, as well as the application of diverse nonlinear functions. To validate the precision of the autoencoder outcomes, we will proceed to perform feature extraction via the PCA method on the original DP configurations. In contrast to the autoencoder neural network, the PCA approach does not necessitate intricate parameter tuning, as its core is primarily grounded on matrix computations.
\begin{figure*}[t]
	\begin{minipage}{0.3\linewidth}
		\centerline{\includegraphics[width=5.6cm]{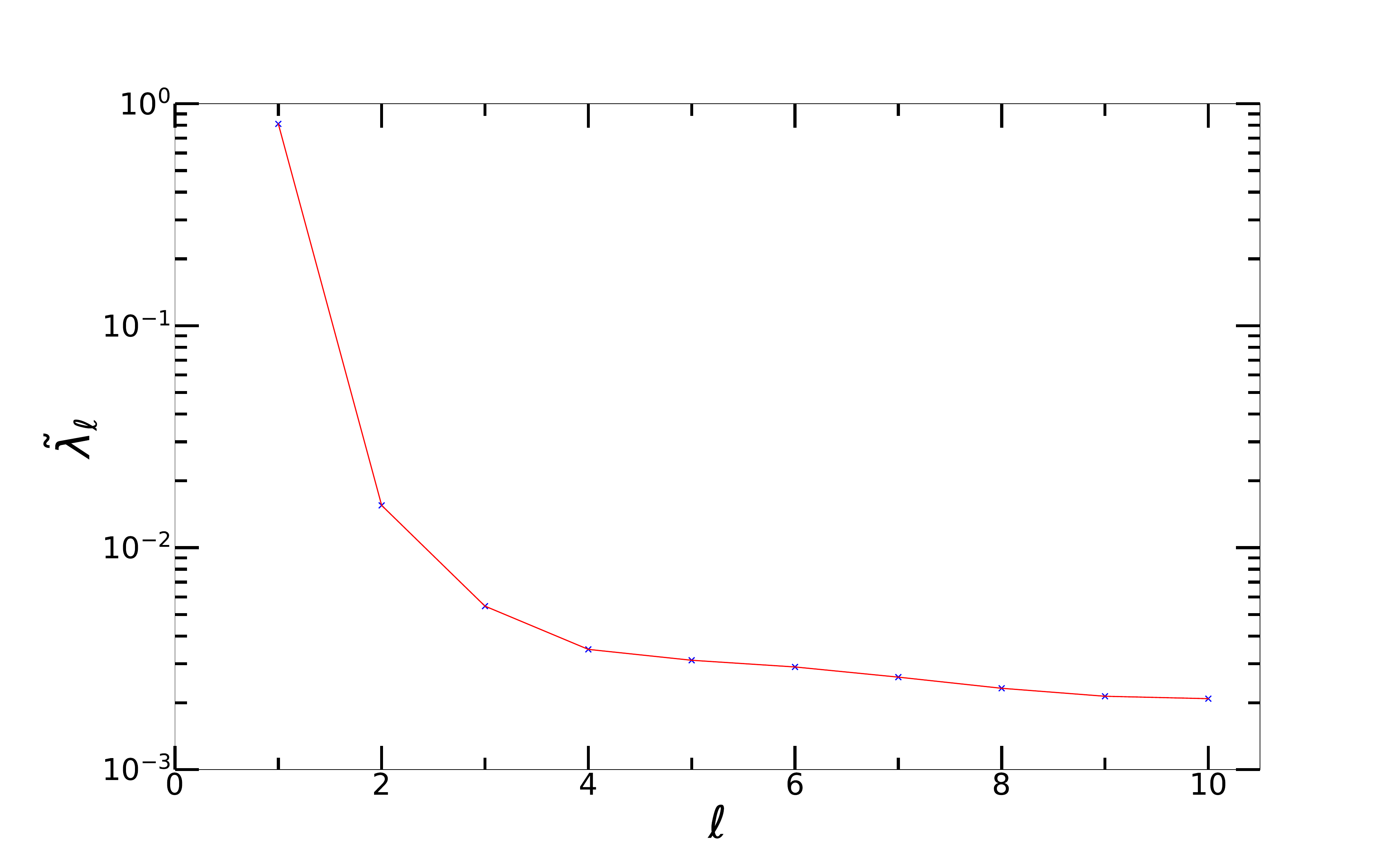}}
		\centerline{(a)}
	\end{minipage}
	\hfill
	\begin{minipage}{0.3\linewidth}
		\centerline{\includegraphics[width=5.6cm]{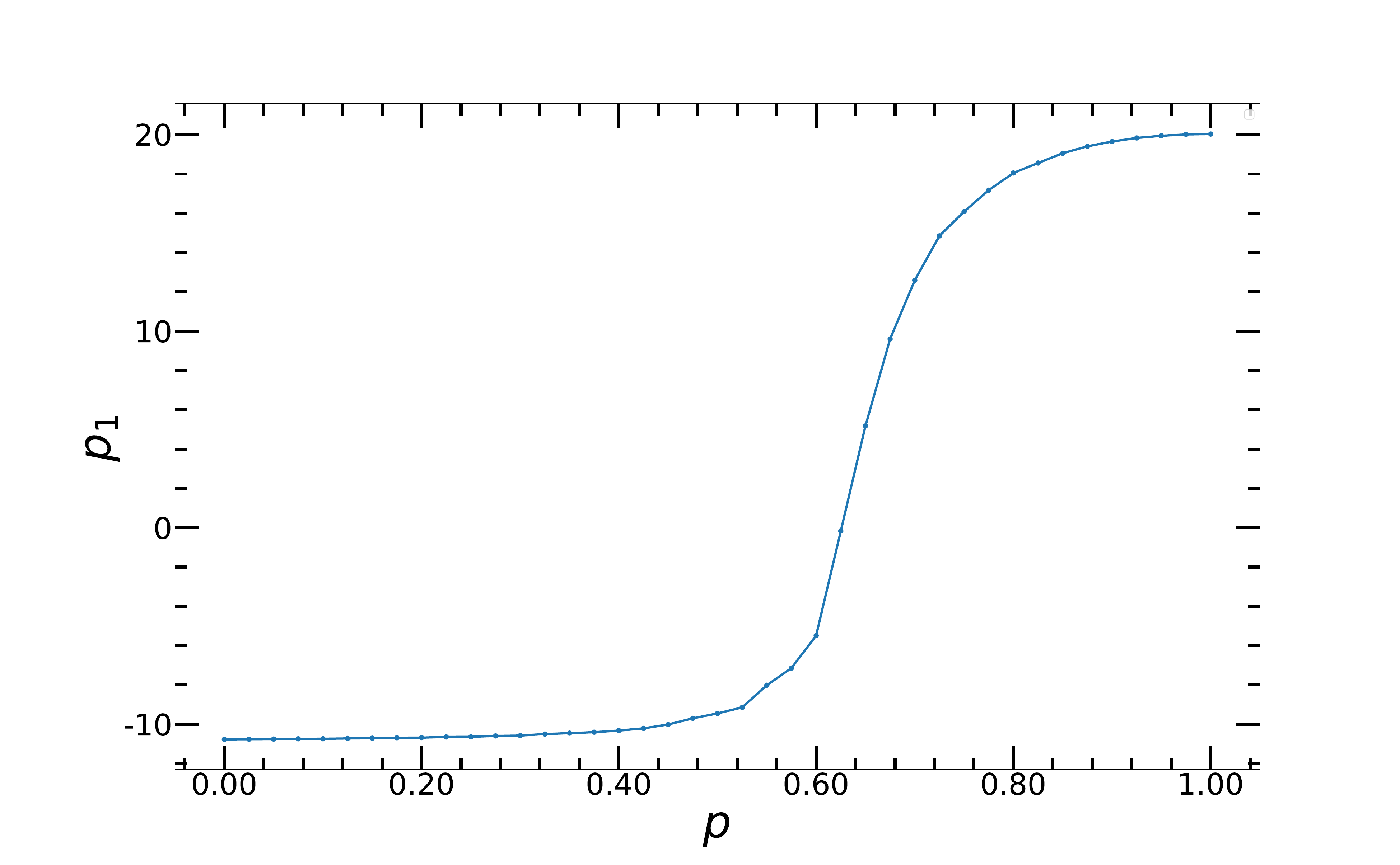}}
		\centerline{(b)}
	\end{minipage}
	\hfill
	\begin{minipage}{0.3\linewidth}
		\centerline{\includegraphics[width=5.6cm]{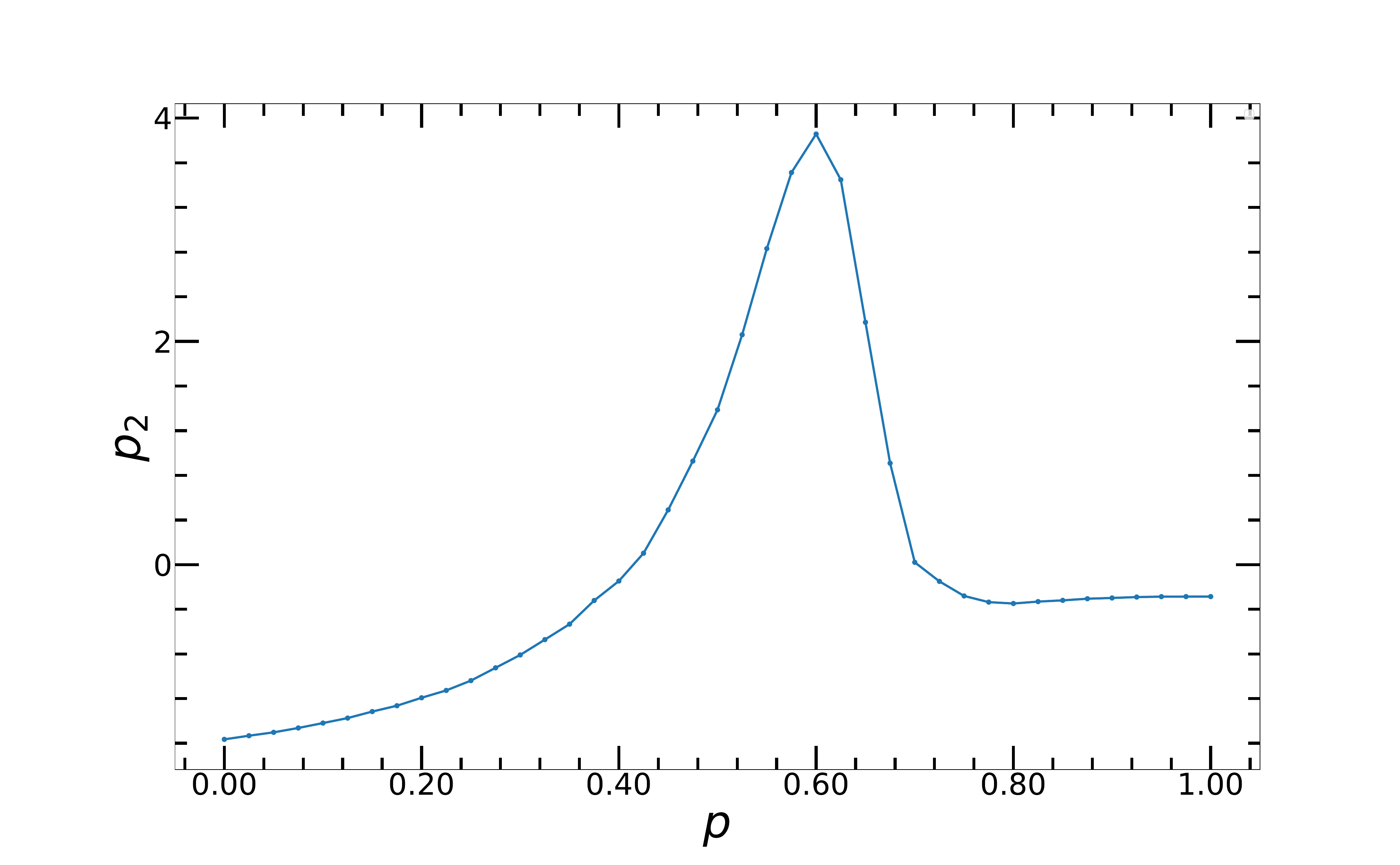}}
		\centerline{(c)}
	\end{minipage}
	\vfill
	\begin{minipage}{0.3\linewidth}
		\centerline{\includegraphics[width=5.6cm]{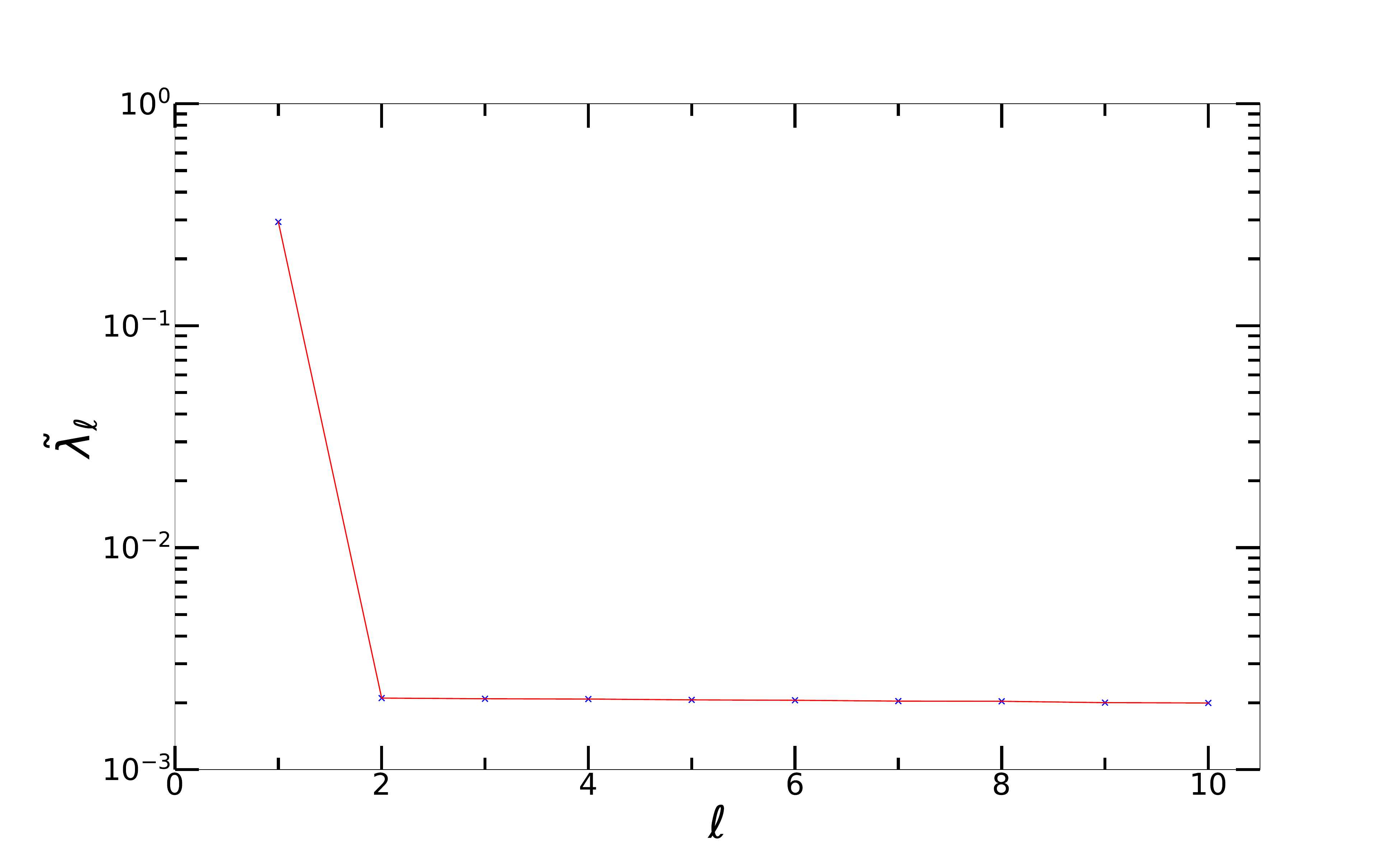}}
		\centerline{(d)}
	\end{minipage}
	\hfill
	\begin{minipage}{0.3\linewidth}
		\centerline{\includegraphics[width=5.6cm]{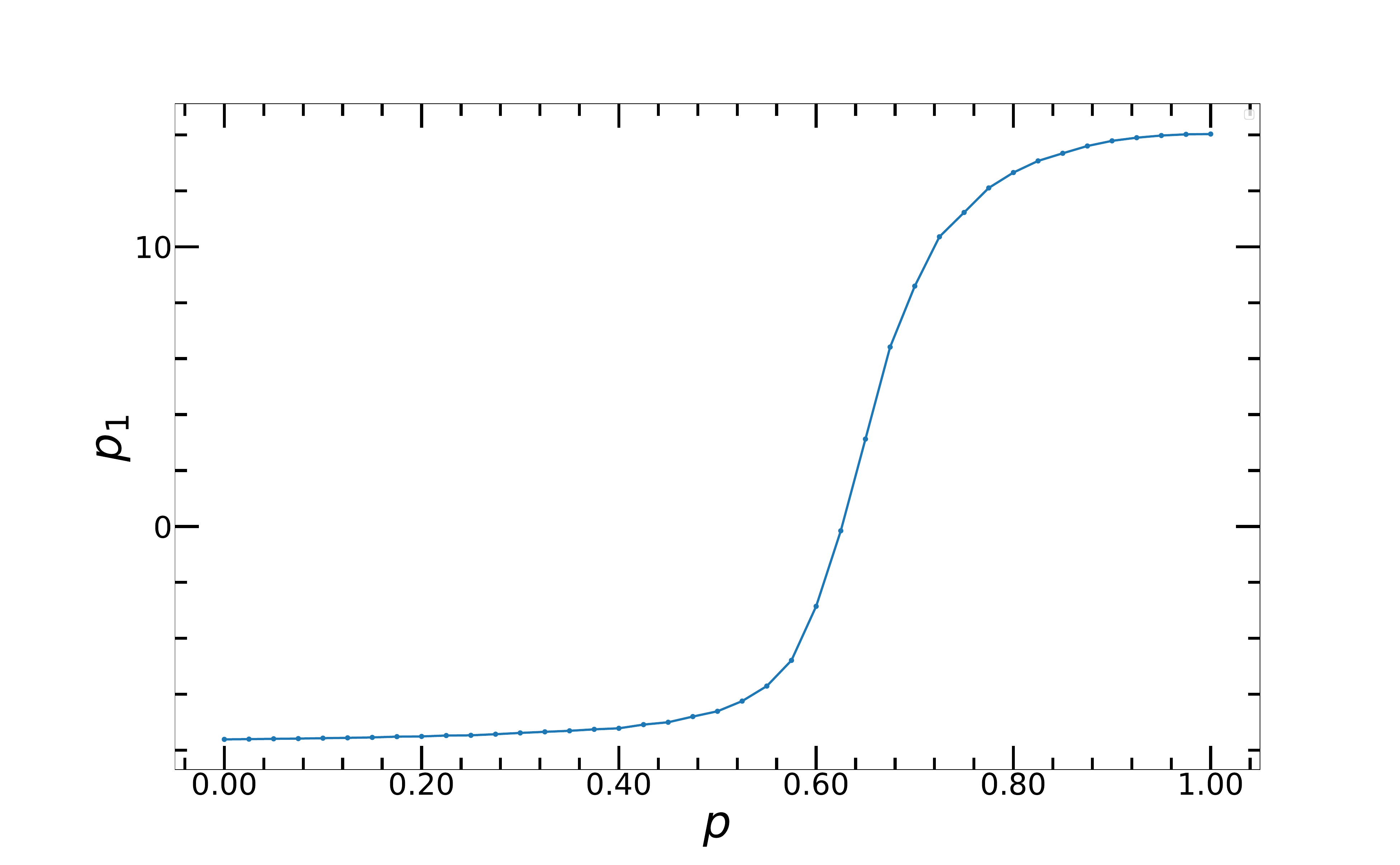}}
		\centerline{(e)}
	\end{minipage}
	\hfill
	\begin{minipage}{0.3\linewidth}
		\centerline{\includegraphics[width=5.6cm]{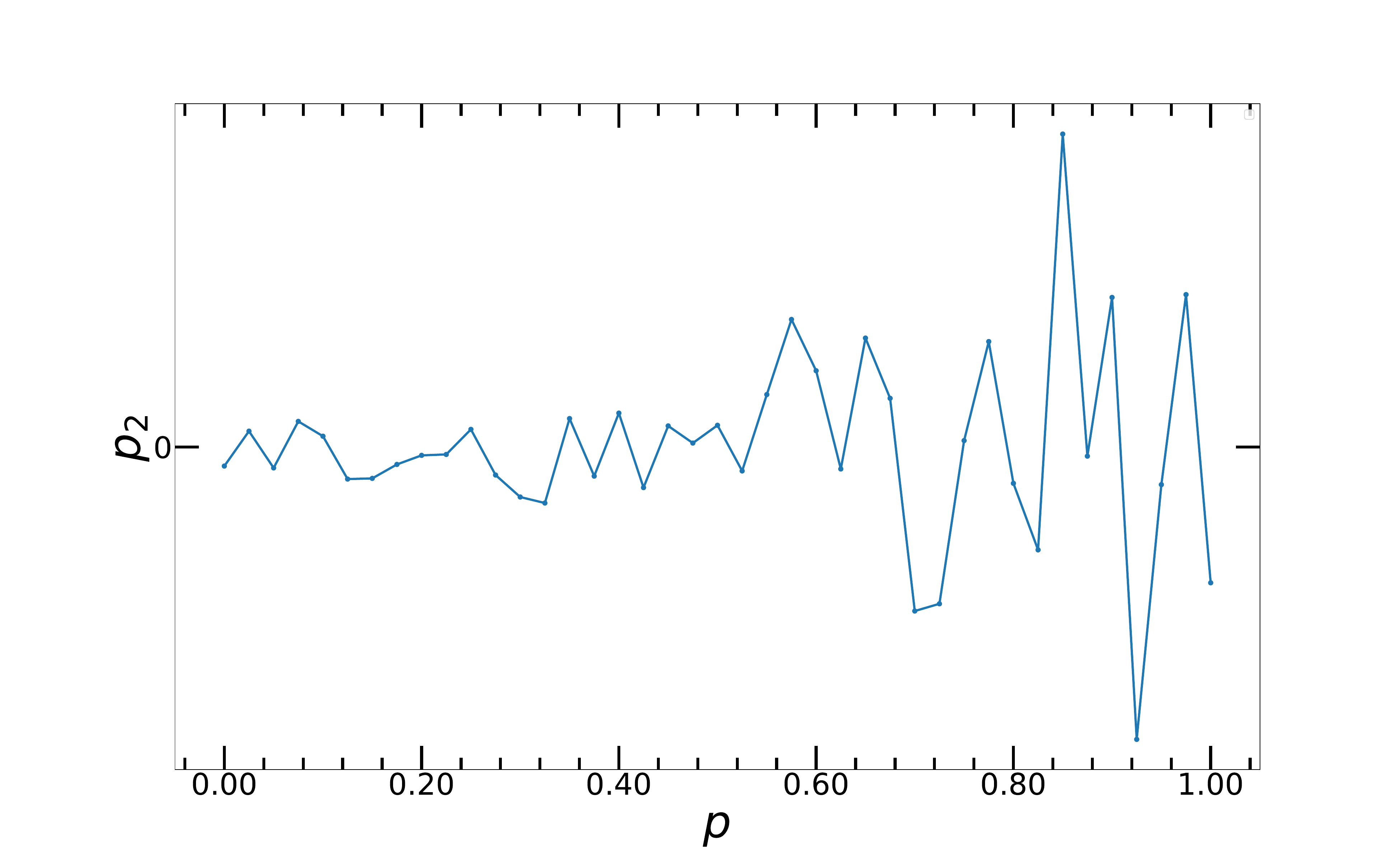}}
		\centerline{(f)}
	\end{minipage}
	\caption{PCA results of original (top panel) and shuffled (bottom panel) configurations of (1+1)-dimensional bond DP. \textbf{a}, The explained variance ratio $\widetilde{\lambda}_{\ell}$ from first ten principal components. \textbf{b}, $p_{1}$ versus $p$, each bond probability corresponding to $100$ samples. \textbf{c}, $p_{2}$ versus $p$, each bond probability corresponding to $100$ samples. \textbf{d}-\textbf{f}, results analogous to \textbf{a}-\textbf{c}.}
	\label{raw_shuffle_pca}
\end{figure*}

The results of PCA for DP are presented in Fig. \ref{raw_shuffle_pca}. The top row of the figure represents the original configuration. The first ten principal components' explained variance ratios are depicted in Fig. \ref{raw_shuffle_pca} (a), indicating that the first principal component exhibits a considerably higher value as compared to the subsequent principal components. This implies that the principal information of the input data is primarily represented by the first principal component, followed by the second one, with diminishing significance for subsequent components. Fig. \ref{raw_shuffle_pca} (b) exhibits the first principal component's bond probability function, and the critical value is found at the jumping location of this function, akin to a single latent variable of autoencoder. The critical value can be determined by fitting to the hyperbolic tangent function. Similarly, Fig. \ref{raw_shuffle_pca} (c) shows the second principal component of PCA as a function of the bond probability, and the critical point can be estimated reasonably well from the peak of this function.

In order to compare the relationship between the single latent variables and the DP' order parameter, we also include the results of the PCA first principal component and MC simulations in Fig. \ref{ae_pca_mc_dp_raw}. Fig. \ref{ae_pca_mc_dp_raw} (c) represents the first principal component results for the (1+1)-dimensional DP' raw configurations, it is clear that this has a similar trend to the autoencoder' single latent variables. To illustrate this in more detail, we compare the re-scaled results of the autoencoder' single latent variables and PCA' first principal components, to the particle density from MC in Fig. \ref{ae_pca_mc_dp_raw} (d). The blue dotted line represents the autoencoder' single latent variables, the cyan one represents the PCA' first principal component, and the red one represents the MC simulation results. Consequently, we notice that the results given by the autoencoder, PCA and MC simulations bear close similarities to one another. As the bond occupation probability increases, both the potential variable $h$ and particle density $\rho$ undergo a gradual increase. When $p \sim p_{c}$, $h$ and $\rho$ exhibit a noticeable jumping behavior, which ultimately saturates at $1$. Based on this observation, as the MC simulation gives the particle density, we can conclude that a single latent variable $h$ and the first principal component is positively correlated with the particle density $\rho$. In the DP model, the autoencoder method could potentially extract particle density information through a single potential variable, so could the first principal component of PCA.

In addition, we assess the concurrence of the results attained from the MC simulations (summation of particle density referred to as $\rho_{all}$) , autoencoder outputs (re-scaled single hidden variable denoted by $h$) and PCA outputs (re-scaled the first principal component denoted by $pca_1$), employing the Pearson Correlation Coefficient \cite{cohen2009pearson} as a measure. The coefficient quantifies the degree of the linear correlation between a pair of variables.
\begin{equation}
r = \dfrac{\sum \limits_{i = 1}^{n} (\rho_{i} - \overline{\rho}) (h_{i} - \overline{h})}{\sqrt{\sum \limits_{i = 1}^{n} (\rho_{i} - \overline{\rho})^{2}} \sqrt{\sum \limits_{i = 1}^{n} (h_{i} - \overline{h})^{2}}},
\end{equation}
\noindent
where $\rho_i$ and $h_i$ are the sum particle density and the rescaled hidden variable at $i$-th bond probability $p$, while $\overline{\rho}$ and $\overline{h}$ are the mean values of $\rho_i$ and $h_i$, respectively.

According to the above equation, the correlation coefficients of $h$, $pca_1$ and $\rho$ are $0.9983$ and $0.9991$, respectively. The high degree of correlation could be another evidence that both the hidden variable and the first principal component can extract the information regarding the particle density.


After conducting two unsupervised learning on DP's raw configurations, we find that the outcomes for the autoencoder's single latent variable and PCA's first principal component are almost identical to those of the MC results. Furthermore, the Pearson correlation coefficient linking these results with the MC results is remarkably high, approaching $1$. These findings may suggest that the particle density of DP is the fundamental data captured by the autoencoder's single latent variable and PCA's first principal component.

\subsubsection{Learning with shuffled configurations}
As a matter of fact, with time evolving, the DP configurations demonstrate certain rules of change. Starting from a fully occupied lattice, under the condition of $p \ll p_{c}$, the number of particles decreases rapidly until the system reaches an absorbing state. When $p \simeq p_{c}$, a percolating cluster emerges. While $p \gg p_{c}$, the system will be completely percolating. To validate the conjecture put forth in the preceding section, we undertake the task of perturbing the original DP configurations. This is executed by randomly shuffling the positions of particles at a specific ratio $r$, whereas keeping the particle density unchanged. The shuffled configurations now form the input for the same feature extraction method. This shuffling approach has the benefit of retaining the particle density, while altering the particle position distribution. 

\begin{figure*}[htbp]
\begin{tabular}{cc}
    \includegraphics[width=0.45\textwidth]{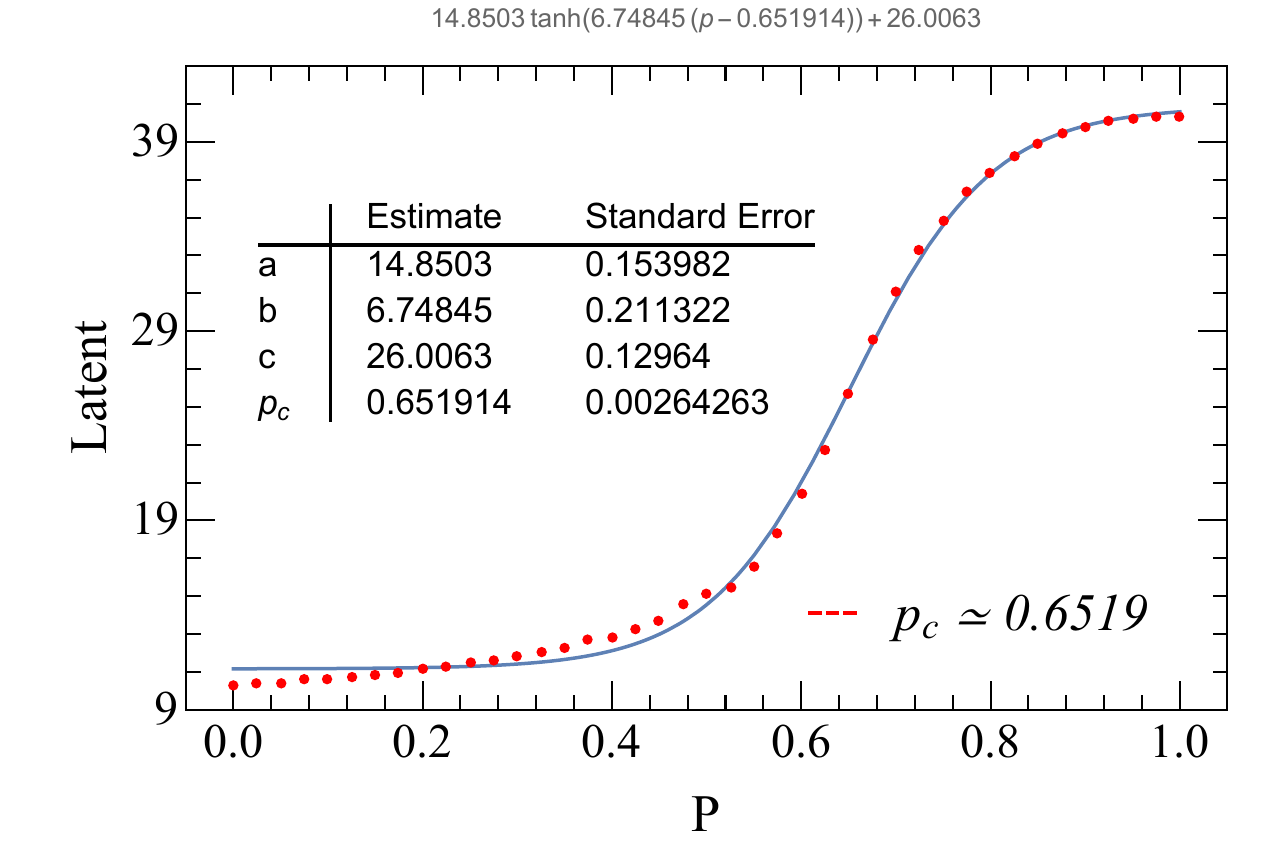} &
    $\qquad$\includegraphics[width=0.45\textwidth]{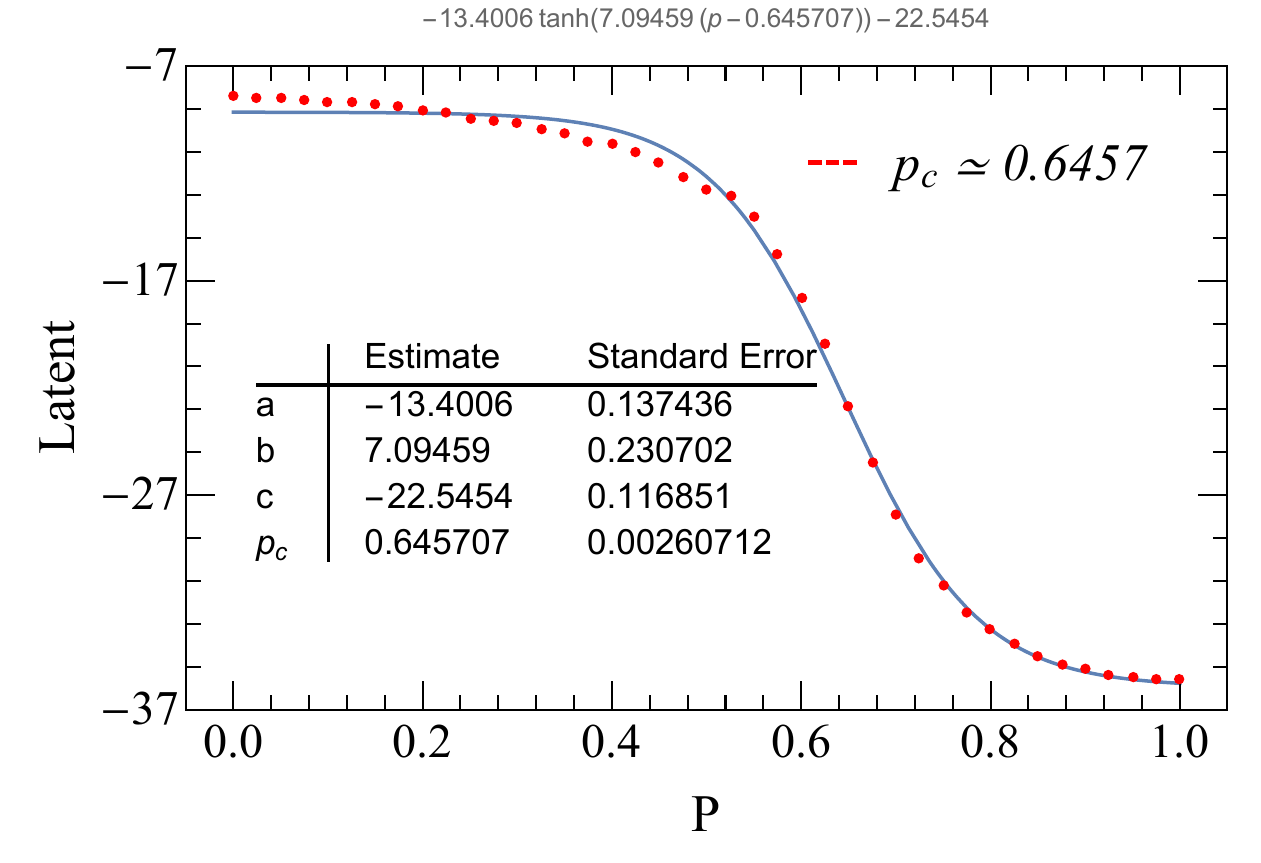} \\
    (a) & $\qquad$ (b)\\
    \includegraphics[width=0.45\textwidth]{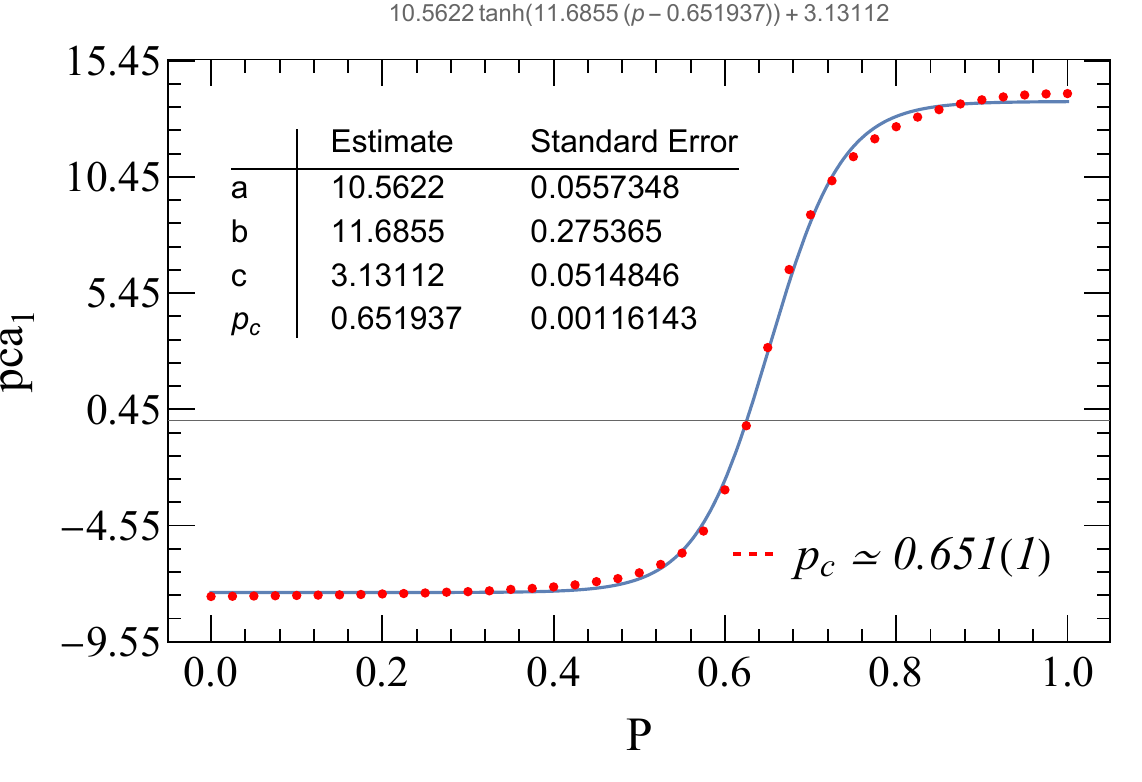} &
    $\qquad$ \includegraphics[width=0.45\textwidth]{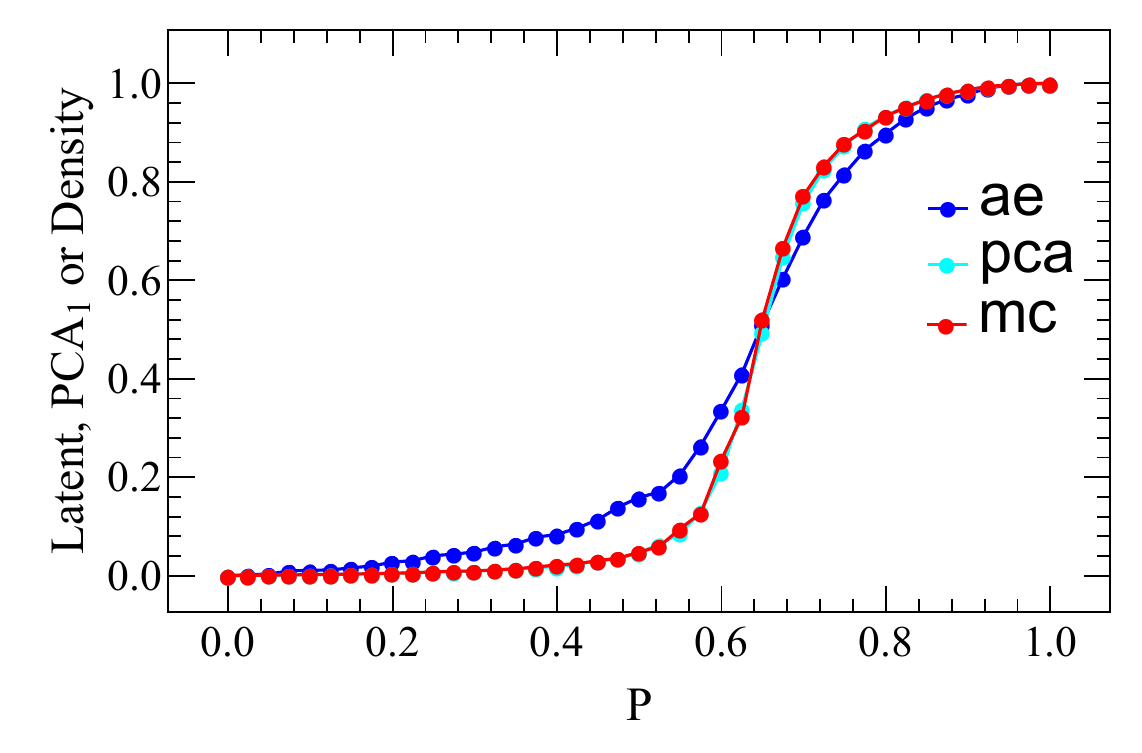} \\
    (c) & $\qquad$(d)
\end{tabular}
\caption{ML shuffled configurations of (1+1)-dimensional DP by 
autoencoder and PCA, where the shuffle ratio $r$ is 1.0. \textbf{a} and \textbf{b} are two similar autoencoder results of (1+1)-dimensional bond DP from the same data set and convolutional autoencoder neural network. Encoding of the raw DP configurations using a single hidden neuron activation gives $Latent$ as a function of the bond probability. Each data point of bond probability is averaged over $200$ testing samples. \textbf{c} and is the PCA result of (1+1)-dimensional DP from the first principal component of the shuffled DP configurations as a function of the bond probability. \textbf{d}, A comparison between the re-scaled hidden variable $h$ of autoencoder, first principal component of PCA and $rho$ of the MC simulations versus $p$, represented by blue, cyan and red colors, respectively. }
\label{ae_pca_mc_dp_shuffle}
\end{figure*}

The long-range (infinitely long) correlations that occur solely at critical states may potentially be disrupted by the act of shuffling, leading to the partitioning of the percolating cluster by forming fragmented pieces. Fig.\ref{raw_shuffle_max_config} (a) depicts a random configuration of (1+1)-dimensional bond DP at the critical state, while Fig. \ref{raw_shuffle_max_config} (b) shows the shuffled one of (a), with a ratio $r = 0.2$. Our results, obtained through utilizing identical data sets, autoencoder neural networks and PCA methods, are presented in the bottom row of Fig. \ref{ae_pca_mc_dp_shuffle}. Figs. \ref{ae_pca_mc_dp_shuffle} (a) and (b) illustrate the critical values as $p_{c} \simeq 0.652(3)$ and $p_{c} \simeq 0.646(3)$, respectively, which coincide with the estimate derived from the MC simulations. Both Figs. \ref{ae_pca_mc_dp_shuffle} (a) and (b) present the comparison between $\rho_{all}$, the sum of particle density, and $h$, the hidden variable, displaying similar behavior. Incidentally, Figs. \ref{ae_pca_mc_dp_shuffle} (c) shows the PCA' first principal component results of shuffled configurations. It is obvious from Figs. \ref{ae_pca_mc_dp_shuffle} (d) that the PCA matches the MC results better than the autoencoder does. Nonetheless, it should be mentioned that there is a slight discrepancy between $\rho$ and $h$ in Fig. \ref{ae_pca_mc_dp_raw} (d) versus Fig. \ref{ae_pca_mc_dp_shuffle} (d), which might be caused by the randomization process during shuffling. 

\begin{figure}
\centering
\includegraphics[width=0.4\textwidth]{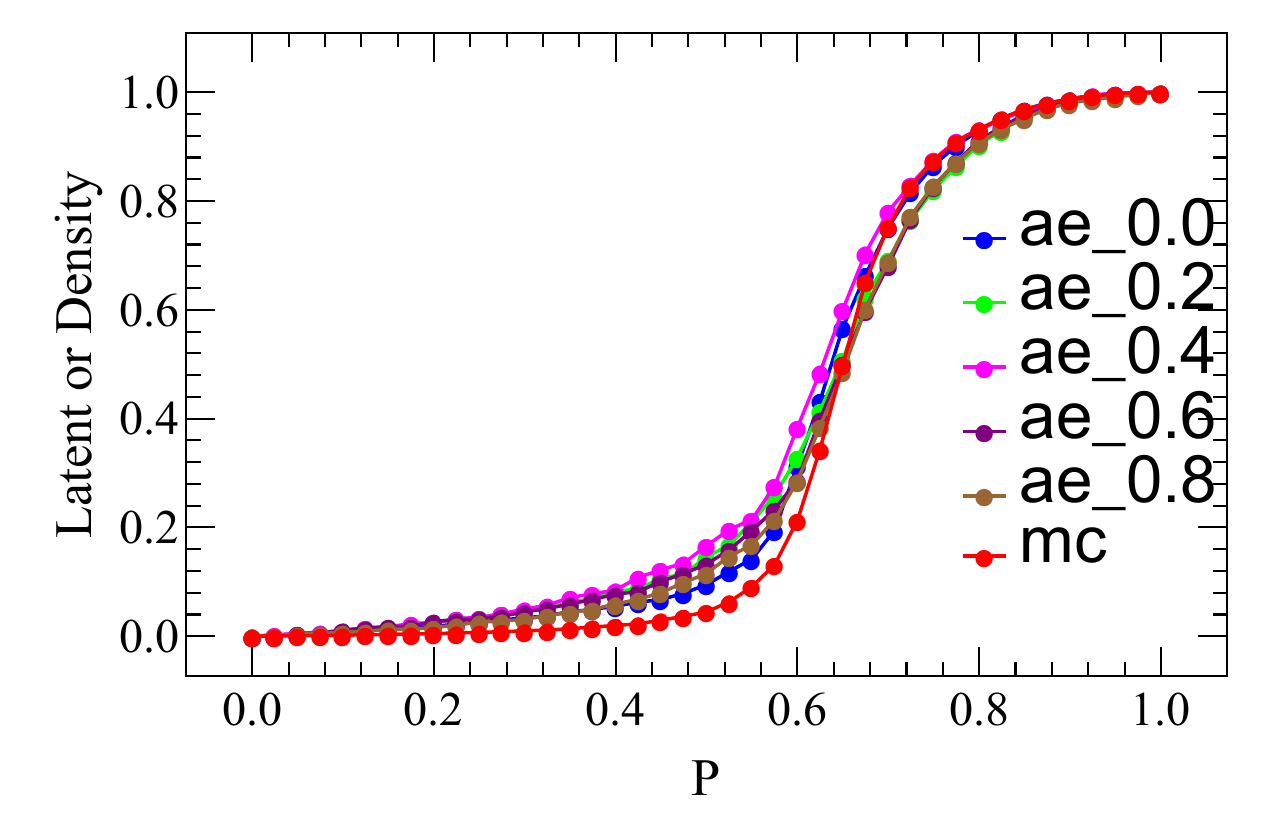}
\caption{ML of original and shuffled configurations of (1+1)-dimensional bond DP by autoencoder, where the lattice size is $N = 16$, the total number of time steps for averaging is $t = 120$, and the number of ensemble average is $200$.}
\label{raw_different_shuffled_ratios_ae_normal}
\end{figure}

The preceding calculation is based on the random shuffling of all lattice sites, with a shuffle ratio of $r=1$. In order to examine the effect of shuffle ratio on the autoencoder single latent variable in greater detail, different ratios are selected to shuffle the raw configurations of DP. Fig. \ref{raw_different_shuffled_ratios_ae_normal} displays the autoencoder outcomes with varying shuffle ratios, where the latent has been re-scaled into the range between 0 and 1. It is apparent from Fig. \ref{raw_different_shuffled_ratios_ae_normal} that the locations of transition points that correspond to different shuffle ratios remain essentially unchanged, and agree well with the MC simulations. This finding excludes the possibility that random shuffling significantly alters the single potential variable. Additionally, a variety of shuffle ratios have benn applied to the original configurations, and the respective autoencoder findings are presented in the second row of Table \ref{ae_shuffle_ratios}. The critical values obtained are in close proximity to the theoretical critical value, $p_c \simeq 0.6447$ \cite{henkel2008non}. 

\begin{table*}[h]
	\centering
	\resizebox{\textwidth}{10mm}{
	\begin{tabular}{|c|c|c|c|c|c|c|c|c|c|c|c|}
        \hline
  $r$(shuffle ratio)   & 0  &0.1 &0.2 &0.3 &0.4 &0.5&0.6&0.7&0.8&0.9&1.0 \\
        \hline
   $p_{c}$(shuffle) & 0.642(1) &0.646(2)  &0.651(2)  &0.634(2)  &  0.630(2)&0.638(2) &0.655(2)&0.649(2)&0.654(1)&0.661(1)&0.651(2) \\
        \hline
   $jumping location$(maximum cluster) & 0.642(1) &0.661(1)  &0.664(1)  &0.678(1)  &  0.689(1)&0.704(1) &0.717(1)&0.774(1)&0.780(1)&0.785(1)&0.790(1) \\
        \hline
   \end{tabular}}
\caption{Autoencoder results of (1+1) dimensional DP with different shuffle ratios, where $L=16$ and $T=120$. The second row in the table represents the single potential variable results after shuffling the configurations, and the third row means the single potential variable results of the maximum cluster after shuffling the configurations with different shuffle ratios.}
\label{ae_shuffle_ratios}
\end{table*}

Furthermore, it is noteworthy that the curves depicted in Fig. \ref{raw_different_shuffled_ratios_ae_normal} display some level of deviation from the MC result. To verify the possible correlation between this observation and the shuffle ratio, we calculate the Euclidean Distance \cite{danielsson1980euclidean} between the corresponding curve and the MC result under different shuffle ratios, which is presented in Table \ref{ae_shuffle_euclidean}. It is evident that the Euclidean Distance remains nearly unchanged as the shuffle ratio increases. Despite the distinct variation in curvature that can be anticipated with varying ratios, it is remarkable that the single latent variable of the autoencoder architecture can reliably identify the critical point.

\begin{table*}[h]
	\centering
	\resizebox{\textwidth}{10mm}{
	\begin{tabular}{|c|c|c|c|c|c|c|c|c|c|c|c|}
        \hline
  shuffle ratio    &0.1 &0.2 &0.3 &0.4 &0.5&0.6&0.7&0.8&0.9&1.0 \\
 
        \hline
   Euclidean Distance & 0.133 &0.137  &0.164  &0.153  &  0.084 &0.123 &0.138 &0.108 &0.159 &0.155\\
        \hline
   \end{tabular}}
\caption{Euclidean Distance between MC and autoencoder results for (1+1) dimensional DP with different shuffle ratios.}
\label{ae_shuffle_euclidean}
\end{table*}

The second row in Fig. \ref{raw_shuffle_pca} presents the PCA results at a shuffle ratio of $r = 1$, similar to the upper row. Notably, the second principal component varies significantly among them. Specifically, in Fig. \ref{raw_shuffle_pca} (f), the second principal component lacks an apparent peak versus bond probability. This observation leads one to speculate that random shuffling might compromise a physics operation in the DP model.

The critical value of data processing obtained through the first principal component of PCA can be observed in the second row of Table \ref{pca1_pc_shuffle_ratio}. Similar to the results obtained via autoencoder, the critical value does not vary with the shuffle ratios. Moreover, the critical values acquired through PCA display strong stability, being closely positioned around $0.651$ and $0.652$. This suggests that linear dimensionality reduction is a highly stable process. There is a slight deviation between these values and the theoretical critical value of $p_c \simeq 0.6447$ \cite{henkel2008non}, which can be attributed to the fact that the first principal component can only capture a portion of the input data's information and does not fully represent it. Nevertheless, the results obtained via PCA at different shuffle ratios align with those from autoencoder, indicating the effectiveness of the single potential variable.

\begin{table*}[h]
	\centering
	\resizebox{\textwidth}{10mm}{
	\begin{tabular}{|c|c|c|c|c|c|c|c|c|c|c|c|}
        \hline
  $r$(shuffle ratio)   & 0  &0.1 &0.2 &0.3 &0.4 &0.5&0.6&0.7&0.8&0.9&1.0 \\
 
        \hline
   $p_{c}$(shuffle) & 0.652(1) &0.652(1)  &0.652(1)  &0.652(1)  &  0.652(1)&0.651(1) &0.651(1)&0.651(1)&0.652(1)&0.651(1)&0.651(1) \\
      \hline
   $jumping location$(maximum cluster) & 0.658(1) &0.666(1)  &0.677(1)  &0.689(1)  &  0.703(1)&0.717(1) &0.730(1)&0.740(1)&0.745(1)&0.750(1)&0.751(1) \\
        \hline
   \end{tabular}}
\caption{PCA results of (1+1) dimensional DP with different shuffle ratios, where $L=16$ and $T=120$. The second row in the table represents the single potential variable results after shuffling the configurations, and the third row represents the single potential variable results of the maximum cluster after shuffling the configurations with different shuffle ratios.}
\label{pca1_pc_shuffle_ratio}
\end{table*}

\begin{figure*}[h]
\begin{tabular}{cccc}
    \includegraphics[width=0.32\textwidth]{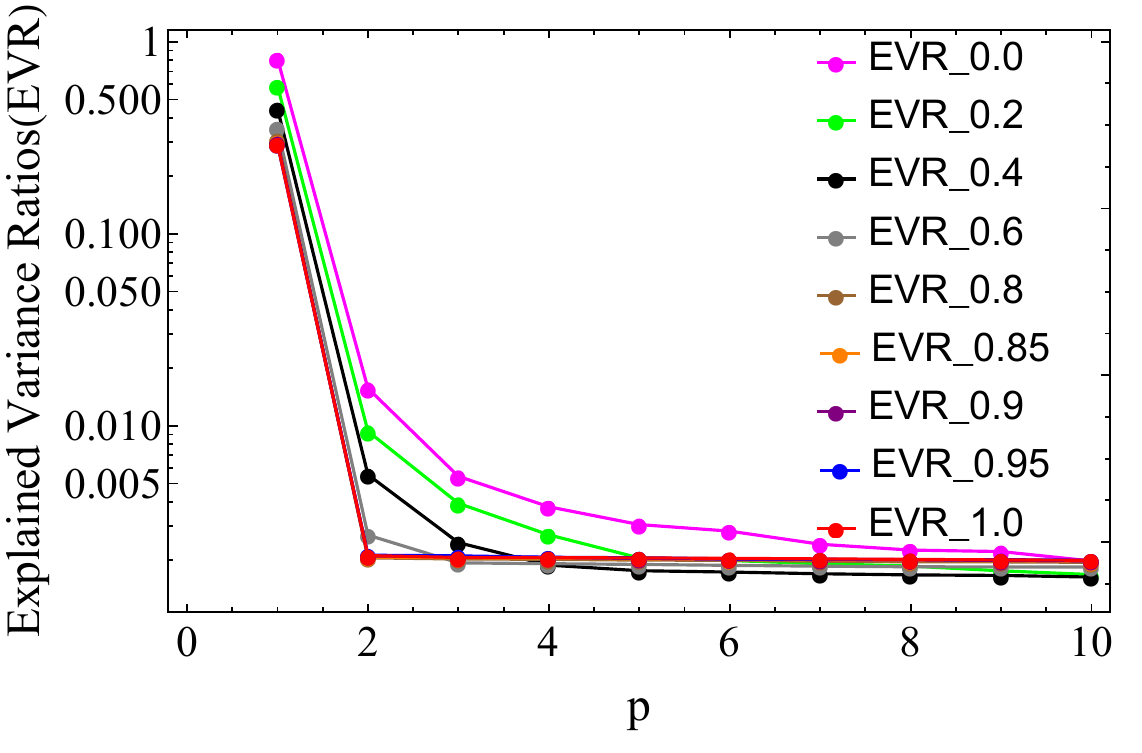} &
    \includegraphics[width=0.32\textwidth]{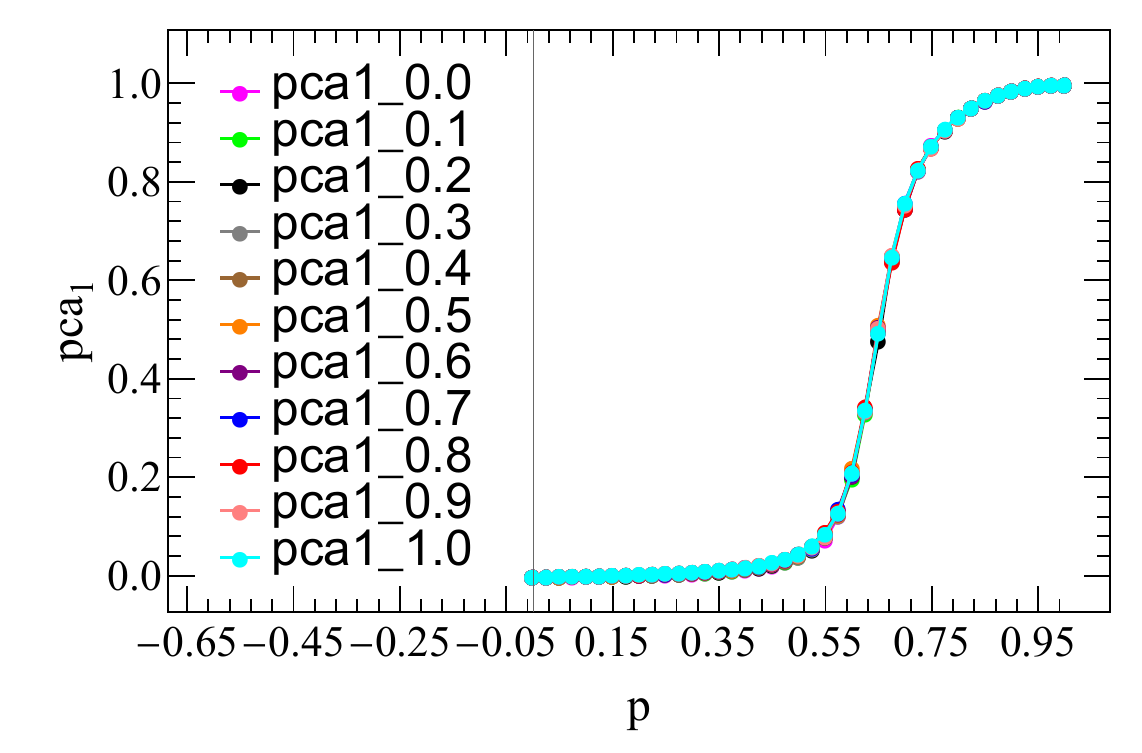} &
    \includegraphics[width=0.32\textwidth]{ 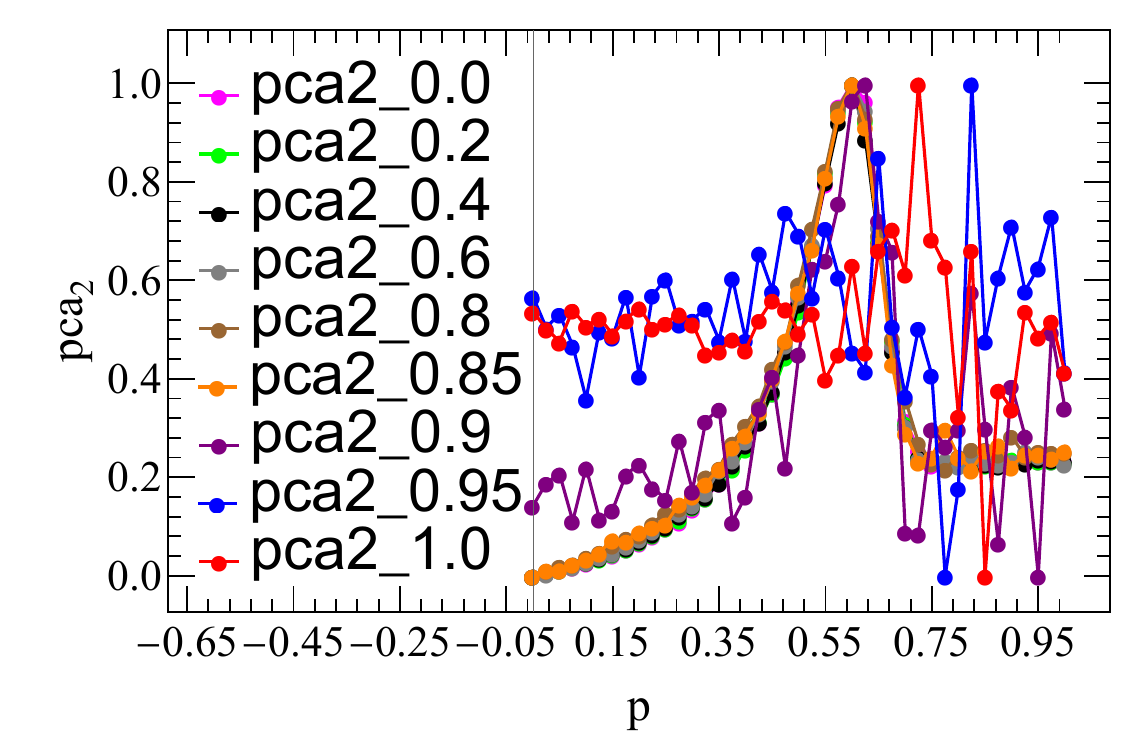} \\
     (a) &  (b)  & (c)
\end{tabular}
\caption{PCA results of (1+1)-dimensional bond DP for different shuffle ratios. Panel (a) corresponds to the explained variance ratio $\widetilde{\lambda}_{\ell}$ from first ten principal components. Panel (b) depicts the first principal component, denoted as $p_{1}$, plotted as a function of the bond probability $p$, while panel (c) shows the second principal component, denoted as $p_{2}$, plotted against the bond probability $p$.}
\label{v_shuffle_ratios}
\end{figure*}

To illustrate the influence of different shuffle ratios on the first and second principal component of PCA, we conduct a more detailed analysis. Fig. \ref{v_shuffle_ratios} (a) shows the explained variance ratios of the first ten principal components of PCA under different shuffle ratios. It can be found that the first principal component can still occupy a large proportion of the explained variance ratios, which can represent the main information of the input data. With the increase of shuffle ratio $r$, the first principal component shows a downward trend, but it is still much higher than other principal components. Fig. \ref{v_shuffle_ratios} (b) presents the first principal component as a function of the bond probability $P$ for different shuffle ratios. It can be found that the first principal component does not seem to vary with the shuffle ratio. And, it can well characterize the location of the critical point. To verify the effect of shuffle ratio on the first principal component, the Euclidean Distance is calculated between the curves corresponding to different shuffle ratios and those of the MC simulations. The results are presented in Table \ref{pca1_shuffle_euclidean}. It can be clearly seen that the Euclidean Distance between the different shuffle ratios is very small, i.e., the curves nearly coincident onto one another. This can explain to some extent the small effect of the shuffle ratio on the first principal component of PCA. In the (1+1)-dimensional DP, the change of the shuffle ratio does not affect the capture of the first principal component of PCA for the critical point.

\begin{figure*}[t]
	\begin{minipage}{0.3\linewidth}
		\centerline{\includegraphics[width=5.6cm]{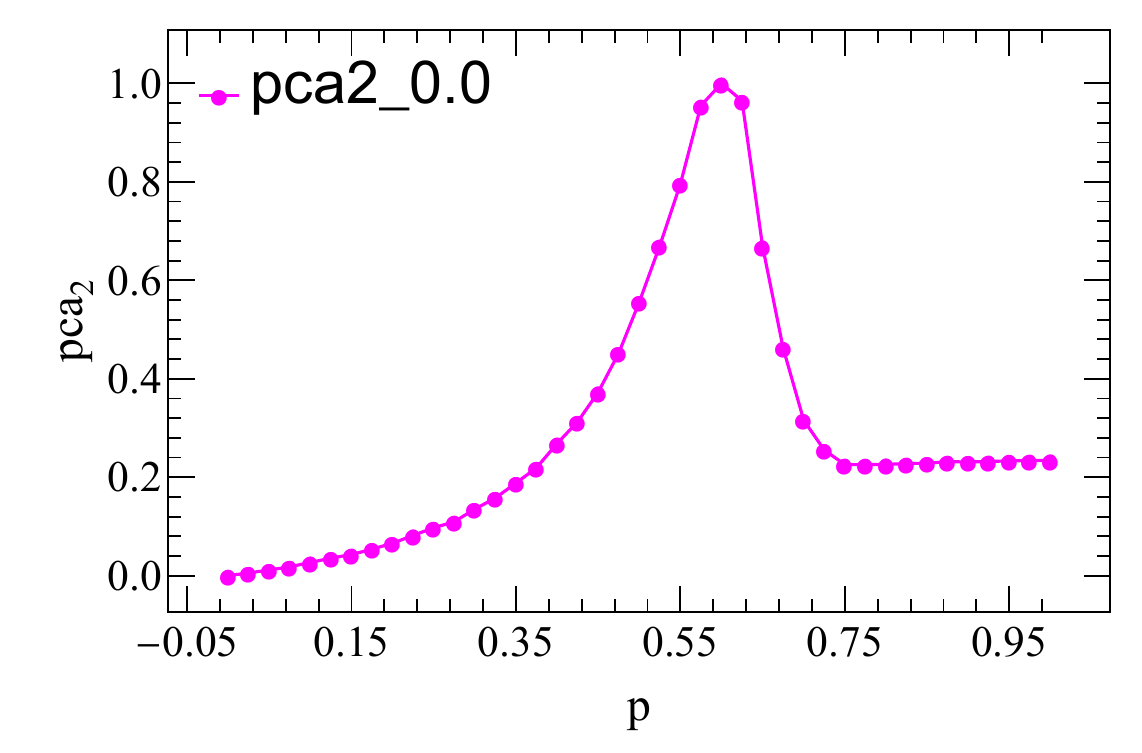}}
		\centerline{(a)}
	\end{minipage}
	\hfill
	\begin{minipage}{0.3\linewidth}
		\centerline{\includegraphics[width=5.6cm]{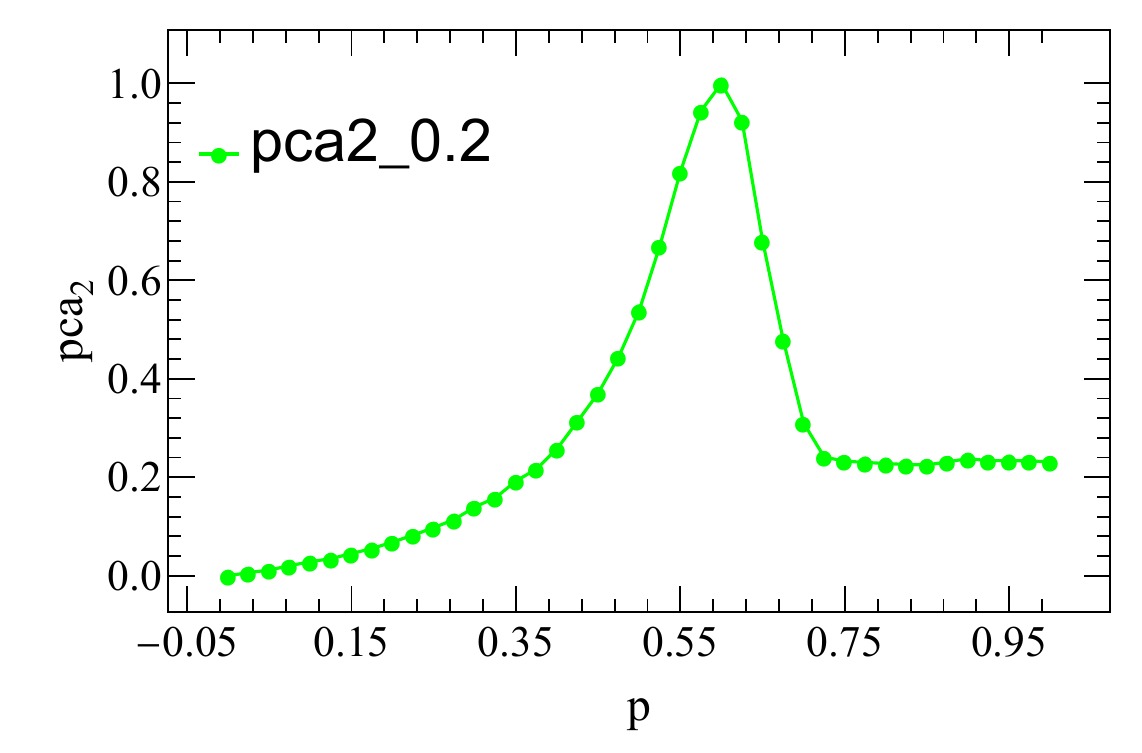}}
		\centerline{(b)}
	\end{minipage}
	\hfill
	\begin{minipage}{0.3\linewidth}
		\centerline{\includegraphics[width=5.6cm]{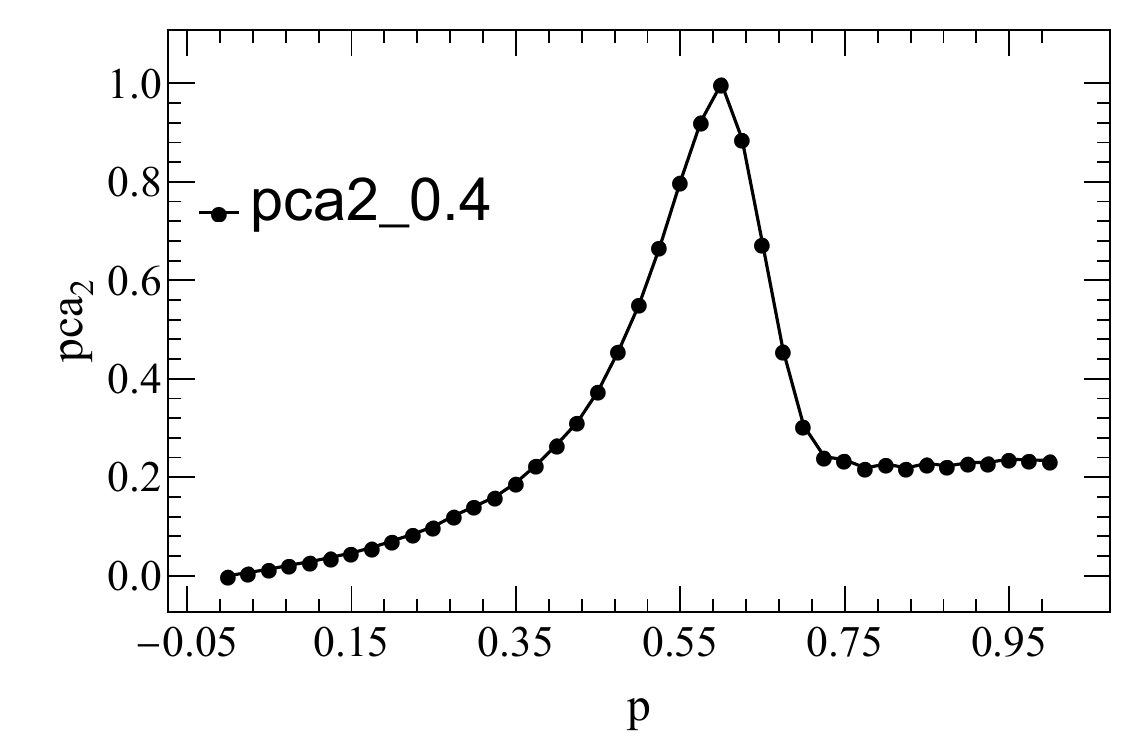}}
		\centerline{(c)}
	\end{minipage}
	\vfill
	\begin{minipage}{0.3\linewidth}
		\centerline{\includegraphics[width=5.6cm]{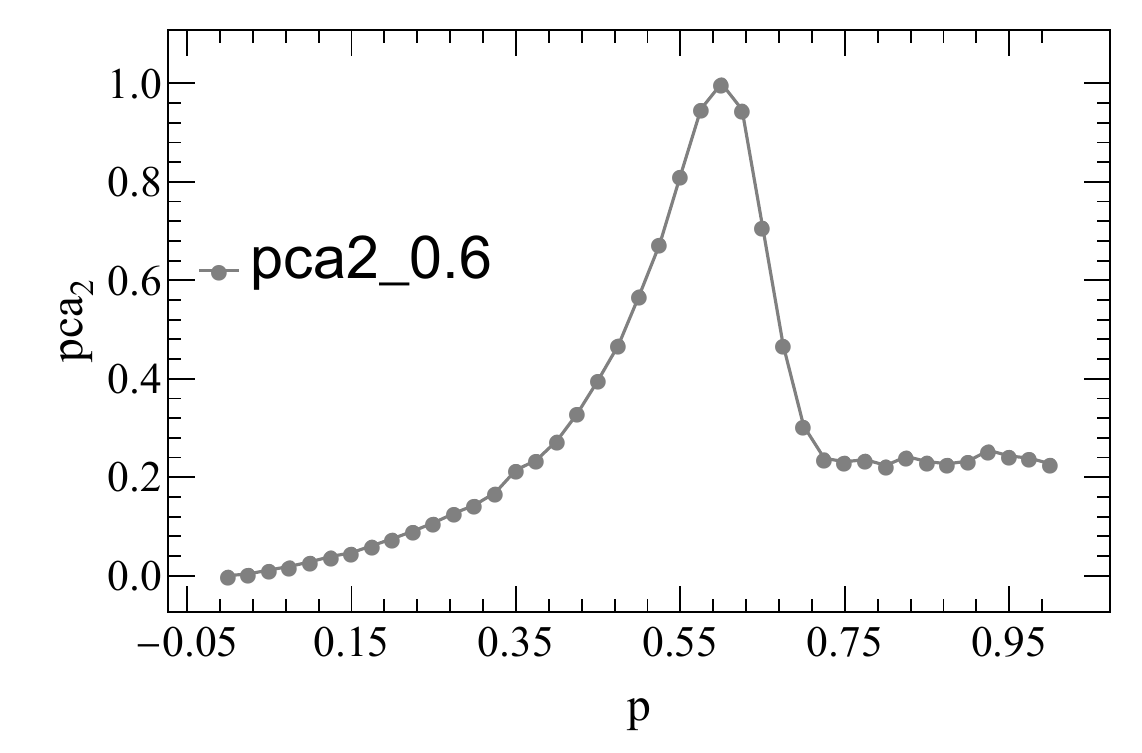}}
		\centerline{(d)}
	\end{minipage}
	\hfill
	\begin{minipage}{0.3\linewidth}
		\centerline{\includegraphics[width=5.6cm]{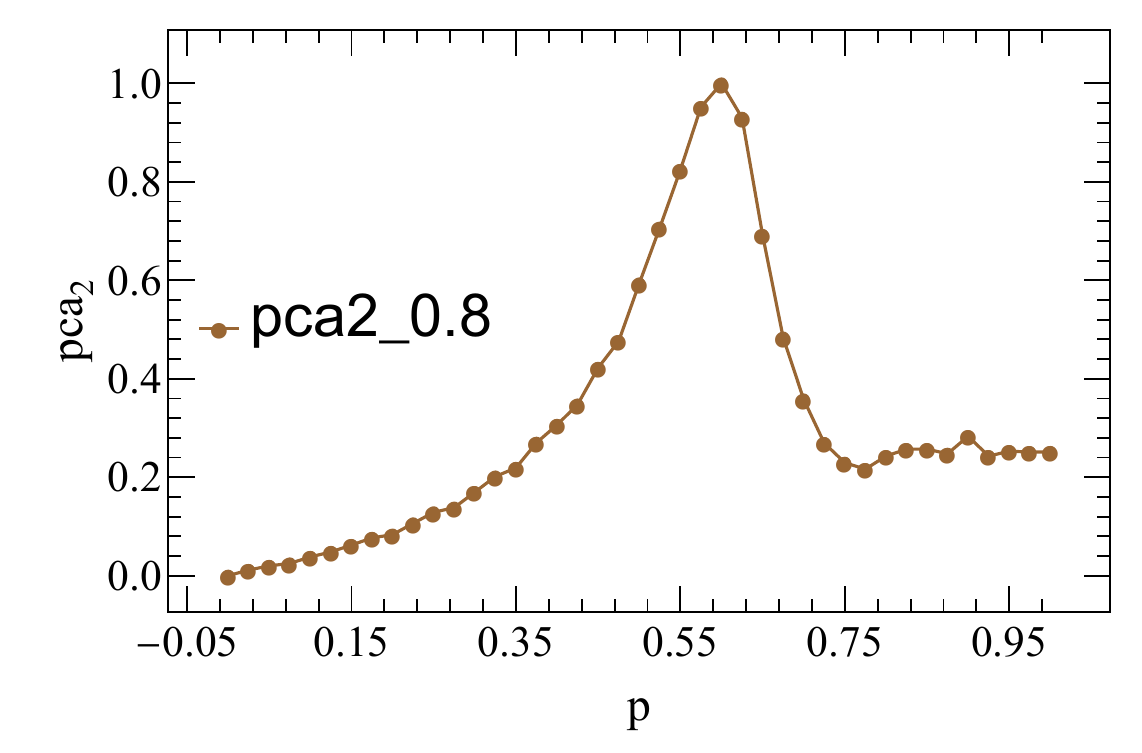}}
		\centerline{(e)}
	\end{minipage}
	\hfill
	\begin{minipage}{0.3\linewidth}
		\centerline{\includegraphics[width=5.6cm]{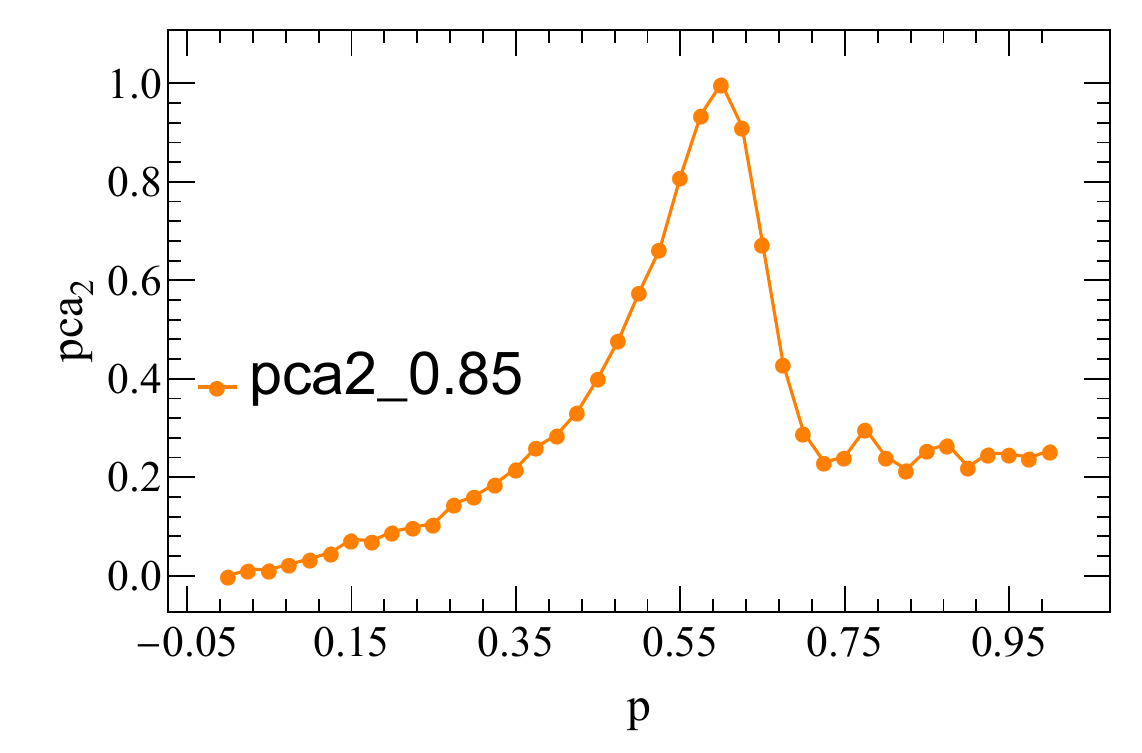}}
		\centerline{(f)}
	\end{minipage}
	\vfill
	\begin{minipage}{0.3\linewidth}
		\centerline{\includegraphics[width=5.6cm]{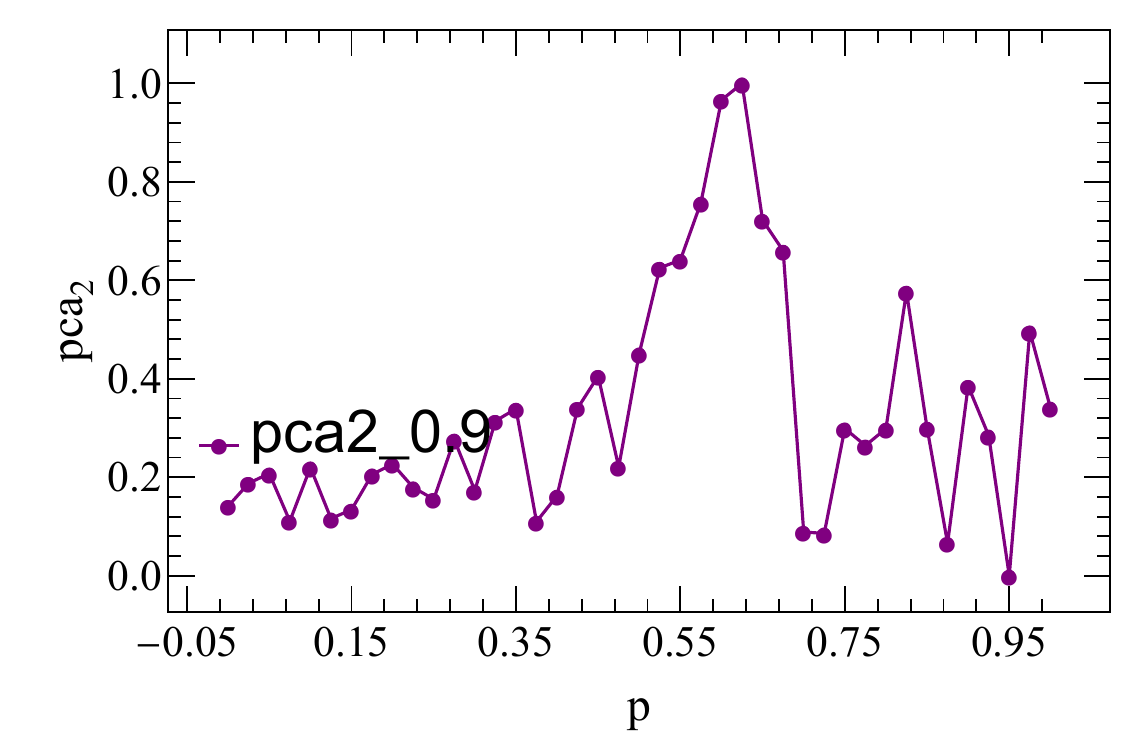}}
		\centerline{(g)}
	\end{minipage}
	\hfill
	\begin{minipage}{0.3\linewidth}
		\centerline{\includegraphics[width=5.6cm]{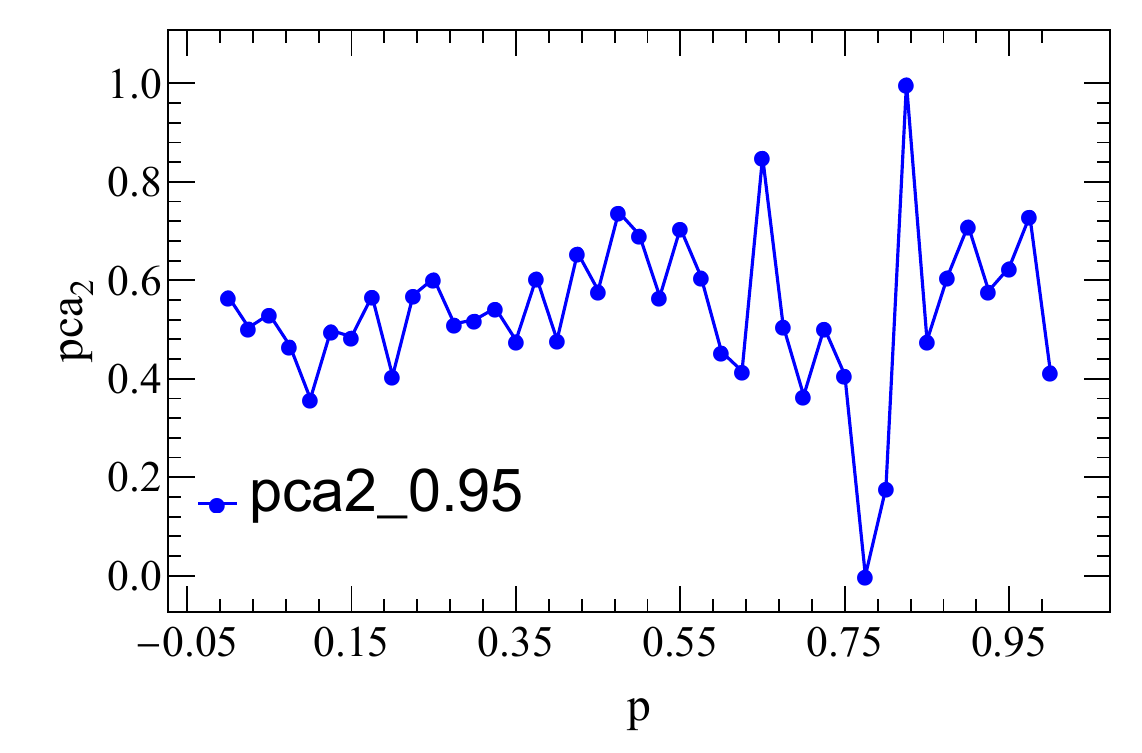}}
		\centerline{(h)}
	\end{minipage}
	\hfill
	\begin{minipage}{0.3\linewidth}
		\centerline{\includegraphics[width=5.6cm]{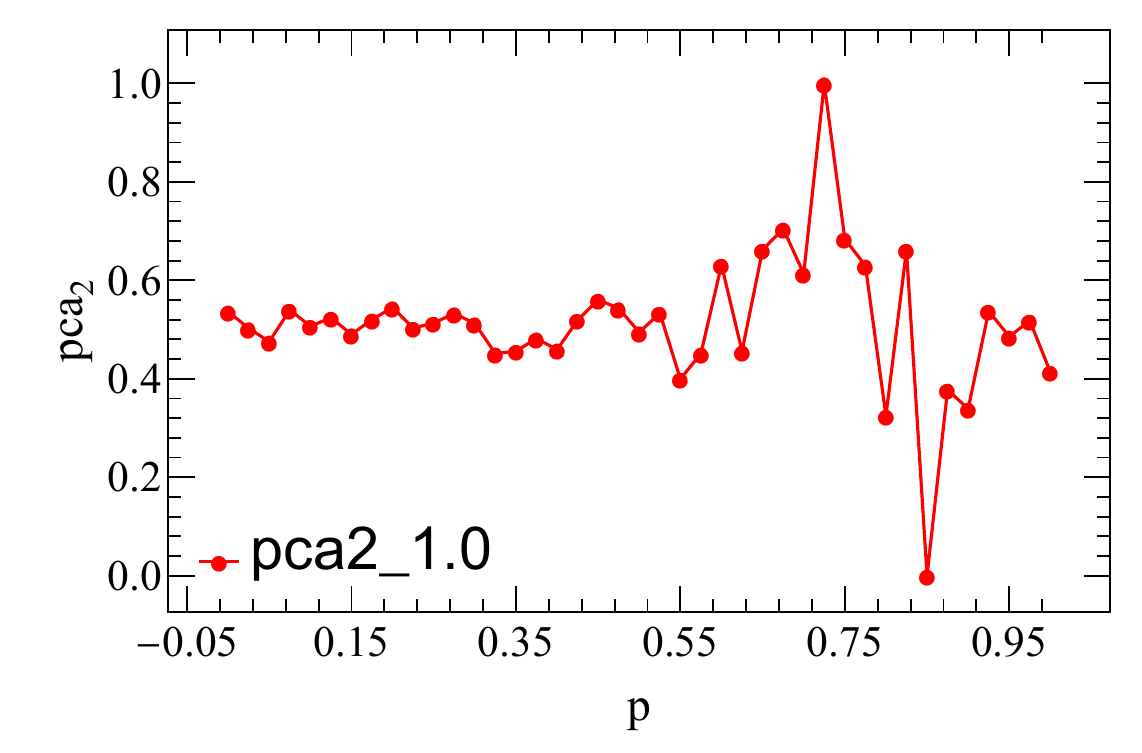}}
		\centerline{(i)}
	\end{minipage}	
	\caption{PCA results of (1+1)-dimensional bond DP for different shuffle ratios. The second principal component $p_{2}$ as a function of the bond probability $p$.}
	\label{ra_shuffle_pca}
\end{figure*}

Fig. \ref{v_shuffle_ratios} (c) shows the second principal component of different shuffle ratios as a function of the bond probability, and its detailed results on individual shuffle ratios can be seen in Fig. \ref{ra_shuffle_pca}. As can be seen from Fig. \ref{ra_shuffle_pca}, when the shuffle ratio increases to $r \sim 0.9$, the second principal component displays a large fluctuation, which indicates that the shuffle ratio can have a certain impact on the second principal component.

The analysis reveals that the PCA first principal component after shuffle is highly robust and its variation with the MC results is negligible. If one combines the shuffle results of the autoencoder's single latent variable and the PCA's first principal component, one finds that a certain rate of random shuffle does not noticeably alter the critical point's location in the system. In essence, with a constant particle density, the autoencoder's single latent variable and the PCA's first principal component consistently capture the system's transition point. This outcome validates our conjecture that both the autoencoder's single latent variable and the PCA's first principal component accurately capture the information related to the particle density.

\begin{table*}[h!]
	\centering
	\resizebox{\textwidth}{10mm}{
	\begin{tabular}{|c|c|c|c|c|c|c|c|c|c|c|c|}
        \hline
  shuffle ratio    &0.1 &0.2 &0.3 &0.4 &0.5&0.6&0.7&0.8&0.9&1.0 \\
 
        \hline
   Euclidean Distance & 0.011 &0.008  &0.006  &0.011  &0.022 &0.015 &0.020 &0.025 &0.020 &0.020 \\
        \hline
   \end{tabular}}
\caption{Euclidean Distance between MC and PCA's first principal component results for (1+1) dimensional DP with different shuffle ratios.}
\label{pca1_shuffle_euclidean}
\end{table*}

Concurrently, the shuffle does change the size of the maximum cluster. This is demonstrated in Fig. \ref{raw_shuffle_pca} (f), where the peak phenomenon is not observable in the PCA' second principal component. This compels us to postulate that the alteration in the PCA' second principal component might be connected to the correlation length.

\subsubsection{Learning with the maximum cluster}
\begin{figure*}[h]
\begin{tabular}{ccc}
    \includegraphics[width=0.45\textwidth]{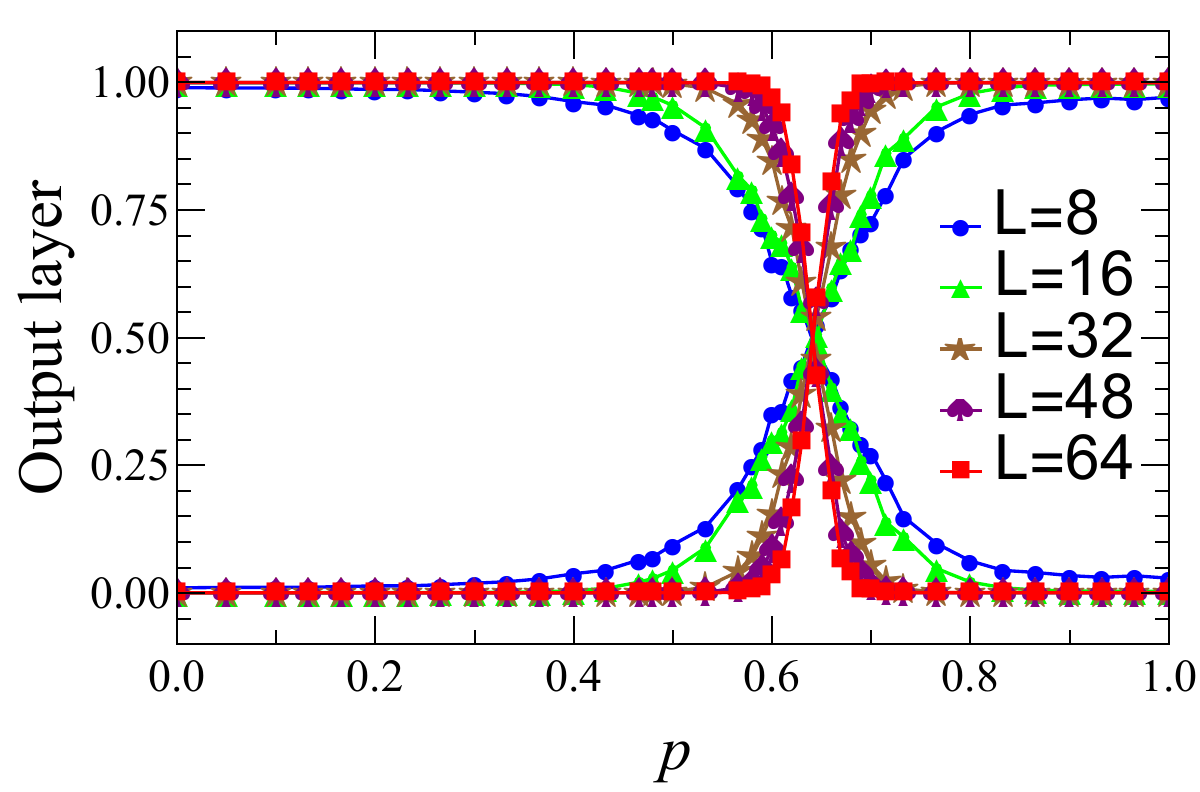} &
    \includegraphics[width=0.45\textwidth]{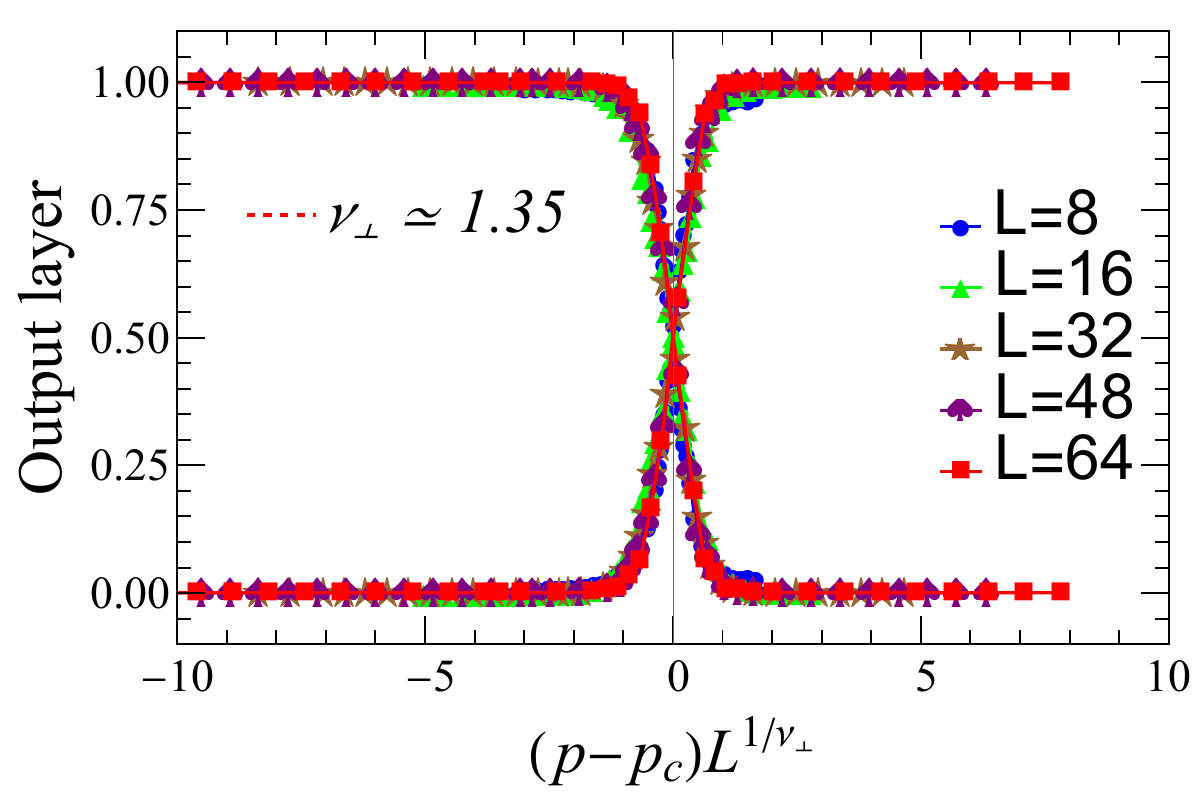} \\
     (a) &  (b)
\end{tabular}
\caption{An example of stochastic critical configurations of bond DP in (1+1) dimensions starting from a fully occupied lattice, as given by the left panel, where $L = 16$, the time length is $120$ and the corresponding bond probability $p$ is $0.6447$. The right panel is a shuffled version of the left one, where the particle density remains unchanged.}
\label{raw_shuffle_superML}
\end{figure*}

\begin{table*}[htbp]
	\centering
	\resizebox{\textwidth}{10mm}{
	\begin{tabular}{|c|c|c|c|c|c|c|c|c|c|c|c|c|}
        \hline
  shuffle ratio   & 0 &0.1 &0.2 &0.3 &0.4 &0.5 &0.6 &0.7 &0.8 &0.9 &1.0 \\
 
        \hline
   $\nu_{\perp}$ & 1.09(1) &1.12(1) & 1.15(1) &1.18(1) &1.20(1) &1.23(1) & 1.25(1)  &  1.27(1) &1.30(1) 
 &1.32(1) &1.35(1)  \\
        \hline
   \end{tabular}}
\caption{The spatial correlation exponent of (1+1)-dimensional DP varied with different shuffle ratios.}
\label{v_exponets_shuffle_ratio_table}
\end{table*}

The prior unsupervised learning of both the original and post-shuffle DP configurations has revealed that a single latent variable can effectively portray the particle density of the input data. It has been observed that the particle density remains unaltered even after shuffling, however, the correlation length is affected. To verify this discovery, supervised learning will be carried out to assess the spatial correlation exponent $\nu_{\perp}$ of DP's post-shuffle configurations at varying ratios.

Fig. \ref{raw_shuffle_superML} (a) displays the binary classification output of five systems with varying sizes, where the intersection of two outputs denotes the predictive critical point of the neural network. By adjusting $\nu_{\perp}$, the abscissa can be rescaled to collapse five sets of curves with different sizes. This approach yields the accurate value of the spatial correlation exponent. Fig. \ref{raw_shuffle_superML} (b) exhibits the scenario of shuffle ratio $r = 1.0$. The same method was employed to determine the spatial correlation exponents corresponding to different shuffle ratios, as demonstrated in Table \ref{v_exponets_shuffle_ratio_table}. Notably, the spatial correlation exponent increases progressively as the shuffle ratio increases, indicating that random shuffling alters the correlation between particles and reduces it.

Thus far, it appears that the single latent variable of the autoencoder neural network and the first principal component of PCA can detect the particle density provided that the particle density of the DP configuration remains constant regardless of the ratio used to shuffle it. We intend to confirm the validity of this perspective through further exploration. In line with this goal, we proceeded to shuffle the original DP configuration at a designated ratio and exclusively utilized the maximum cluster of each shuffled configuration for autoencoder and PCA learning. We emphasized the use of the maximum cluster since it relates to long-range correlation. Fig.\ref{raw_shuffle_max_config} (c) illustrates the maximum cluster in Fig.\ref{raw_shuffle_max_config} (b), i.e., the other points are removed.

\begin{figure}
\centering
\includegraphics[width=0.4\textwidth]{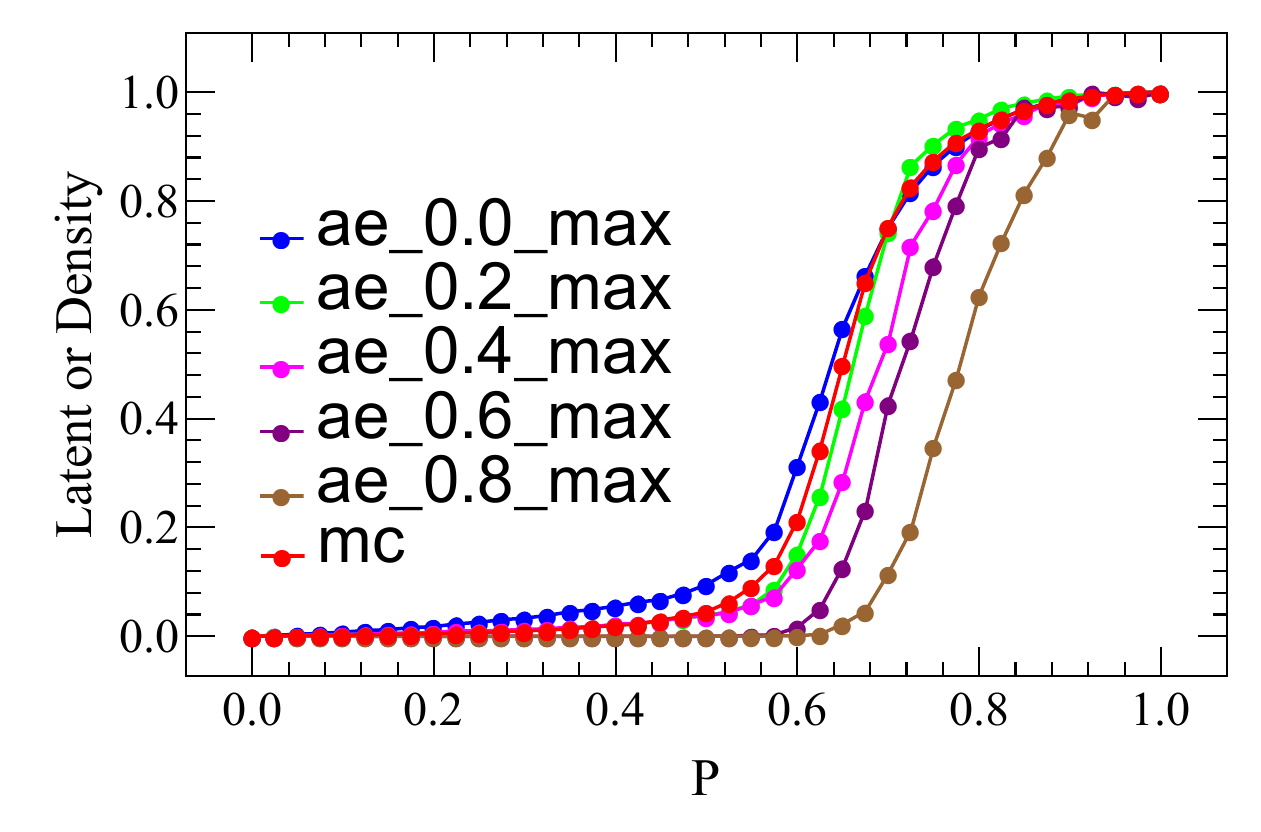}
\caption{Autoencoder results of the maximum cluster after shuffling the raw configurations for (1+1)-dimensional DP, where the lattice size is $N = 16$, the total number of time steps for averaging is $t = 120$, and the number of ensemble average is $200$.}
\label{raw_different_shuffled_ratios_max_ae_normal}
\end{figure}

The autoencoder result of the maximum cluster of DP is displayed in Fig. \ref{raw_different_shuffled_ratios_max_ae_normal}. It is observed that the estimated critical point location shifts continuously with an increase in the shuffle ratio. Specifically, the critical value increases. The third row of Table \ref{ae_shuffle_ratios} shows the detailed critical values. For instance, the critical point of the maximum cluster associated with the original configuration is $0.642(1)$. When the shuffle ratio is $1$, the critical point increases to $0.790(1)$. This finding is notable as it implies that changing the input configuration of autoencoder learning, which involves only the maximum cluster, alters the particle density. Additionally, after shuffling the original configuration using different shuffle ratios, the spatial correlation changes. As a result, the critical value of the new configuration is transferred, which aligns well with the outcomes obtained through supervised learning. To be more specific, it can be inferred that the original configurations demonstrate a rise in spatial correlation exponent $\nu_{\perp}$ and a decrease in correlation length subsequent to shuffle processing.

\begin{figure*}[h]
\begin{tabular}{ccc}
    \includegraphics[width=0.45\textwidth]{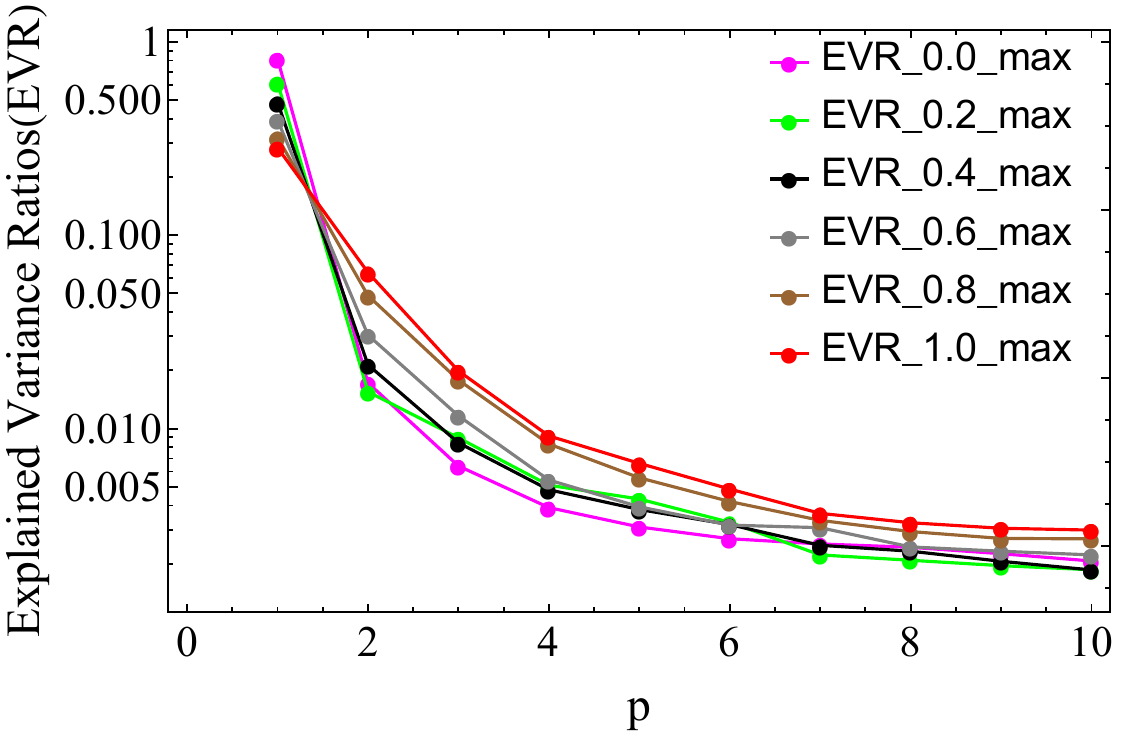} &
    \includegraphics[width=0.45\textwidth]{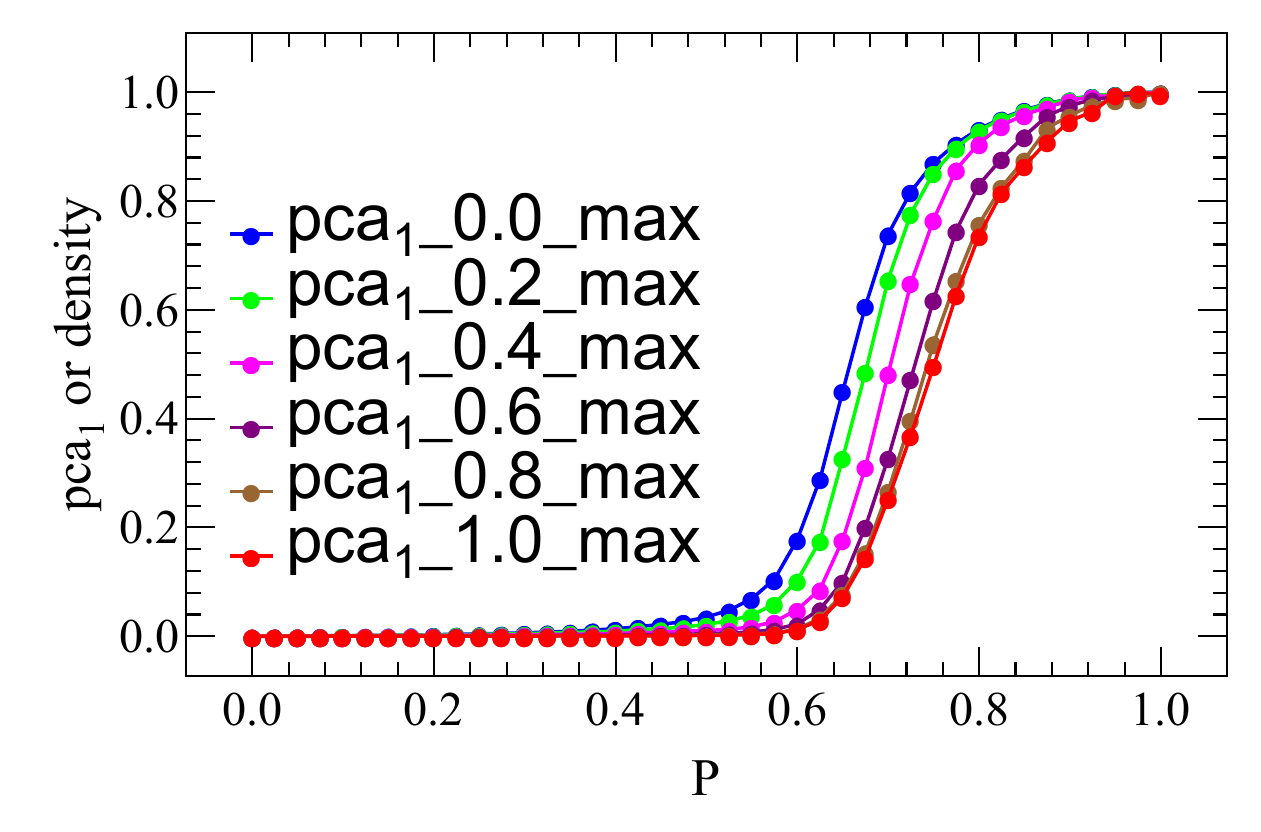} \\
     (a) &  (b)
\end{tabular}
\caption{PCA results of the maximum cluster of (1+1)-dimensional bond DP for different shuffle ratios. The left panel corresponds to the explained variance ratio $\widetilde{\lambda}_{\ell}$ from first ten principal components. The right panel is the first principal component $p_{1}$ as a function of the bond probability $p$.}
\label{v_shuffle_ratios_max}
\end{figure*}

\begin{figure*}[t]
	\begin{minipage}{0.3\linewidth}
		\centerline{\includegraphics[width=5.6cm]{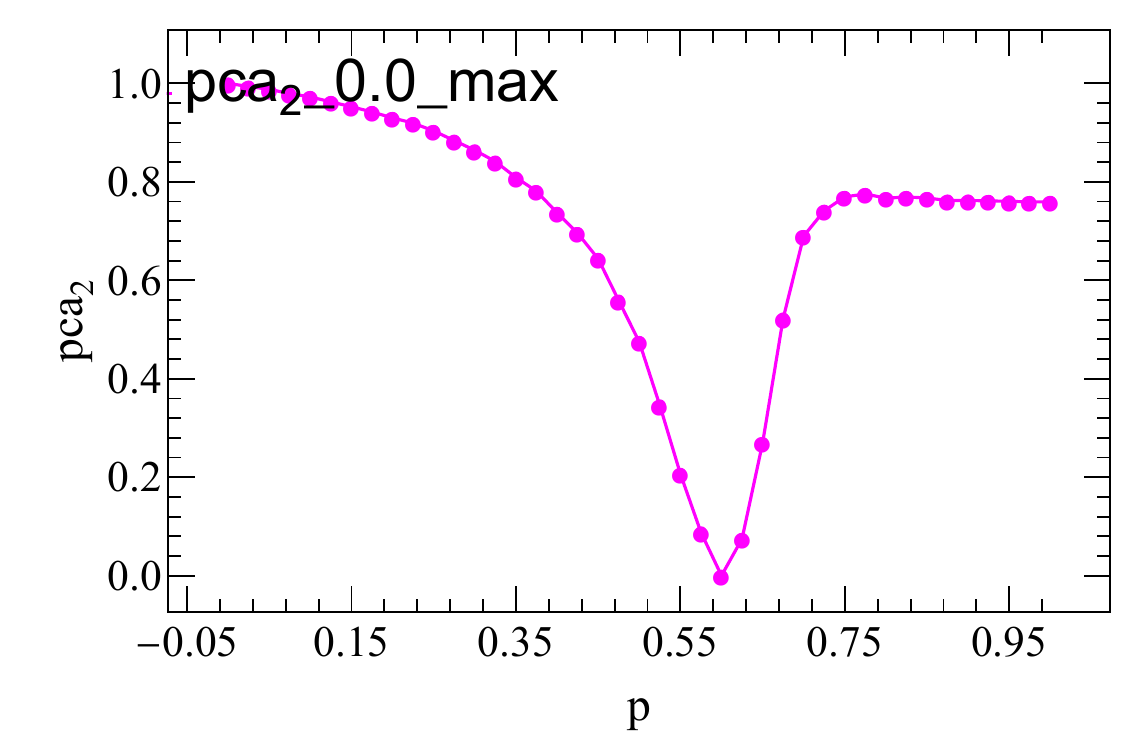}}
		\centerline{(a)}
	\end{minipage}
	\hfill
	\begin{minipage}{0.3\linewidth}
		\centerline{\includegraphics[width=5.6cm]{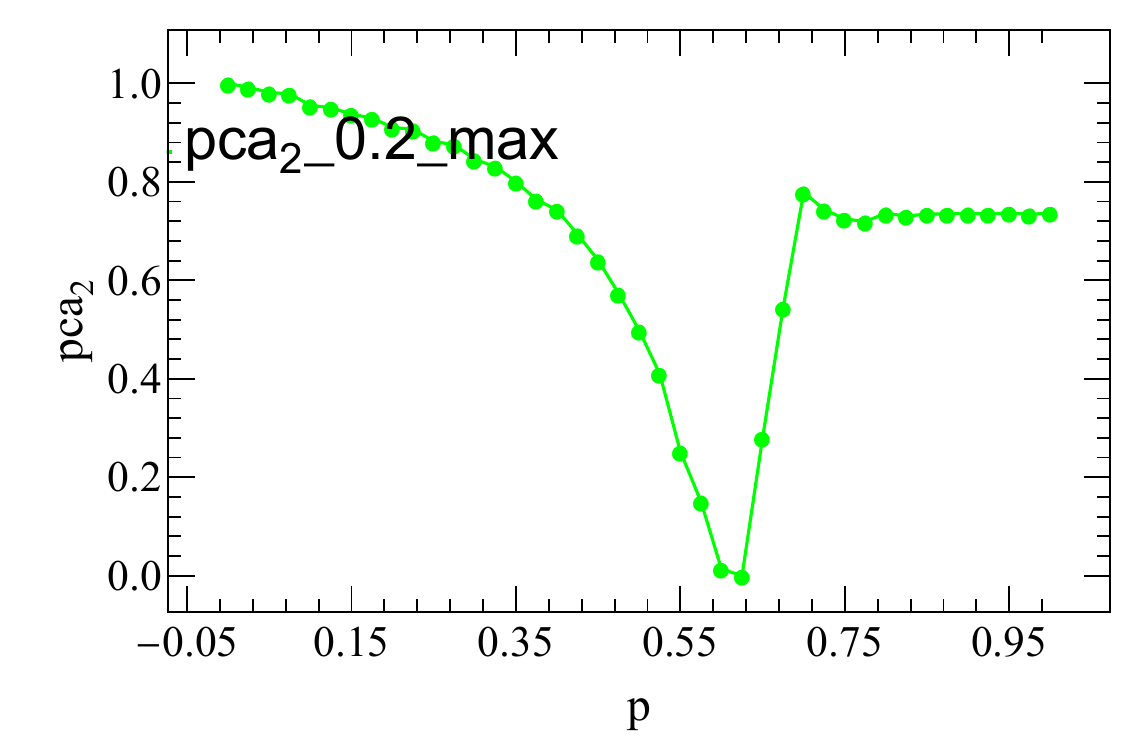}}
		\centerline{(b)}
	\end{minipage}
	\hfill
	\begin{minipage}{0.3\linewidth}
		\centerline{\includegraphics[width=5.6cm]{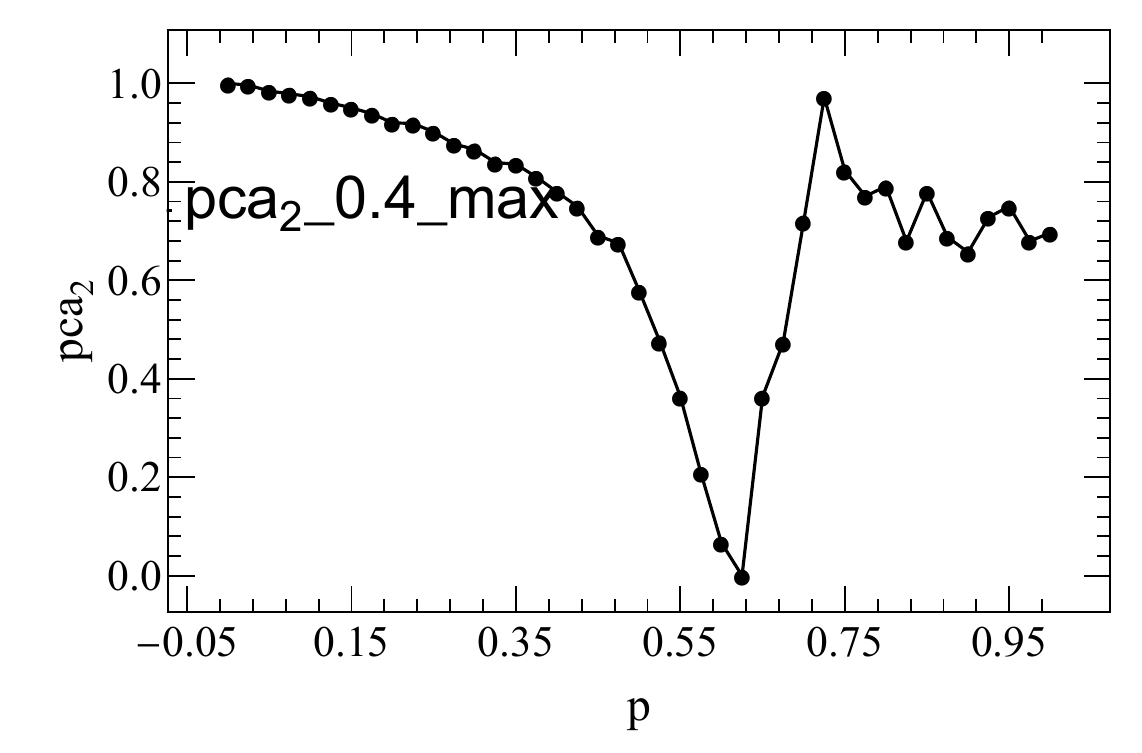}}
		\centerline{(c)}
	\end{minipage}
	\vfill
	\begin{minipage}{0.3\linewidth}
		\centerline{\includegraphics[width=5.6cm]{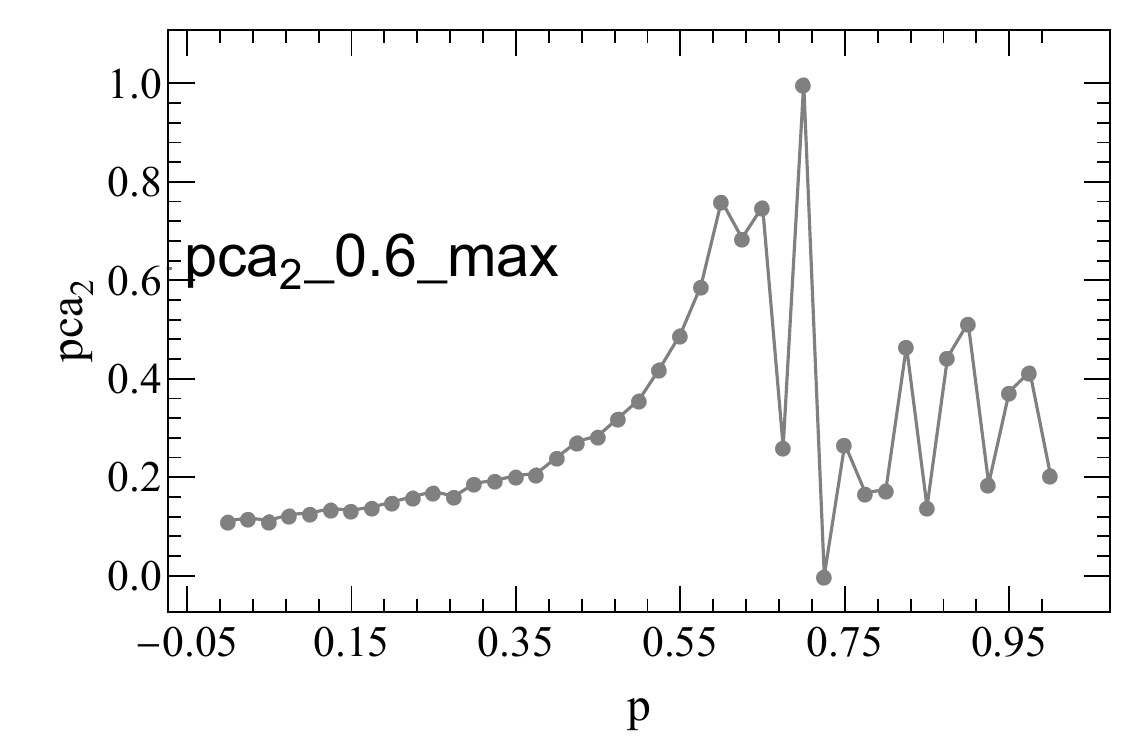}}
		\centerline{(d)}
	\end{minipage}
	\hfill
	\begin{minipage}{0.3\linewidth}
		\centerline{\includegraphics[width=5.6cm]{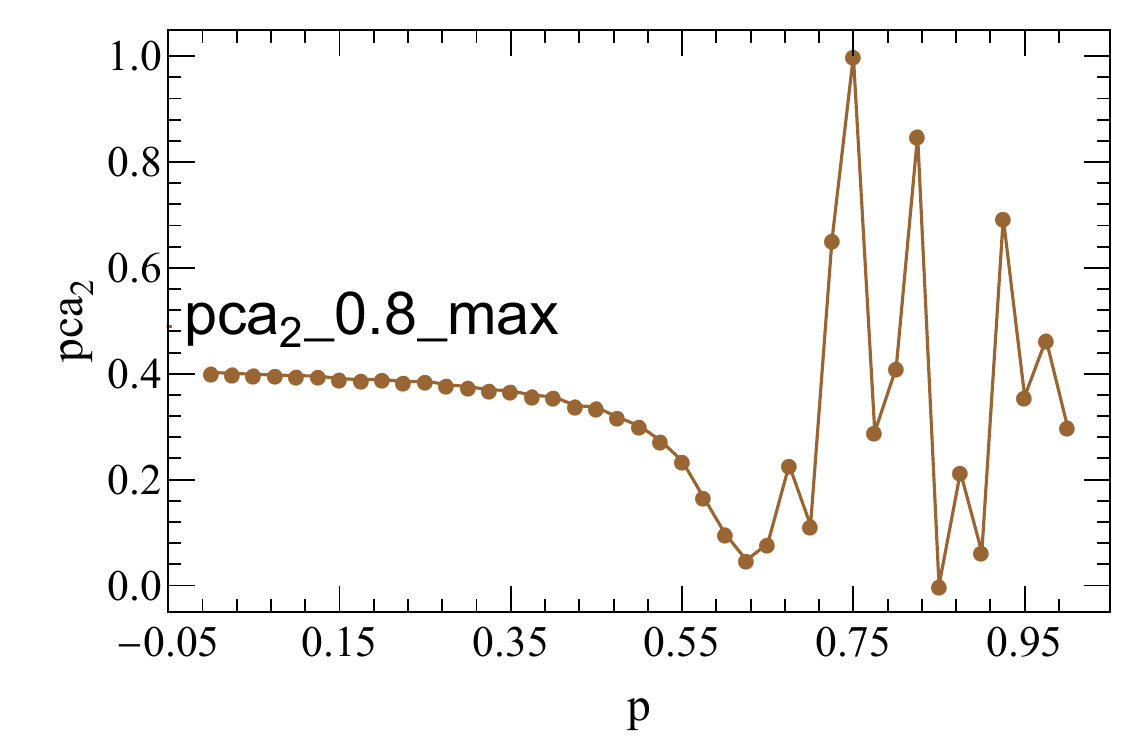}}
		\centerline{(e)}
	\end{minipage}
	\hfill
	\begin{minipage}{0.3\linewidth}
		\centerline{\includegraphics[width=5.6cm]{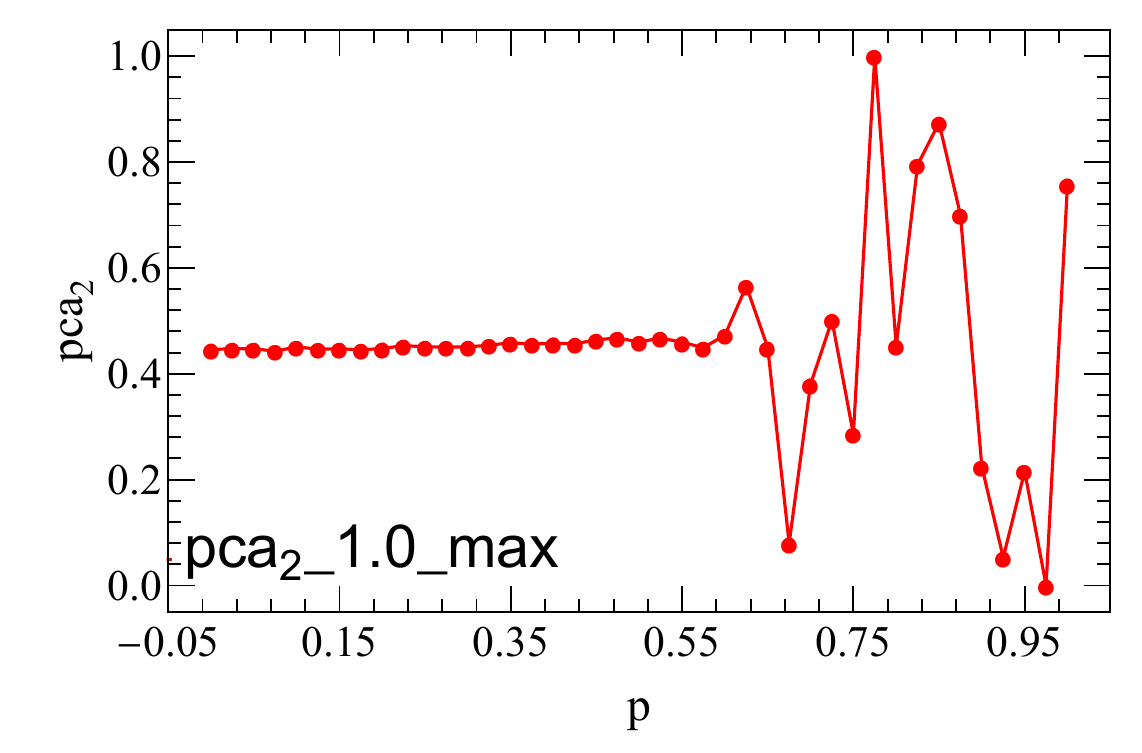}}
		\centerline{(f)}
	\end{minipage}
	\caption{PCA results of the maximum cluster of (1+1)-dimensional bond DP for different shuffle ratios. The second principal component $p_{2}$ as a function of the bond probability $p$.}
	\label{raw_shuffle_pca2_max}
\end{figure*}

Similar to the previous autoencoder learning, we also select the maximum cluster after shuffling with different ratios to perform PCA dimensionality reduction. Fig. \ref{v_shuffle_ratios_max} (a) shows the explained variance ratios of the corresponding maximum clusters. It can be found from Fig. \ref{v_shuffle_ratios_max} that with the increase of the shuffle ratio, the first principal component gradually decreases, and the gap between the first and second principal components becomes smaller. This indicates that the shuffle ratio has an impact on the proportion of each principal component, making the first principal component represent the reduction of the proportion of the input data. Fig. \ref{v_shuffle_ratios_max} (b) displays the first principal component as a function of the bond probability. The detailed results of the critical values are shown in the third row of Table \ref{pca1_pc_shuffle_ratio} based on the preserved maximum cluster. With the increase of shuffle ratio, the critical value obtained by the first principal component of PCA increases continuously. It is found that this is highly consistent with the result of a single latent variable of autoencoder. 

However, for the second principal component, a significant fluctuation occurs at $r \simeq 0.6$, as shown in Fig. \ref{raw_shuffle_pca2_max}. This leads us to conclude that both the first and second principal components of PCA have the capacity to learn the quantity associated with the correlation length. We note that alterations to the shuffle ratio result in changes to the amount of correlation length-related quantities acquired by PCA.

In Wang Lei's paper, it is conjectured that the first principal component of PCA has the ability to effectively represent the order parameter information of Ising model. Specifically, the order parameter of the Ising model is expressed by the mean magnetization, which is formulated as $M = \sum_i  s_{i}/N$. Similarly, for the DP model, the order parameter is the particle density. Upon performing PCA analysis on DP, it has been verified that the first principal component of PCA indeed serves as the order parameter, representing the particle density. The research on the single potential variable of autoencoder also suggests that a single potential variable is capable of representing the particle density of DP and is considered as the order parameter. Furthermore, by shuffling the configurations, it has been found that the single latent variable is not directly correlated to the particle position. Rather, the process of converting the particle density through shuffling and selecting the maximum cluster has an impact on the critical value of the system.

\section{Conclusion}
\label{Conclusion}
In this paper, the non-equilibrium DP model is mainly studied by using autoencoder neural networks and PCA, aiming to explore the physical significance of a single latent variable and the first principal component. Based on the detailed study of DP original configuration, we conjecture that autoencoder' single latent variable and PCA' first principal component can represent the order parameter of the model, that is, the particle density. 

To examine this conjecture, we apply a particular degree of random shuffling to the original configurations of DP and subsequently conducted organized single latent variable extraction through the autoencoder and first principal component extraction through PCA on these shuffled configurations. Through the unsupervised learning of shuffling, we determined that the autoencoder's single latent variables and PCA's first principal component could significantly represent the particle density of the DP model.

However, it is noted that the shuffling of the system has an effect on the maximum cluster. As a result, we hypothesize that the PCA second principal component may be linked to the maximal cluster, given that manipulating the shuffle ratio significantly altered the PCA second principal component. To explore this phenomenon, we carry out a maximum cluster extraction on the DP model and conducted an autoencoder and PCA analysis on it. Additionally, by maintaining the maximum cluster from various shuffle ratios, we discover that this operation altered the correlation of DP's configuration, leading to a certain degree of shift in the critical point position of the system.

The results show that the single latent variable of autoencoder can effectively represent the order parameter of the DP model, while the first principal component of PCA can represent the same physical meaning to a large extent. This research clears the way for the study of DP-liked systems of nonequilibrium phase transitions, which allows us to use unsupervised learning to effectively extract the critical points of phase transition systems and perform phase classification. On a larger scale, this is also useful for many phase transition systems in statistical physics and condensed matter physics if their order parameters can be expressed in terms of particle density, or if their order parameters can be converted into quantities dependent on particle density.

\section{Acknowledgements}
This work was supported in part by Key Laboratory of Quark and Lepton Physics (MOE), Central China Normal University (Grant No.QLPL2022P01), National Research Incubation Fund of Baoshan University(BYPY202216), Startup Fund for Phd of Baoshan University(BSKY202305), Wen Bangchun Academician Workstation(202205AF150032), the Fundamental Research Funds for the Central Universities, China (Grant No. CCNU19QN029), the National Natural Science Foundation of China (Grant No. 61873104 and 11947063), the Natural Science Foundation of Hubei Province of China through the project 2023AFB721, and the 111 Project 2.0 with Grant No. BP0820038.

\bibliographystyle{apsrev4-2}
\bibliography{apssamp}

\begin{thebibliography}{46}%
\makeatletter
\providecommand \@ifxundefined [1]{%
 \@ifx{#1\undefined}
}%
\providecommand \@ifnum [1]{%
 \ifnum #1\expandafter \@firstoftwo
 \else \expandafter \@secondoftwo
 \fi
}%
\providecommand \@ifx [1]{%
 \ifx #1\expandafter \@firstoftwo
 \else \expandafter \@secondoftwo
 \fi
}%
\providecommand \natexlab [1]{#1}%
\providecommand \enquote  [1]{``#1''}%
\providecommand \bibnamefont  [1]{#1}%
\providecommand \bibfnamefont [1]{#1}%
\providecommand \citenamefont [1]{#1}%
\providecommand \href@noop [0]{\@secondoftwo}%
\providecommand \href [0]{\begingroup \@sanitize@url \@href}%
\providecommand \@href[1]{\@@startlink{#1}\@@href}%
\providecommand \@@href[1]{\endgroup#1\@@endlink}%
\providecommand \@sanitize@url [0]{\catcode `\\12\catcode `\$12\catcode
  `\&12\catcode `\#12\catcode `\^12\catcode `\_12\catcode `\%12\relax}%
\providecommand \@@startlink[1]{}%
\providecommand \@@endlink[0]{}%
\providecommand \url  [0]{\begingroup\@sanitize@url \@url }%
\providecommand \@url [1]{\endgroup\@href {#1}{\urlprefix }}%
\providecommand \urlprefix  [0]{URL }%
\providecommand \Eprint [0]{\href }%
\providecommand \doibase [0]{https://doi.org/}%
\providecommand \selectlanguage [0]{\@gobble}%
\providecommand \bibinfo  [0]{\@secondoftwo}%
\providecommand \bibfield  [0]{\@secondoftwo}%
\providecommand \translation [1]{[#1]}%
\providecommand \BibitemOpen [0]{}%
\providecommand \bibitemStop [0]{}%
\providecommand \bibitemNoStop [0]{.\EOS\space}%
\providecommand \EOS [0]{\spacefactor3000\relax}%
\providecommand \BibitemShut  [1]{\csname bibitem#1\endcsname}%
\let\auto@bib@innerbib\@empty
\bibitem [{\citenamefont {Jordan}\ and\ \citenamefont
  {Mitchell}(2015)}]{jordan2015machine}%
  \BibitemOpen
  \bibfield  {author} {\bibinfo {author} {\bibfnamefont {M.~I.}\ \bibnamefont
  {Jordan}}\ and\ \bibinfo {author} {\bibfnamefont {T.~M.}\ \bibnamefont
  {Mitchell}},\ }\href@noop {} {\bibfield  {journal} {\bibinfo  {journal}
  {Science}\ }\textbf {\bibinfo {volume} {349}},\ \bibinfo {pages} {255}
  (\bibinfo {year} {2015})}\BibitemShut {NoStop}%
\bibitem [{\citenamefont {Goodfellow}\ \emph
  {et~al.}(2016{\natexlab{a}})\citenamefont {Goodfellow}, \citenamefont
  {Bengio},\ and\ \citenamefont {Courville}}]{goodfellow2016machine}%
  \BibitemOpen
  \bibfield  {author} {\bibinfo {author} {\bibfnamefont {I.}~\bibnamefont
  {Goodfellow}}, \bibinfo {author} {\bibfnamefont {Y.}~\bibnamefont {Bengio}},\
  and\ \bibinfo {author} {\bibfnamefont {A.}~\bibnamefont {Courville}},\
  }\href@noop {} {\bibfield  {journal} {\bibinfo  {journal} {Deep learning}\
  }\textbf {\bibinfo {volume} {1}},\ \bibinfo {pages} {98} (\bibinfo {year}
  {2016}{\natexlab{a}})}\BibitemShut {NoStop}%
\bibitem [{\citenamefont {Carleo}\ \emph {et~al.}(2019)\citenamefont {Carleo},
  \citenamefont {Cirac}, \citenamefont {Cranmer}, \citenamefont {Daudet},
  \citenamefont {Schuld}, \citenamefont {Tishby}, \citenamefont
  {Vogt-Maranto},\ and\ \citenamefont {Zdeborov{\'a}}}]{carleo2019machine}%
  \BibitemOpen
  \bibfield  {author} {\bibinfo {author} {\bibfnamefont {G.}~\bibnamefont
  {Carleo}}, \bibinfo {author} {\bibfnamefont {I.}~\bibnamefont {Cirac}},
  \bibinfo {author} {\bibfnamefont {K.}~\bibnamefont {Cranmer}}, \bibinfo
  {author} {\bibfnamefont {L.}~\bibnamefont {Daudet}}, \bibinfo {author}
  {\bibfnamefont {M.}~\bibnamefont {Schuld}}, \bibinfo {author} {\bibfnamefont
  {N.}~\bibnamefont {Tishby}}, \bibinfo {author} {\bibfnamefont
  {L.}~\bibnamefont {Vogt-Maranto}},\ and\ \bibinfo {author} {\bibfnamefont
  {L.}~\bibnamefont {Zdeborov{\'a}}},\ }\href@noop {} {\bibfield  {journal}
  {\bibinfo  {journal} {Reviews of Modern Physics}\ }\textbf {\bibinfo {volume}
  {91}},\ \bibinfo {pages} {045002} (\bibinfo {year} {2019})}\BibitemShut
  {NoStop}%
\bibitem [{\citenamefont {Mehta}\ \emph {et~al.}(2019)\citenamefont {Mehta},
  \citenamefont {Bukov}, \citenamefont {Wang}, \citenamefont {Day},
  \citenamefont {Richardson}, \citenamefont {Fisher},\ and\ \citenamefont
  {Schwab}}]{mehta2019high}%
  \BibitemOpen
  \bibfield  {author} {\bibinfo {author} {\bibfnamefont {P.}~\bibnamefont
  {Mehta}}, \bibinfo {author} {\bibfnamefont {M.}~\bibnamefont {Bukov}},
  \bibinfo {author} {\bibfnamefont {C.-H.}\ \bibnamefont {Wang}}, \bibinfo
  {author} {\bibfnamefont {A.~G.}\ \bibnamefont {Day}}, \bibinfo {author}
  {\bibfnamefont {C.}~\bibnamefont {Richardson}}, \bibinfo {author}
  {\bibfnamefont {C.~K.}\ \bibnamefont {Fisher}},\ and\ \bibinfo {author}
  {\bibfnamefont {D.~J.}\ \bibnamefont {Schwab}},\ }\href@noop {} {\bibfield
  {journal} {\bibinfo  {journal} {Physics reports}\ }\textbf {\bibinfo {volume}
  {810}},\ \bibinfo {pages} {1} (\bibinfo {year} {2019})}\BibitemShut {NoStop}%
\bibitem [{\citenamefont {Carrasquilla}\ and\ \citenamefont
  {Melko}(2017)}]{carrasquilla2017machine}%
  \BibitemOpen
  \bibfield  {author} {\bibinfo {author} {\bibfnamefont {J.}~\bibnamefont
  {Carrasquilla}}\ and\ \bibinfo {author} {\bibfnamefont {R.~G.}\ \bibnamefont
  {Melko}},\ }\href@noop {} {\bibfield  {journal} {\bibinfo  {journal} {Nature
  Physics}\ }\textbf {\bibinfo {volume} {13}},\ \bibinfo {pages} {431}
  (\bibinfo {year} {2017})}\BibitemShut {NoStop}%
\bibitem [{\citenamefont {Wang}(2016)}]{wang2016discovering}%
  \BibitemOpen
  \bibfield  {author} {\bibinfo {author} {\bibfnamefont {L.}~\bibnamefont
  {Wang}},\ }\href@noop {} {\bibfield  {journal} {\bibinfo  {journal} {Physical
  Review B}\ }\textbf {\bibinfo {volume} {94}},\ \bibinfo {pages} {195105}
  (\bibinfo {year} {2016})}\BibitemShut {NoStop}%
\bibitem [{\citenamefont {Zhang}\ \emph {et~al.}(2019)\citenamefont {Zhang},
  \citenamefont {Liu},\ and\ \citenamefont {Wei}}]{zhang2019machine}%
  \BibitemOpen
  \bibfield  {author} {\bibinfo {author} {\bibfnamefont {W.}~\bibnamefont
  {Zhang}}, \bibinfo {author} {\bibfnamefont {J.}~\bibnamefont {Liu}},\ and\
  \bibinfo {author} {\bibfnamefont {T.-C.}\ \bibnamefont {Wei}},\ }\href@noop
  {} {\bibfield  {journal} {\bibinfo  {journal} {Physical Review E}\ }\textbf
  {\bibinfo {volume} {99}},\ \bibinfo {pages} {032142} (\bibinfo {year}
  {2019})}\BibitemShut {NoStop}%
\bibitem [{\citenamefont {Li}\ \emph {et~al.}(2018)\citenamefont {Li},
  \citenamefont {Tan},\ and\ \citenamefont {Jiang}}]{li2018applications}%
  \BibitemOpen
  \bibfield  {author} {\bibinfo {author} {\bibfnamefont {C.-D.}\ \bibnamefont
  {Li}}, \bibinfo {author} {\bibfnamefont {D.-R.}\ \bibnamefont {Tan}},\ and\
  \bibinfo {author} {\bibfnamefont {F.-J.}\ \bibnamefont {Jiang}},\ }\href@noop
  {} {\bibfield  {journal} {\bibinfo  {journal} {Annals of Physics}\ }\textbf
  {\bibinfo {volume} {391}},\ \bibinfo {pages} {312} (\bibinfo {year}
  {2018})}\BibitemShut {NoStop}%
\bibitem [{\citenamefont {Tang}\ \emph {et~al.}(2024)\citenamefont {Tang},
  \citenamefont {Liu}, \citenamefont {Zhang},\ and\ \citenamefont
  {Zhang}}]{tang2024learning}%
  \BibitemOpen
  \bibfield  {author} {\bibinfo {author} {\bibfnamefont {Y.}~\bibnamefont
  {Tang}}, \bibinfo {author} {\bibfnamefont {J.}~\bibnamefont {Liu}}, \bibinfo
  {author} {\bibfnamefont {J.}~\bibnamefont {Zhang}},\ and\ \bibinfo {author}
  {\bibfnamefont {P.}~\bibnamefont {Zhang}},\ }\href@noop {} {\bibfield
  {journal} {\bibinfo  {journal} {Nature Communications}\ }\textbf {\bibinfo
  {volume} {15}},\ \bibinfo {pages} {1117} (\bibinfo {year}
  {2024})}\BibitemShut {NoStop}%
\bibitem [{\citenamefont {Jo}\ \emph {et~al.}(2021)\citenamefont {Jo},
  \citenamefont {Lee}, \citenamefont {Choi},\ and\ \citenamefont
  {Kahng}}]{jo2021absorbing}%
  \BibitemOpen
  \bibfield  {author} {\bibinfo {author} {\bibfnamefont {M.}~\bibnamefont
  {Jo}}, \bibinfo {author} {\bibfnamefont {J.}~\bibnamefont {Lee}}, \bibinfo
  {author} {\bibfnamefont {K.}~\bibnamefont {Choi}},\ and\ \bibinfo {author}
  {\bibfnamefont {B.}~\bibnamefont {Kahng}},\ }\href@noop {} {\bibfield
  {journal} {\bibinfo  {journal} {Physical Review Research}\ }\textbf {\bibinfo
  {volume} {3}},\ \bibinfo {pages} {013238} (\bibinfo {year}
  {2021})}\BibitemShut {NoStop}%
\bibitem [{\citenamefont {Shen}\ \emph {et~al.}(2021)\citenamefont {Shen},
  \citenamefont {Li}, \citenamefont {Deng},\ and\ \citenamefont
  {Zhang}}]{shen2021supervised}%
  \BibitemOpen
  \bibfield  {author} {\bibinfo {author} {\bibfnamefont {J.}~\bibnamefont
  {Shen}}, \bibinfo {author} {\bibfnamefont {W.}~\bibnamefont {Li}}, \bibinfo
  {author} {\bibfnamefont {S.}~\bibnamefont {Deng}},\ and\ \bibinfo {author}
  {\bibfnamefont {T.}~\bibnamefont {Zhang}},\ }\href@noop {} {\bibfield
  {journal} {\bibinfo  {journal} {Physical Review E}\ }\textbf {\bibinfo
  {volume} {103}},\ \bibinfo {pages} {052140} (\bibinfo {year}
  {2021})}\BibitemShut {NoStop}%
\bibitem [{\citenamefont {Shen}\ \emph
  {et~al.}(2022{\natexlab{a}})\citenamefont {Shen}, \citenamefont {Liu},
  \citenamefont {Chen}, \citenamefont {Xu}, \citenamefont {Chen}, \citenamefont
  {Deng}, \citenamefont {Li}, \citenamefont {Papp},\ and\ \citenamefont
  {Yang}}]{shen2022transfer}%
  \BibitemOpen
  \bibfield  {author} {\bibinfo {author} {\bibfnamefont {J.}~\bibnamefont
  {Shen}}, \bibinfo {author} {\bibfnamefont {F.}~\bibnamefont {Liu}}, \bibinfo
  {author} {\bibfnamefont {S.}~\bibnamefont {Chen}}, \bibinfo {author}
  {\bibfnamefont {D.}~\bibnamefont {Xu}}, \bibinfo {author} {\bibfnamefont
  {X.}~\bibnamefont {Chen}}, \bibinfo {author} {\bibfnamefont {S.}~\bibnamefont
  {Deng}}, \bibinfo {author} {\bibfnamefont {W.}~\bibnamefont {Li}}, \bibinfo
  {author} {\bibfnamefont {G.}~\bibnamefont {Papp}},\ and\ \bibinfo {author}
  {\bibfnamefont {C.}~\bibnamefont {Yang}},\ }\href@noop {} {\bibfield
  {journal} {\bibinfo  {journal} {Physical Review E}\ }\textbf {\bibinfo
  {volume} {105}},\ \bibinfo {pages} {064139} (\bibinfo {year}
  {2022}{\natexlab{a}})}\BibitemShut {NoStop}%
\bibitem [{\citenamefont {Shen}\ \emph
  {et~al.}(2022{\natexlab{b}})\citenamefont {Shen}, \citenamefont {Li},
  \citenamefont {Deng}, \citenamefont {Xu}, \citenamefont {Chen},\ and\
  \citenamefont {Liu}}]{shen2022machine}%
  \BibitemOpen
  \bibfield  {author} {\bibinfo {author} {\bibfnamefont {J.}~\bibnamefont
  {Shen}}, \bibinfo {author} {\bibfnamefont {W.}~\bibnamefont {Li}}, \bibinfo
  {author} {\bibfnamefont {S.}~\bibnamefont {Deng}}, \bibinfo {author}
  {\bibfnamefont {D.}~\bibnamefont {Xu}}, \bibinfo {author} {\bibfnamefont
  {S.}~\bibnamefont {Chen}},\ and\ \bibinfo {author} {\bibfnamefont
  {F.}~\bibnamefont {Liu}},\ }\href@noop {} {\bibfield  {journal} {\bibinfo
  {journal} {Scientific Reports}\ }\textbf {\bibinfo {volume} {12}},\ \bibinfo
  {pages} {19728} (\bibinfo {year} {2022}{\natexlab{b}})}\BibitemShut {NoStop}%
\bibitem [{\citenamefont {Carrasquilla}(2020)}]{carrasquilla2020machine}%
  \BibitemOpen
  \bibfield  {author} {\bibinfo {author} {\bibfnamefont {J.}~\bibnamefont
  {Carrasquilla}},\ }\href@noop {} {\bibfield  {journal} {\bibinfo  {journal}
  {Advances in Physics: X}\ }\textbf {\bibinfo {volume} {5}},\ \bibinfo {pages}
  {1797528} (\bibinfo {year} {2020})}\BibitemShut {NoStop}%
\bibitem [{\citenamefont {Sarma}\ \emph {et~al.}(2019)\citenamefont {Sarma},
  \citenamefont {Deng},\ and\ \citenamefont {Duan}}]{sarma2019machine}%
  \BibitemOpen
  \bibfield  {author} {\bibinfo {author} {\bibfnamefont {S.~D.}\ \bibnamefont
  {Sarma}}, \bibinfo {author} {\bibfnamefont {D.-L.}\ \bibnamefont {Deng}},\
  and\ \bibinfo {author} {\bibfnamefont {L.-M.}\ \bibnamefont {Duan}},\
  }\href@noop {} {\bibfield  {journal} {\bibinfo  {journal} {arXiv preprint
  arXiv:1903.03516}\ } (\bibinfo {year} {2019})}\BibitemShut {NoStop}%
\bibitem [{\citenamefont {Rodriguez-Nieva}\ and\ \citenamefont
  {Scheurer}(2019)}]{rodriguez2019identifying}%
  \BibitemOpen
  \bibfield  {author} {\bibinfo {author} {\bibfnamefont {J.~F.}\ \bibnamefont
  {Rodriguez-Nieva}}\ and\ \bibinfo {author} {\bibfnamefont {M.~S.}\
  \bibnamefont {Scheurer}},\ }\href@noop {} {\bibfield  {journal} {\bibinfo
  {journal} {Nature Physics}\ }\textbf {\bibinfo {volume} {15}},\ \bibinfo
  {pages} {790} (\bibinfo {year} {2019})}\BibitemShut {NoStop}%
\bibitem [{\citenamefont {Deng}\ \emph {et~al.}(2017)\citenamefont {Deng},
  \citenamefont {Li},\ and\ \citenamefont {Sarma}}]{deng2017machine}%
  \BibitemOpen
  \bibfield  {author} {\bibinfo {author} {\bibfnamefont {D.-L.}\ \bibnamefont
  {Deng}}, \bibinfo {author} {\bibfnamefont {X.}~\bibnamefont {Li}},\ and\
  \bibinfo {author} {\bibfnamefont {S.~D.}\ \bibnamefont {Sarma}},\ }\href@noop
  {} {\bibfield  {journal} {\bibinfo  {journal} {Physical Review B}\ }\textbf
  {\bibinfo {volume} {96}},\ \bibinfo {pages} {195145} (\bibinfo {year}
  {2017})}\BibitemShut {NoStop}%
\bibitem [{\citenamefont {Huang}(2021)}]{huang2021statistical}%
  \BibitemOpen
  \bibfield  {author} {\bibinfo {author} {\bibfnamefont {H.}~\bibnamefont
  {Huang}},\ }\href@noop {} {\emph {\bibinfo {title} {Statistical mechanics of
  neural networks}}}\ (\bibinfo  {publisher} {Springer},\ \bibinfo {year}
  {2021})\BibitemShut {NoStop}%
\bibitem [{\citenamefont {Karniadakis}\ \emph {et~al.}(2021)\citenamefont
  {Karniadakis}, \citenamefont {Kevrekidis}, \citenamefont {Lu}, \citenamefont
  {Perdikaris}, \citenamefont {Wang},\ and\ \citenamefont
  {Yang}}]{karniadakis2021physics}%
  \BibitemOpen
  \bibfield  {author} {\bibinfo {author} {\bibfnamefont {G.~E.}\ \bibnamefont
  {Karniadakis}}, \bibinfo {author} {\bibfnamefont {I.~G.}\ \bibnamefont
  {Kevrekidis}}, \bibinfo {author} {\bibfnamefont {L.}~\bibnamefont {Lu}},
  \bibinfo {author} {\bibfnamefont {P.}~\bibnamefont {Perdikaris}}, \bibinfo
  {author} {\bibfnamefont {S.}~\bibnamefont {Wang}},\ and\ \bibinfo {author}
  {\bibfnamefont {L.}~\bibnamefont {Yang}},\ }\href@noop {} {\bibfield
  {journal} {\bibinfo  {journal} {Nature Reviews Physics}\ }\textbf {\bibinfo
  {volume} {3}},\ \bibinfo {pages} {422} (\bibinfo {year} {2021})}\BibitemShut
  {NoStop}%
\bibitem [{\citenamefont {Wetzel}(2017)}]{wetzel2017unsupervised}%
  \BibitemOpen
  \bibfield  {author} {\bibinfo {author} {\bibfnamefont {S.~J.}\ \bibnamefont
  {Wetzel}},\ }\href@noop {} {\bibfield  {journal} {\bibinfo  {journal}
  {Physical Review E}\ }\textbf {\bibinfo {volume} {96}},\ \bibinfo {pages}
  {022140} (\bibinfo {year} {2017})}\BibitemShut {NoStop}%
\bibitem [{\citenamefont {Ponte}\ and\ \citenamefont
  {Melko}(2017)}]{ponte2017kernel}%
  \BibitemOpen
  \bibfield  {author} {\bibinfo {author} {\bibfnamefont {P.}~\bibnamefont
  {Ponte}}\ and\ \bibinfo {author} {\bibfnamefont {R.~G.}\ \bibnamefont
  {Melko}},\ }\href@noop {} {\bibfield  {journal} {\bibinfo  {journal}
  {Physical Review B}\ }\textbf {\bibinfo {volume} {96}},\ \bibinfo {pages}
  {205146} (\bibinfo {year} {2017})}\BibitemShut {NoStop}%
\bibitem [{\citenamefont {Wetzel}\ and\ \citenamefont
  {Scherzer}(2017)}]{wetzel2017machine}%
  \BibitemOpen
  \bibfield  {author} {\bibinfo {author} {\bibfnamefont {S.~J.}\ \bibnamefont
  {Wetzel}}\ and\ \bibinfo {author} {\bibfnamefont {M.}~\bibnamefont
  {Scherzer}},\ }\href@noop {} {\bibfield  {journal} {\bibinfo  {journal}
  {Physical Review B}\ }\textbf {\bibinfo {volume} {96}},\ \bibinfo {pages}
  {184410} (\bibinfo {year} {2017})}\BibitemShut {NoStop}%
\bibitem [{\citenamefont {Jadrich}\ \emph {et~al.}(2018)\citenamefont
  {Jadrich}, \citenamefont {Lindquist},\ and\ \citenamefont
  {Truskett}}]{jadrich2018unsupervised}%
  \BibitemOpen
  \bibfield  {author} {\bibinfo {author} {\bibfnamefont {R.}~\bibnamefont
  {Jadrich}}, \bibinfo {author} {\bibfnamefont {B.}~\bibnamefont {Lindquist}},\
  and\ \bibinfo {author} {\bibfnamefont {T.}~\bibnamefont {Truskett}},\
  }\href@noop {} {\bibfield  {journal} {\bibinfo  {journal} {The Journal of
  chemical physics}\ }\textbf {\bibinfo {volume} {149}},\ \bibinfo {pages}
  {194109} (\bibinfo {year} {2018})}\BibitemShut {NoStop}%
\bibitem [{\citenamefont {Bourlard}\ and\ \citenamefont
  {Kamp}(1988)}]{bourlard1988auto}%
  \BibitemOpen
  \bibfield  {author} {\bibinfo {author} {\bibfnamefont {H.}~\bibnamefont
  {Bourlard}}\ and\ \bibinfo {author} {\bibfnamefont {Y.}~\bibnamefont
  {Kamp}},\ }\href@noop {} {\bibfield  {journal} {\bibinfo  {journal}
  {Biological cybernetics}\ }\textbf {\bibinfo {volume} {59}},\ \bibinfo
  {pages} {291} (\bibinfo {year} {1988})}\BibitemShut {NoStop}%
\bibitem [{\citenamefont {Hinton}\ and\ \citenamefont
  {Zemel}(1994)}]{hinton1994autoencoders}%
  \BibitemOpen
  \bibfield  {author} {\bibinfo {author} {\bibfnamefont {G.~E.}\ \bibnamefont
  {Hinton}}\ and\ \bibinfo {author} {\bibfnamefont {R.~S.}\ \bibnamefont
  {Zemel}},\ }\href@noop {} {\bibfield  {journal} {\bibinfo  {journal}
  {Advances in neural information processing systems}\ }\textbf {\bibinfo
  {volume} {6}},\ \bibinfo {pages} {3} (\bibinfo {year} {1994})}\BibitemShut
  {NoStop}%
\bibitem [{\citenamefont {Hinton}\ and\ \citenamefont
  {Salakhutdinov}(2006)}]{hinton2006reducing}%
  \BibitemOpen
  \bibfield  {author} {\bibinfo {author} {\bibfnamefont {G.~E.}\ \bibnamefont
  {Hinton}}\ and\ \bibinfo {author} {\bibfnamefont {R.~R.}\ \bibnamefont
  {Salakhutdinov}},\ }\href@noop {} {\bibfield  {journal} {\bibinfo  {journal}
  {science}\ }\textbf {\bibinfo {volume} {313}},\ \bibinfo {pages} {504}
  (\bibinfo {year} {2006})}\BibitemShut {NoStop}%
\bibitem [{\citenamefont {Pearson}(1901)}]{pearson1901liii}%
  \BibitemOpen
  \bibfield  {author} {\bibinfo {author} {\bibfnamefont {K.}~\bibnamefont
  {Pearson}},\ }\href@noop {} {\bibfield  {journal} {\bibinfo  {journal} {The
  London, Edinburgh, and Dublin Philosophical Magazine and Journal of Science}\
  }\textbf {\bibinfo {volume} {2}},\ \bibinfo {pages} {559} (\bibinfo {year}
  {1901})}\BibitemShut {NoStop}%
\bibitem [{\citenamefont {Abdi}\ and\ \citenamefont
  {Williams}(2010)}]{abdi2010principal}%
  \BibitemOpen
  \bibfield  {author} {\bibinfo {author} {\bibfnamefont {H.}~\bibnamefont
  {Abdi}}\ and\ \bibinfo {author} {\bibfnamefont {L.~J.}\ \bibnamefont
  {Williams}},\ }\href@noop {} {\bibfield  {journal} {\bibinfo  {journal}
  {Wiley interdisciplinary reviews: computational statistics}\ }\textbf
  {\bibinfo {volume} {2}},\ \bibinfo {pages} {433} (\bibinfo {year}
  {2010})}\BibitemShut {NoStop}%
\bibitem [{\citenamefont {Hinrichsen}(2000)}]{hinrichsen2000non}%
  \BibitemOpen
  \bibfield  {author} {\bibinfo {author} {\bibfnamefont {H.}~\bibnamefont
  {Hinrichsen}},\ }\href@noop {} {\bibfield  {journal} {\bibinfo  {journal}
  {Advances in physics}\ }\textbf {\bibinfo {volume} {49}},\ \bibinfo {pages}
  {815} (\bibinfo {year} {2000})}\BibitemShut {NoStop}%
\bibitem [{\citenamefont {L{\"u}beck}(2004)}]{lubeck2004universal}%
  \BibitemOpen
  \bibfield  {author} {\bibinfo {author} {\bibfnamefont {S.}~\bibnamefont
  {L{\"u}beck}},\ }\href@noop {} {\bibfield  {journal} {\bibinfo  {journal}
  {International Journal of Modern Physics B}\ }\textbf {\bibinfo {volume}
  {18}},\ \bibinfo {pages} {3977} (\bibinfo {year} {2004})}\BibitemShut
  {NoStop}%
\bibitem [{\citenamefont {Henkel}\ \emph {et~al.}(2008)\citenamefont {Henkel},
  \citenamefont {Hinrichsen}, \citenamefont {L{\"u}beck},\ and\ \citenamefont
  {Pleimling}}]{henkel2008non}%
  \BibitemOpen
  \bibfield  {author} {\bibinfo {author} {\bibfnamefont {M.}~\bibnamefont
  {Henkel}}, \bibinfo {author} {\bibfnamefont {H.}~\bibnamefont {Hinrichsen}},
  \bibinfo {author} {\bibfnamefont {S.}~\bibnamefont {L{\"u}beck}},\ and\
  \bibinfo {author} {\bibfnamefont {M.}~\bibnamefont {Pleimling}},\ }\href@noop
  {} {\emph {\bibinfo {title} {Non-equilibrium phase transitions}}},\
  Vol.~\bibinfo {volume} {1}\ (\bibinfo  {publisher} {Springer},\ \bibinfo
  {year} {2008})\BibitemShut {NoStop}%
\bibitem [{\citenamefont {Dickman}\ and\ \citenamefont
  {Maia}(2008)}]{dickman2008nature}%
  \BibitemOpen
  \bibfield  {author} {\bibinfo {author} {\bibfnamefont {R.}~\bibnamefont
  {Dickman}}\ and\ \bibinfo {author} {\bibfnamefont {D.~S.}\ \bibnamefont
  {Maia}},\ }\href@noop {} {\bibfield  {journal} {\bibinfo  {journal} {Journal
  of Physics A: Mathematical and Theoretical}\ }\textbf {\bibinfo {volume}
  {41}},\ \bibinfo {pages} {405002} (\bibinfo {year} {2008})}\BibitemShut
  {NoStop}%
\bibitem [{\citenamefont {Mata}(2021)}]{mata2021overview}%
  \BibitemOpen
  \bibfield  {author} {\bibinfo {author} {\bibfnamefont {A.~S.}\ \bibnamefont
  {Mata}},\ }\href@noop {} {\bibfield  {journal} {\bibinfo  {journal} {Chaos:
  An Interdisciplinary Journal of Nonlinear Science}\ }\textbf {\bibinfo
  {volume} {31}},\ \bibinfo {pages} {012101} (\bibinfo {year}
  {2021})}\BibitemShut {NoStop}%
\bibitem [{\citenamefont {Ziff}\ \emph {et~al.}(1986)\citenamefont {Ziff},
  \citenamefont {Gulari},\ and\ \citenamefont {Barshad}}]{ziff1986kinetic}%
  \BibitemOpen
  \bibfield  {author} {\bibinfo {author} {\bibfnamefont {R.~M.}\ \bibnamefont
  {Ziff}}, \bibinfo {author} {\bibfnamefont {E.}~\bibnamefont {Gulari}},\ and\
  \bibinfo {author} {\bibfnamefont {Y.}~\bibnamefont {Barshad}},\ }\href@noop
  {} {\bibfield  {journal} {\bibinfo  {journal} {Physical review letters}\
  }\textbf {\bibinfo {volume} {56}},\ \bibinfo {pages} {2553} (\bibinfo {year}
  {1986})}\BibitemShut {NoStop}%
\bibitem [{\citenamefont {Grinstein}\ \emph {et~al.}(1989)\citenamefont
  {Grinstein}, \citenamefont {Lai},\ and\ \citenamefont
  {Browne}}]{grinstein1989critical}%
  \BibitemOpen
  \bibfield  {author} {\bibinfo {author} {\bibfnamefont {G.}~\bibnamefont
  {Grinstein}}, \bibinfo {author} {\bibfnamefont {Z.-W.}\ \bibnamefont {Lai}},\
  and\ \bibinfo {author} {\bibfnamefont {D.~A.}\ \bibnamefont {Browne}},\
  }\href@noop {} {\bibfield  {journal} {\bibinfo  {journal} {Physical Review
  A}\ }\textbf {\bibinfo {volume} {40}},\ \bibinfo {pages} {4820} (\bibinfo
  {year} {1989})}\BibitemShut {NoStop}%
\bibitem [{\citenamefont {Hammersley}(2013)}]{hammersley2013monte}%
  \BibitemOpen
  \bibfield  {author} {\bibinfo {author} {\bibfnamefont {J.}~\bibnamefont
  {Hammersley}},\ }\href@noop {} {\emph {\bibinfo {title} {Monte carlo
  methods}}}\ (\bibinfo  {publisher} {Springer Science \& Business Media},\
  \bibinfo {year} {2013})\BibitemShut {NoStop}%
\bibitem [{\citenamefont {Voulodimos}\ \emph {et~al.}(2018)\citenamefont
  {Voulodimos}, \citenamefont {Doulamis}, \citenamefont {Doulamis},
  \citenamefont {Protopapadakis} \emph {et~al.}}]{voulodimos2018deep}%
  \BibitemOpen
  \bibfield  {author} {\bibinfo {author} {\bibfnamefont {A.}~\bibnamefont
  {Voulodimos}}, \bibinfo {author} {\bibfnamefont {N.}~\bibnamefont
  {Doulamis}}, \bibinfo {author} {\bibfnamefont {A.}~\bibnamefont {Doulamis}},
  \bibinfo {author} {\bibfnamefont {E.}~\bibnamefont {Protopapadakis}}, \emph
  {et~al.},\ }\href@noop {} {\bibfield  {journal} {\bibinfo  {journal}
  {Computational intelligence and neuroscience}\ }\textbf {\bibinfo {volume}
  {2018}} (\bibinfo {year} {2018})}\BibitemShut {NoStop}%
\bibitem [{\citenamefont {Freitag}\ and\ \citenamefont
  {Roy}(2018)}]{freitag2018unsupervised}%
  \BibitemOpen
  \bibfield  {author} {\bibinfo {author} {\bibfnamefont {M.}~\bibnamefont
  {Freitag}}\ and\ \bibinfo {author} {\bibfnamefont {S.}~\bibnamefont {Roy}},\
  }\href@noop {} {\bibfield  {journal} {\bibinfo  {journal} {arXiv preprint
  arXiv:1804.07899}\ } (\bibinfo {year} {2018})}\BibitemShut {NoStop}%
\bibitem [{\citenamefont {Chen}\ \emph {et~al.}(2018)\citenamefont {Chen},
  \citenamefont {Yeo}, \citenamefont {Lee},\ and\ \citenamefont
  {Lau}}]{chen2018autoencoder}%
  \BibitemOpen
  \bibfield  {author} {\bibinfo {author} {\bibfnamefont {Z.}~\bibnamefont
  {Chen}}, \bibinfo {author} {\bibfnamefont {C.~K.}\ \bibnamefont {Yeo}},
  \bibinfo {author} {\bibfnamefont {B.~S.}\ \bibnamefont {Lee}},\ and\ \bibinfo
  {author} {\bibfnamefont {C.~T.}\ \bibnamefont {Lau}},\ }in\ \href@noop {}
  {\emph {\bibinfo {booktitle} {2018 Wireless telecommunications symposium
  (WTS)}}}\ (\bibinfo {organization} {IEEE},\ \bibinfo {year} {2018})\ pp.\
  \bibinfo {pages} {1--5}\BibitemShut {NoStop}%
\bibitem [{\citenamefont {Alexandrou}\ \emph {et~al.}(2019)\citenamefont
  {Alexandrou}, \citenamefont {Athenodorou}, \citenamefont {Chrysostomou},\
  and\ \citenamefont {Paul}}]{alexandrou2019unsupervised}%
  \BibitemOpen
  \bibfield  {author} {\bibinfo {author} {\bibfnamefont {C.}~\bibnamefont
  {Alexandrou}}, \bibinfo {author} {\bibfnamefont {A.}~\bibnamefont
  {Athenodorou}}, \bibinfo {author} {\bibfnamefont {C.}~\bibnamefont
  {Chrysostomou}},\ and\ \bibinfo {author} {\bibfnamefont {S.}~\bibnamefont
  {Paul}},\ }\href@noop {} {\bibfield  {journal} {\bibinfo  {journal} {arXiv
  preprint arXiv:1903.03506}\ } (\bibinfo {year} {2019})}\BibitemShut {NoStop}%
\bibitem [{\citenamefont {Goodfellow}\ \emph
  {et~al.}(2016{\natexlab{b}})\citenamefont {Goodfellow}, \citenamefont
  {Bengio},\ and\ \citenamefont {Courville}}]{goodfellow2016deep}%
  \BibitemOpen
  \bibfield  {author} {\bibinfo {author} {\bibfnamefont {I.}~\bibnamefont
  {Goodfellow}}, \bibinfo {author} {\bibfnamefont {Y.}~\bibnamefont {Bengio}},\
  and\ \bibinfo {author} {\bibfnamefont {A.}~\bibnamefont {Courville}},\
  }\href@noop {} {\emph {\bibinfo {title} {Deep learning}}}\ (\bibinfo
  {publisher} {MIT press},\ \bibinfo {year} {2016})\BibitemShut {NoStop}%
\bibitem [{\citenamefont {Hochreiter}\ and\ \citenamefont
  {Schmidhuber}(1997)}]{hochreiter1997long}%
  \BibitemOpen
  \bibfield  {author} {\bibinfo {author} {\bibfnamefont {S.}~\bibnamefont
  {Hochreiter}}\ and\ \bibinfo {author} {\bibfnamefont {J.}~\bibnamefont
  {Schmidhuber}},\ }\href@noop {} {\bibfield  {journal} {\bibinfo  {journal}
  {Neural computation}\ }\textbf {\bibinfo {volume} {9}},\ \bibinfo {pages}
  {1735} (\bibinfo {year} {1997})}\BibitemShut {NoStop}%
\bibitem [{\citenamefont {Glassner}(2018)}]{glassner2018deep}%
  \BibitemOpen
  \bibfield  {author} {\bibinfo {author} {\bibfnamefont {A.}~\bibnamefont
  {Glassner}},\ }\href@noop {} {\bibfield  {journal} {\bibinfo  {journal}
  {Seattle, WA, USA: The Imaginary Institute}\ } (\bibinfo {year}
  {2018})}\BibitemShut {NoStop}%
\bibitem [{\citenamefont {Hu}\ \emph {et~al.}(2017)\citenamefont {Hu},
  \citenamefont {Singh},\ and\ \citenamefont {Scalettar}}]{hu2017discovering}%
  \BibitemOpen
  \bibfield  {author} {\bibinfo {author} {\bibfnamefont {W.}~\bibnamefont
  {Hu}}, \bibinfo {author} {\bibfnamefont {R.~R.}\ \bibnamefont {Singh}},\ and\
  \bibinfo {author} {\bibfnamefont {R.~T.}\ \bibnamefont {Scalettar}},\
  }\href@noop {} {\bibfield  {journal} {\bibinfo  {journal} {Physical Review
  E}\ }\textbf {\bibinfo {volume} {95}},\ \bibinfo {pages} {062122} (\bibinfo
  {year} {2017})}\BibitemShut {NoStop}%
\bibitem [{\citenamefont {Cohen}\ \emph {et~al.}(2009)\citenamefont {Cohen},
  \citenamefont {Huang}, \citenamefont {Chen}, \citenamefont {Benesty},
  \citenamefont {Benesty}, \citenamefont {Chen}, \citenamefont {Huang},\ and\
  \citenamefont {Cohen}}]{cohen2009pearson}%
  \BibitemOpen
  \bibfield  {author} {\bibinfo {author} {\bibfnamefont {I.}~\bibnamefont
  {Cohen}}, \bibinfo {author} {\bibfnamefont {Y.}~\bibnamefont {Huang}},
  \bibinfo {author} {\bibfnamefont {J.}~\bibnamefont {Chen}}, \bibinfo {author}
  {\bibfnamefont {J.}~\bibnamefont {Benesty}}, \bibinfo {author} {\bibfnamefont
  {J.}~\bibnamefont {Benesty}}, \bibinfo {author} {\bibfnamefont
  {J.}~\bibnamefont {Chen}}, \bibinfo {author} {\bibfnamefont {Y.}~\bibnamefont
  {Huang}},\ and\ \bibinfo {author} {\bibfnamefont {I.}~\bibnamefont {Cohen}},\
  }\href@noop {} {\bibfield  {journal} {\bibinfo  {journal} {Noise reduction in
  speech processing}\ ,\ \bibinfo {pages} {1}} (\bibinfo {year}
  {2009})}\BibitemShut {NoStop}%
\bibitem [{\citenamefont {Danielsson}(1980)}]{danielsson1980euclidean}%
  \BibitemOpen
  \bibfield  {author} {\bibinfo {author} {\bibfnamefont {P.-E.}\ \bibnamefont
  {Danielsson}},\ }\href@noop {} {\bibfield  {journal} {\bibinfo  {journal}
  {Computer Graphics and image processing}\ }\textbf {\bibinfo {volume} {14}},\
  \bibinfo {pages} {227} (\bibinfo {year} {1980})}\BibitemShut {NoStop}%
\end{thebibliography}%

\end{document}